\documentstyle[aps,epsfig,floats]{revtex}

\tighten
\newcommand{\beq}{\begin{equation}}
\newcommand{\eeq}{\end{equation}}
\newcommand{\ba}{\begin{eqnarray}}
\newcommand{\ea}{\end{eqnarray}}

\newcommand{\xbj}{x_{\scriptscriptstyle B}}

\newcommand{\bm}[1]{\bbox{#1}}

\newcommand{\st}{{\scriptscriptstyle T}}
\newcommand{\sL}{{\scriptscriptstyle L}}
\newcommand{\be}{\begin{equation}}
\newcommand{\ee}{\end{equation}}
\newcommand{\bea}{\begin{eqnarray}}
\newcommand{\eea}{\end{eqnarray}}

\def\slash{\rlap{/}}

\begin{document}

\draft
\title{
\begin{flushright}
\begin{minipage}{4 cm}
\small
hep-ph/0303034
\end{minipage}
\end{flushright}
Universality of T-odd effects in single spin and azimuthal asymmetries}

\author{
D. Boer,
P.J. Mulders
and F.\ Pijlman}
\address{\mbox{}\\
Department of Physics and Astronomy, Vrije Universiteit \\
De Boelelaan 1081, NL-1081 HV Amsterdam, the Netherlands\\
}

\maketitle

\begin{abstract}
We analyze the transverse momentum dependent distribution and
fragmentation functions in space-like and time-like
hard processes involving at least two
hadrons, in particular 1-particle inclusive leptoproduction, the Drell-Yan
process and two-particle inclusive hadron production in electron-positron
annihilation. As is well-known, transverse momentum dependence allows for the 
appearance
of unsuppressed single spin azimuthal asymmetries, such as Sivers and
Collins asymmetries.
Recently, Belitsky, Ji and Yuan obtained fully color gauge invariant 
expressions for
the relevant matrix elements appearing in these asymmetries at leading order
in an expansion in the inverse hard scale.
We rederive these results and extend them to observables at the next order 
in this expansion.
We observe that at leading order one retains a probability
interpretation, contrary to a claim in the literature and
show the direct relation between
the Sivers effect in single spin asymmetries and the Qiu-Sterman
mechanism. We also study fragmentation functions, where the process
dependent gauge link structure of the correlators is not the only
source of T-odd observables and discuss the implications for universality.
\end{abstract}


\section{Introduction}

The study of polarization and transverse momentum dependent distribution
functions was initiated by Ralston and Soper~\cite{RS-79}. Their study of the
Drell-Yan process was performed at tree level and did not address the
color gauge (non-)invariance of the distribution functions.
At leading order (tree level) no single spin asymmetries were obtained
in the Drell-Yan process (see also \cite{Tangerman-Mulders-95a}).
Sivers~\cite{s90} proposed a specific non-trivial correlation involving
polarization and transverse momentum, that would lead to unsuppressed
single spin azimuthal asymmetries.
For distribution functions such a correlation seemed to entail a violation of
time reversal invariance. Collins~\cite{Collins-93b} showed that this was
not the case for similar correlations in the fragmentation process. 
Nevertheless, phenomenological studies of the
consequences of the Sivers effect were performed
\cite{Anselmino,Boer-Mulders-98}. Recently,
Brodsky, Hwang and Schmidt~\cite{BHS} (BHS) demonstrated in an explicit model
calculation that the Sivers asymmetry can in principle arise, after which
Collins~\cite{Collins-02} demonstrated that it is the presence of a
path-ordered exponential in the definition of transverse momentum dependent
distribution functions that allows for the Sivers effect without a violation
of time reversal invariance.

This generated renewed interest in the proper gauge invariant definition
of transverse momentum dependent correlators. The definitions of
transverse momentum dependent parton densities of Ref.~\cite{CS82,CSS83} did 
contain path-ordered exponentials (links) to ensure
color gauge invariance, but these were not closed paths (each quark field has
a straight link to infinity attached to it, but pieces at
infinity are missing). 
If one includes such links by hand, it is no problem to consider also closed
paths, but Efremov and Radyushkin~\cite{Rad-Efr} had
demonstrated that the path of the link can be derived in
transverse momentum integrated parton densities and this can also be done
when the transverse momentum is not integrated over. In this way
different processes can yield different paths \cite{Boer-Mulders-00}, 
but no physical observable effects were expected from such links.
However, until recently these derivations were incomplete, since the obtained 
paths were not closed. The missing piece would have to involve transverse
gluon fields at lightcone infinity, which were thought not to affect physical
matrix elements or at the very least lead to contributions suppressed
compared to the leading order. 
Recently, the derivation of fully color gauge invariant matrix elements,
with paths closed at light-cone infinity, was completed
by Belitsky, Ji and Yuan~\cite{Ji-Yuan,Belitsky}. They observed that
transverse gluon fields do not always lead to suppression, contrary to
common belief, formalizing the model results of BHS.
The resulting fully color gauge invariant matrix elements strengthen the
observation of Collins~\cite{Collins-02} that the presence of the link
invalidated the earlier proof of the absence of the Sivers function due to
time reversal invariance.

With all these technical details clarified, the justification of the
phenomenological studies of the Sivers (and similar) effects was provided.
Next, however, the question of observable process-dependence arose.
Collins~\cite{Collins-02} demonstrated that the Sivers asymmetry in
(semi-inclusive)
deep inelastic scattering (DIS) and the Drell-Yan process must occur with
opposite signs. This has been confirmed in the BHS model calculation
\cite{BHS2}, but
still awaits experimental verification. It would be the first
demonstration of an observable effect due to the presence of a
path-ordered exponential in the hadron correlators and thereby
would show the intrinsic non-locality of the operators occuring in these
semi-inclusive processes.

Given this process dependence it is relevant to study which color gauge
invariant distribution and fragmentation functions appear in different
processes.
In this paper we study color gauge invariance of transverse momentum
dependent distribution and fragmentation functions appearing
in hard processes, in particular in semi-inclusive deep inelastic
leptoproduction (SIDIS), the Drell-Yan process (DY) and $e^+e^-$-annihilation.

We will employ a field theoretical approach to these hard processes and
follow the notation and derivation of Ref.~\cite{Boer-Mulders-00}, now taking
into account the additional contributions uncovered by Belitsky, Ji and
Yuan~\cite{Belitsky}.
Our analysis is different at several points, but we confirm their
results. In addition, we obtain new results for the first sub-leading order
results
in an expansion in inverse powers of the hard scale.  For instance, we
demonstrate for the first time a direct relation between the Sivers effect
in single spin asymmetries and the Qiu-Sterman mechanism. Also, we study
transverse momentum dependent fragmentation functions,
where the process dependence of the gauge link
structure of the correlators is not simply an overall sign.
Rather one finds that two different (but universal) matrix elements enter
in different combinations.

In this paper 
`leading' and `sub-leading' always refer to
the expansion in inverse powers of the hard scale.
Perturbative QCD corrections beyond tree level
(next-to-leading order in $\alpha_s$) will need to be taken into account
as well in further studies (see Refs.\
\cite{CSS83,Collins-93b,Boer-Sudakov} 
for discussions of additional complications beyond tree level). In this
paper we present a new way of isolating the leading and first sub-leading
order parts of the cross sections in terms of correlators including the
proper gauge links {\em before\/} evaluating them explicitly. These separate
color gauge invariant expressions for each order have not been presented
before. They facilitate
the evaluation of asymmetries arising at a given order. 

\begin{figure}[b]
\begin{center}
\epsfig{file=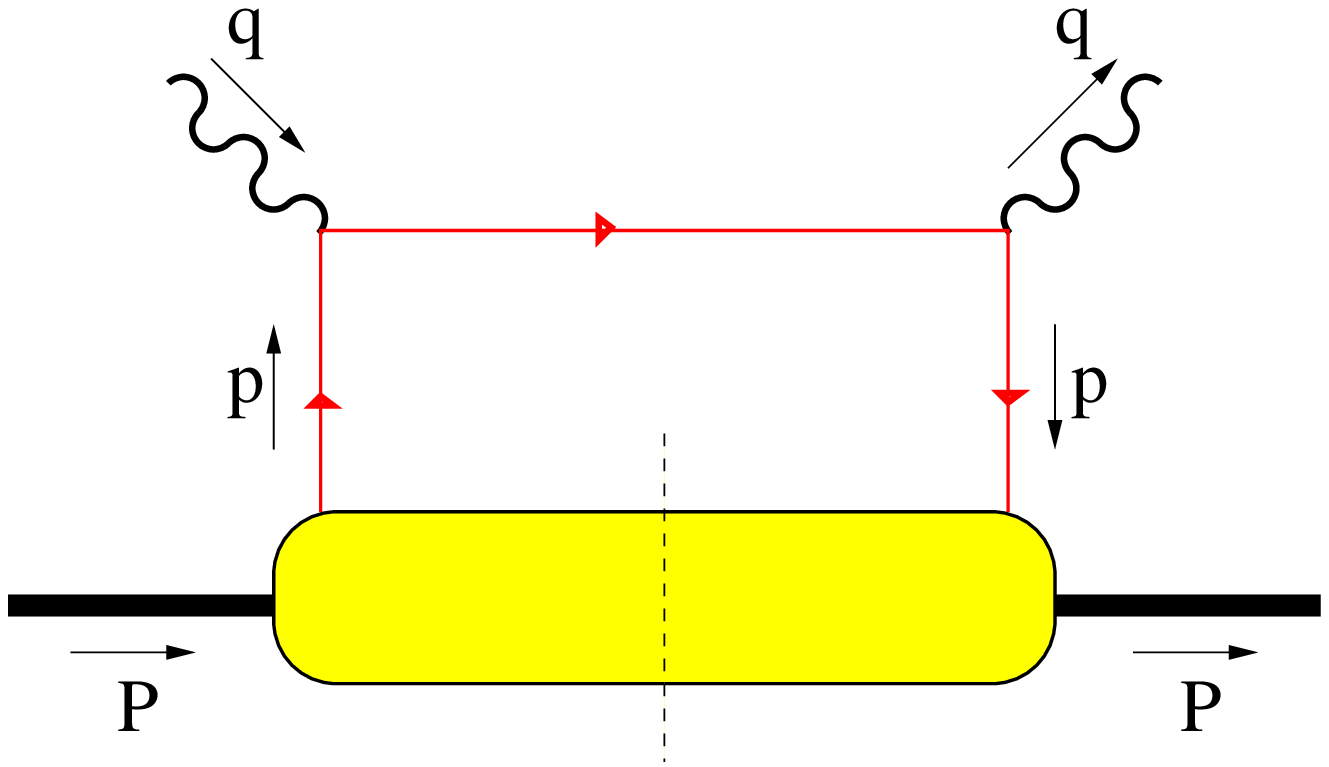, width=5cm}
\hspace{2cm}
\epsfig{file=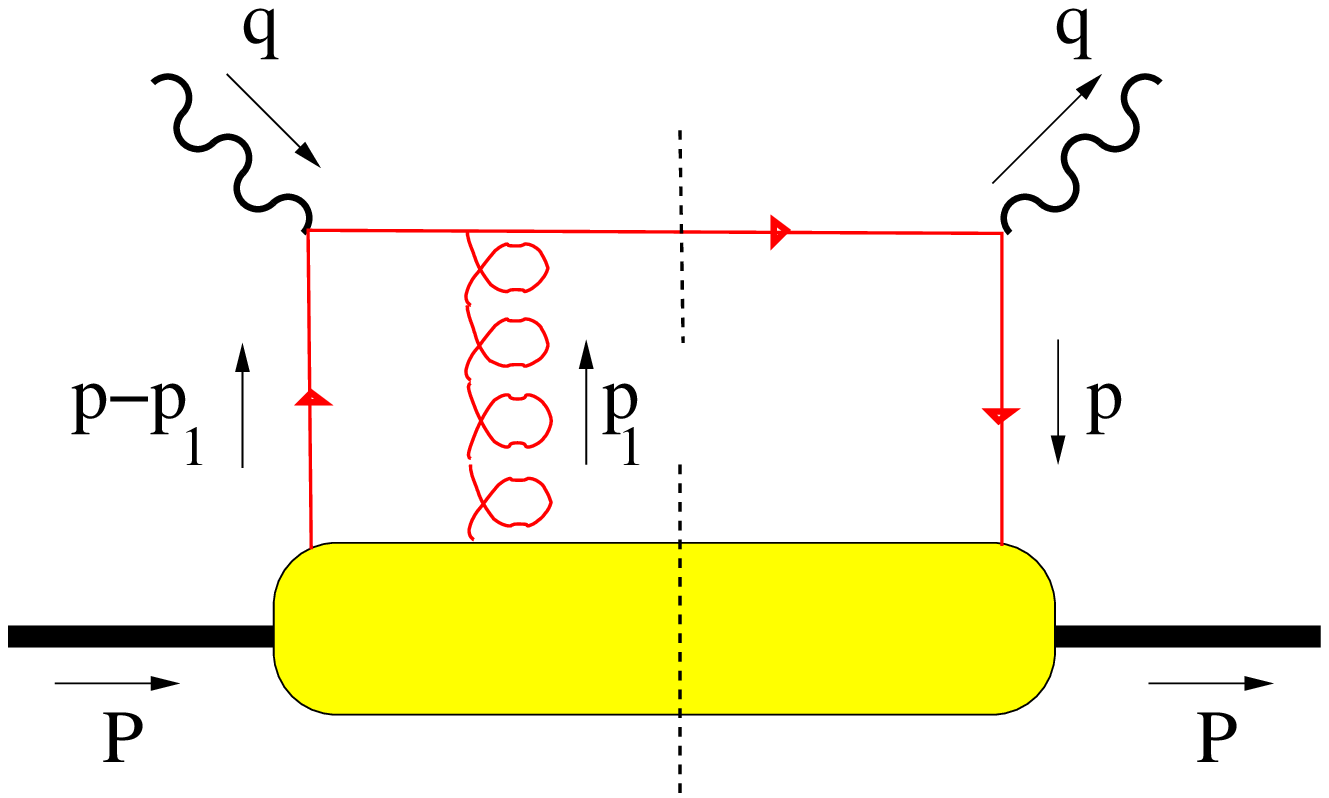, width=5cm}
\\
(a) \hspace{7cm} (b)
\\ \mbox{}
\caption{
Diagrams contributing in inclusive deep inelastic scattering}
\label{fig1}
\end{center}
\end{figure}

\begin{figure}[t]
\begin{center}
\epsfig{file=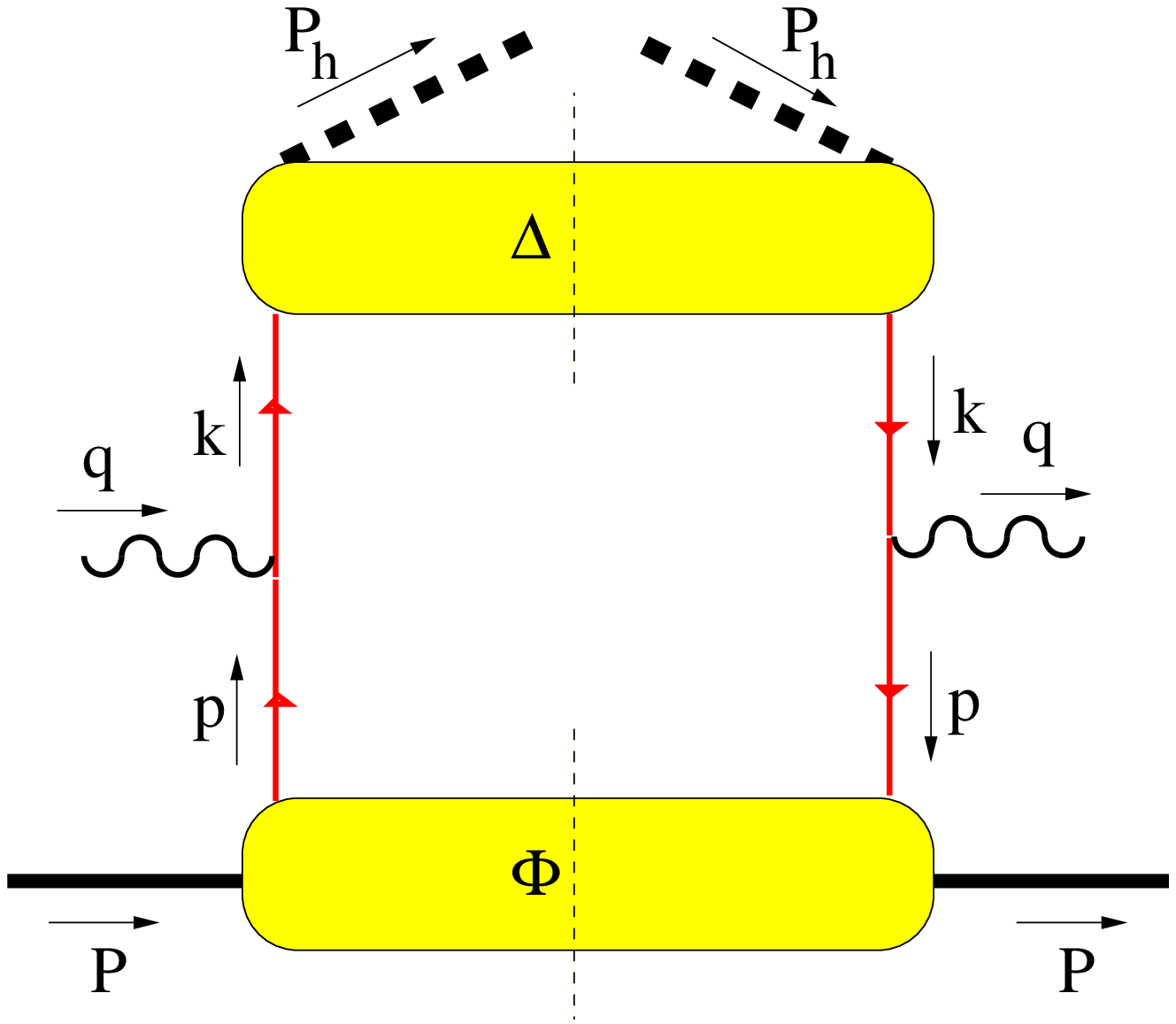, width=5cm}
\hspace{0.5cm}
\epsfig{file=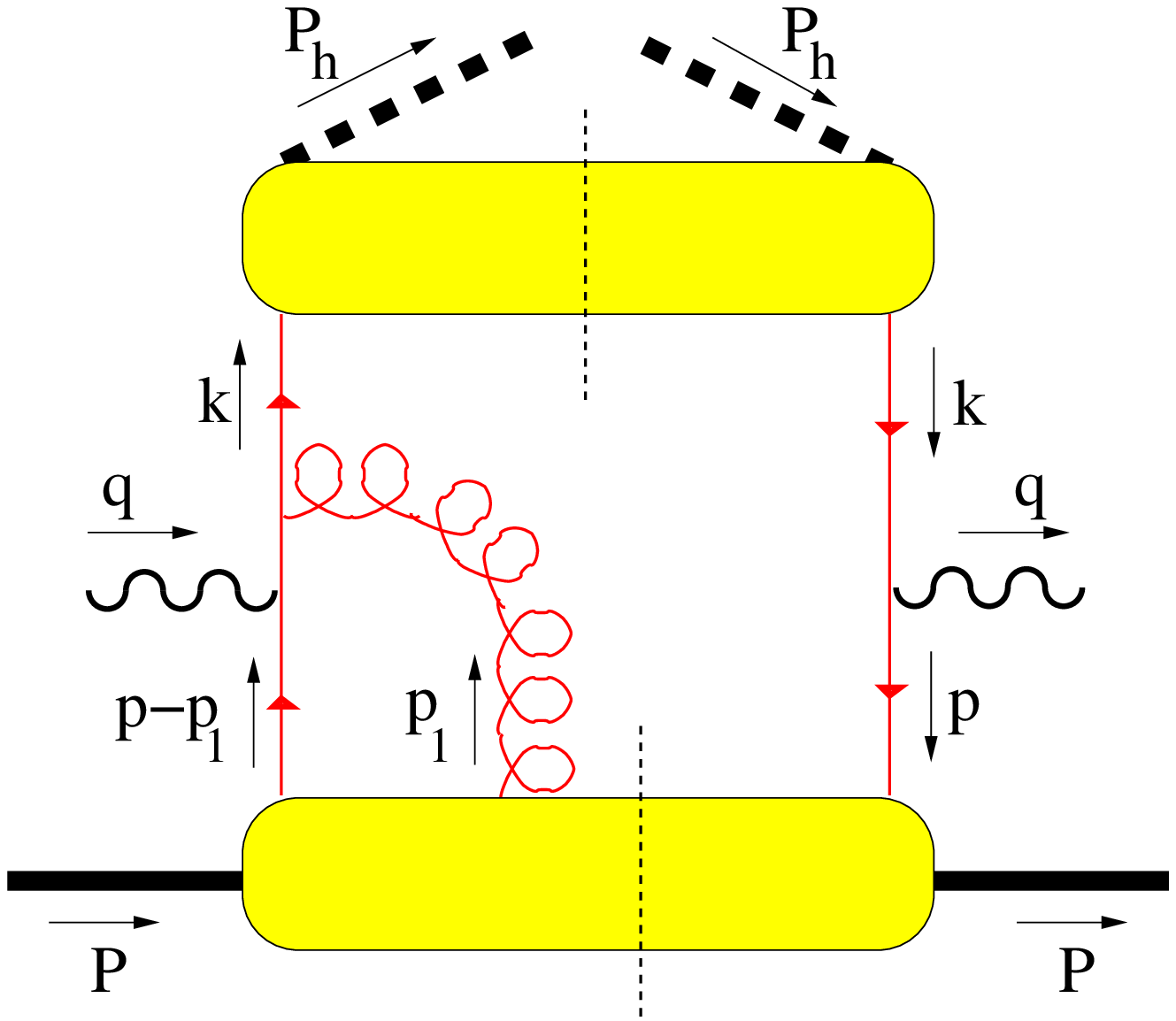, width=5cm}
\hspace{0.5cm}
\epsfig{file=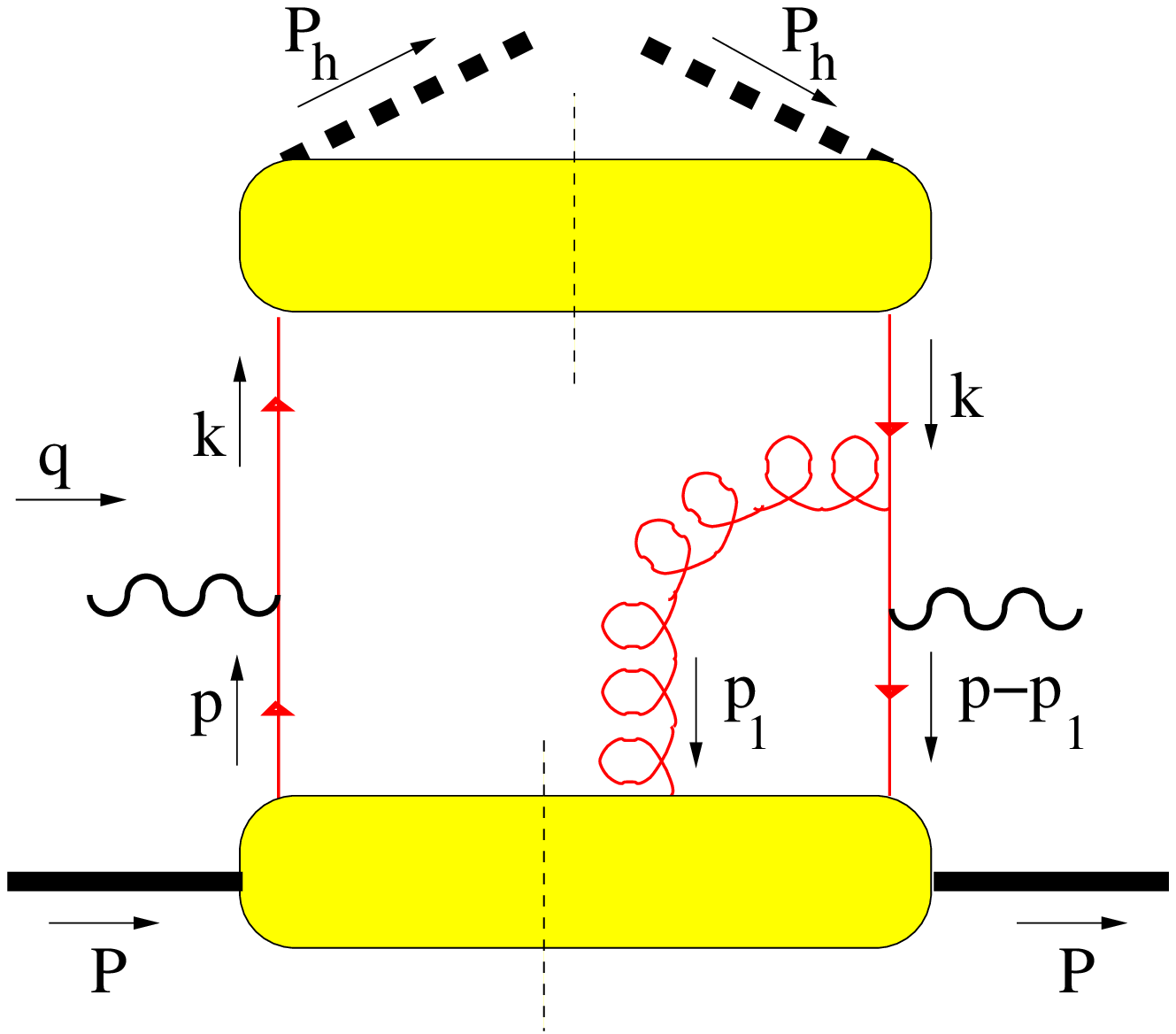, width=5cm}
\\
(a) \hspace{5cm} (b) \hspace{5cm} (c)
\\[0.3cm]
\epsfig{file=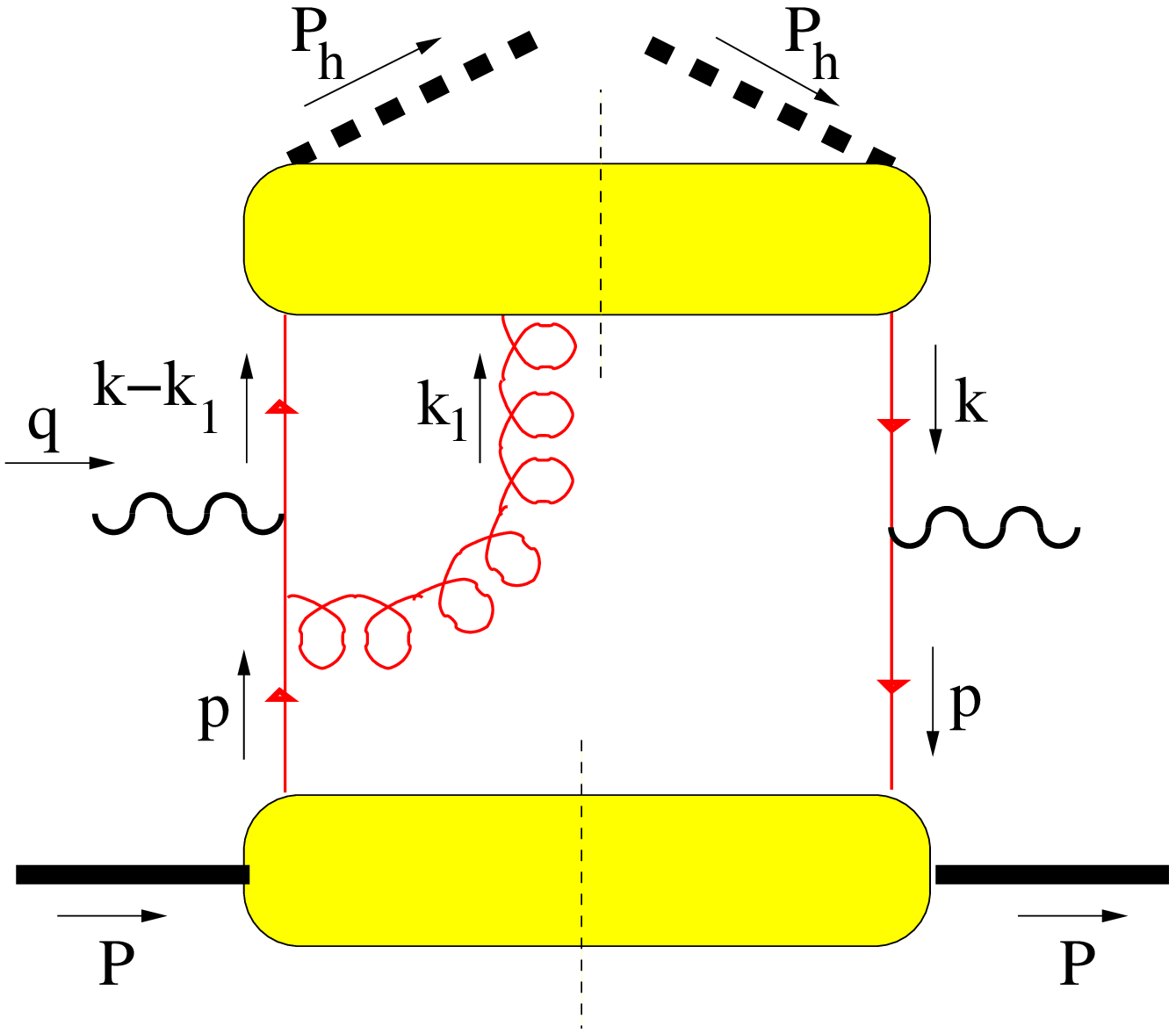, width=5cm}
\hspace{0.5cm}
\epsfig{file=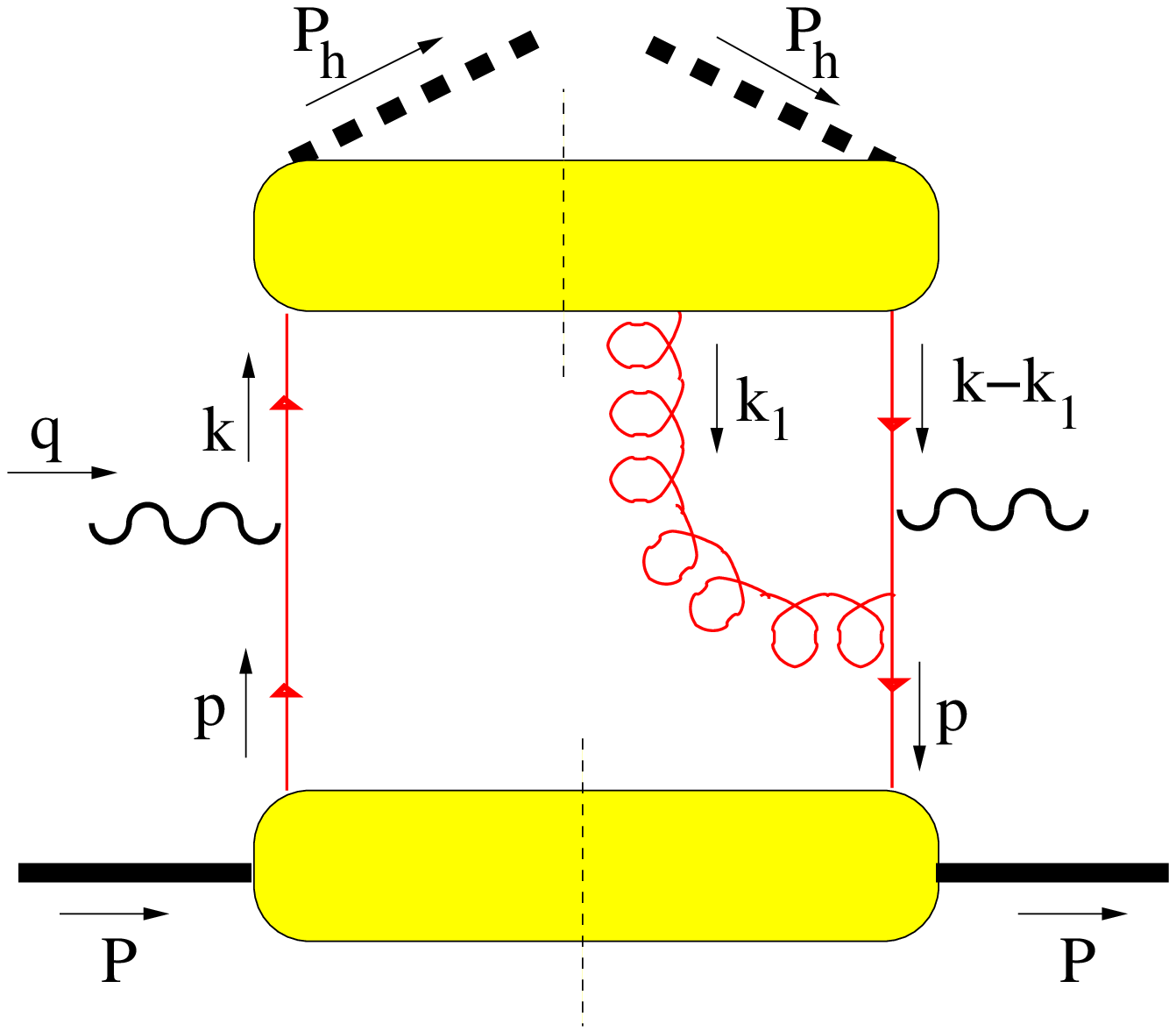, width=5cm}
\hspace{0.5cm}
\epsfig{file=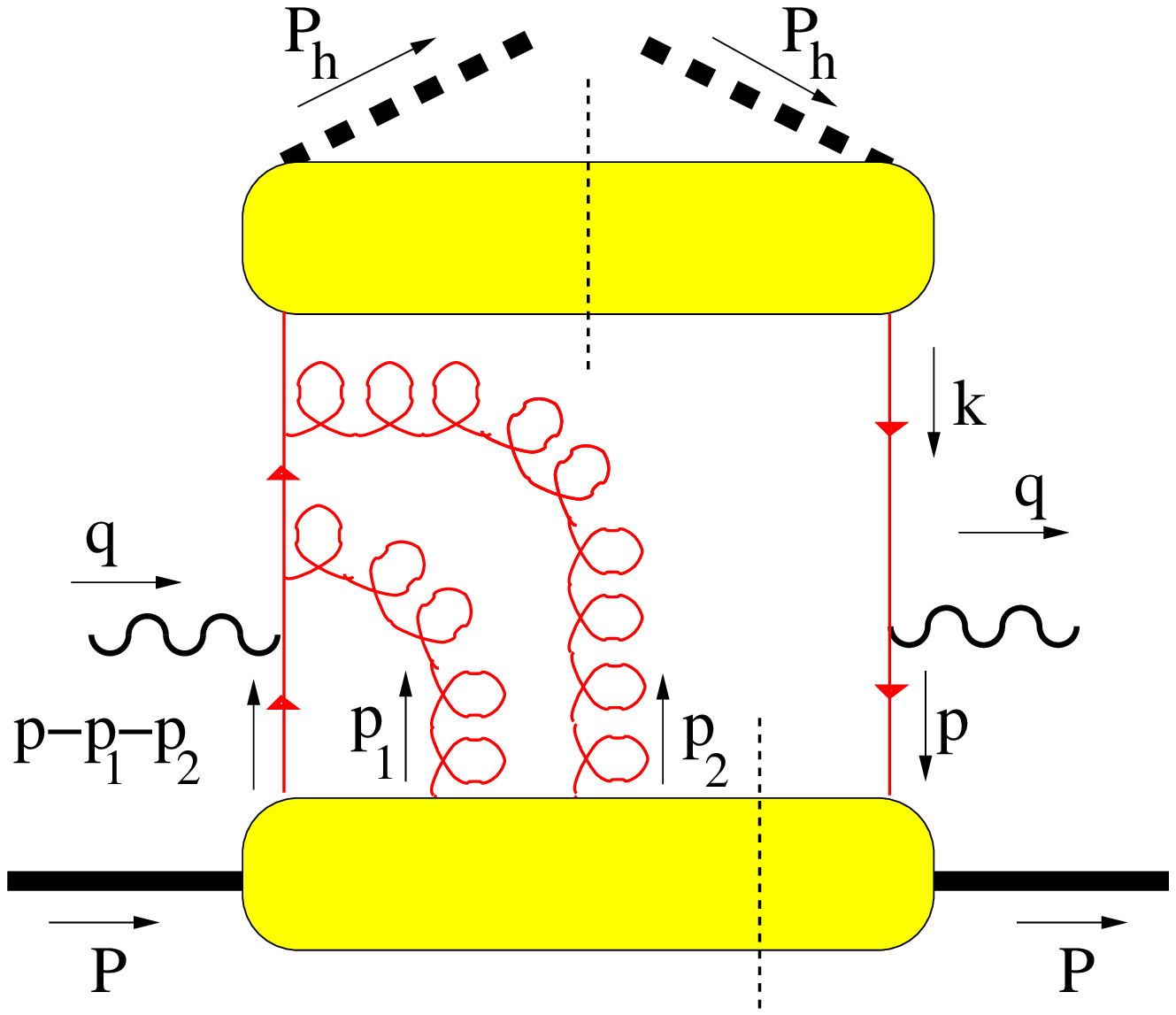, width=5cm}
\\
(d) \hspace{5cm} (e) \hspace{5cm} (f)
\\ \mbox{}
\caption{
Diagrams contributing in 1-particle inclusive deep inelastic scattering}
\label{fig2}
\end{center}
\end{figure}

Now we will outline the more technical steps to be followed in this paper.
In the field theoretical approach, expressions for the structure functions 
of inclusive
deep inelastic scattering (DIS) are obtained from diagrams as shown in
Fig.~\ref{fig1}~\cite{EFP-83,Efremov-Teryaev-84}.
In these diagrams soft parts appear that
represent matrix elements of the fields corresponding to the
quark and gluon legs connecting the hard and soft parts of a diagram.
The expressions for SIDIS structure functions are obtained from diagrams
as shown in Fig.~\ref{fig2}. Again soft parts represent specific
matrix elements. In this paper we will only consider tree-level results,
which means that if gluons appear, they are in essence legs of the
soft parts. In other words, their (soft) couplings to the hard scattering part
are included in the definition of the matrix elements.
This approach requires a careful treatment of
the diagrams involving quark-quark-gluon
matrix elements such as those in Fig.~1b or Figs~2b - 2f.
This is important in order to arrive at color gauge invariant matrix
elements that form the universal quantities, the distribution and
fragmentation functions, appearing in cross sections.

In section II we outline the diagrammatic approach in
a number of steps, using two complementary lightcone directions $n_+$ and
$n_-$, which in the presence of a hard scale (in DIS or SIDIS,
the photon momentum) are fixed by the hadron momenta.
The simplest (handbag) diagrams in Figs~1a and 2a
only involve quark-quark matrix elements. In DIS
the hadron momentum defines the lightcone direction $n_+$
and the nonlocality in the matrix elements is restricted along the
lightcone direction $n_-$ (for which $n_+\cdot n_-$ = 1).
As is well-known, diagrams as in Fig.~1b with any
number of $A^+ = A\cdot n_-$ gluons yield the
necessary gauge link connecting the two quark fields \cite{Rad-Efr}. 
The nonlocal quark-quark
operator combination with a gauge link can be expanded into a tower
of local twist-two operators with different spins. Their matrix elements
appear in the cross section as leading terms in an expansion in inverse 
powers of the hard scale. Diagrams with
(transverse) $A_\st^\alpha$ gluons or with $A^-$ gluons appear in
matrix elements of higher twist operators, which appear in the cross
section in terms suppressed by inverse powers of the hard scale.

The situation in SIDIS (Figs.~\ref{fig2}), discussed in section III,
differs in a subtle way from
that of DIS, because the nonlocality in the operator combinations is not
restricted to the lightcone, but involves also transverse separations.
The kinematics only constrain the nonlocality to the light{\em front\/}. 
In our
analysis we first consider the $A^+$ gluon legs in diagrams as in
Fig.~\ref{fig2}b and \ref{fig2}f.
These diagrams, as in DIS, will give rise to gauge links, but in this case
connecting a quark field along the $n_-$ direction to $\pm \infty$
(cf.~\cite{CSS83,Boer-Mulders-00}), where the sign depends on the type
of process.
By including diagrams of the type in Fig.~\ref{fig2}c as well, one can
absorb all $A^+$ gluons into the lower blob. Diagrams like
Figs.~\ref{fig2}d and \ref{fig2}e
allow one to absorb all $A^-$ gluons into the upper blob, resulting again in
gauge links running along $n_+$ to infinity. Effectively one
then considers only Fig.~\ref{fig2}a, but now with a $\Phi$ and $\Delta$
that contain the gauge links.

Diagrams with transverse $A_\st^\alpha$-legs, lead to quark-quark-gluon
matrix elements, which will turn out to be suppressed, except for the boundary
terms at lightcone infinity, recently discussed by Belitsky {\em
et al.}~\cite{Belitsky}. We outline an alternative for this procedure
and show that the latter appear when one expresses these
fields in the appropriate field strength tensor $G^{+\alpha}$.
The boundary terms that arise in this way combine into
the transverse piece that completes the gauge link connecting the two
quark fields (running via lightcone infinity). Upon integration over
transverse momenta the result reduces to the correct gauge invariant operator
of ordinary inclusive DIS. 

In sections IV-VI a comparison is made between different processes involving
at least two hadrons, in particular between 1-particle inclusive
leptoproduction, the Drell-Yan 
process and two-particle inclusive hadron production in electron-positron
annihilation.
For instance, in Drell-Yan the links run in opposite directions along the 
$n_-$ direction compared to SIDIS (as noticed in 
\cite{Boer-Mulders-00,Collins-02}).
Because the two situations can be connected via a time reversal operation,
one can define T-even and T-odd functions that appear in the parametrization
of the color gauge invariant matrix elements. For these functions
factorization in principle should hold, although they appear with different
signs in SIDIS and DY \cite{Collins-02}. 
The T-odd functions appear in single spin asymmetries in these
processes~\cite{s90,Collins-93b,MT96,Boer-Mulders-98}
or they appear in pairs in unpolarized
azimuthal asymmetries~\cite{Boer-97,Boer-Brodsky-Hwang,Gamberg}.
In section VII we study the
time reversal properties of distribution and fragmentation functions and
present explicit parametrizations. 

\section{Hadron tensor and correlators in SIDIS}

The hadron tensor for 1-particle inclusive leptoproduction is given by
\begin{equation}
2 M {\cal W}^{(lH)}_{\mu\nu}(q; P,S;P_h,S_h) = \frac{1}{(2\pi)^4}
\int \frac{{\rm d}^3 P_X}{(2\pi)^3 2P_X^0}
(2\pi)^4 \delta^4(q+P-P_X-P_h) H^{lH}_{\mu\nu}(P_X;PS;P_h S_h).
\end{equation}
with $H_{\mu\nu}$ being the product of current expectation values
\bea
H_{\mu\nu}^{(lH)}(P_X;P,S;P_h, S_h) &=&
\langle P,S|J_\mu (0) |P_X;P_h S_h\rangle
\langle P_X;P_h S_h| J_\nu (0) | P,S\rangle .
\eea
Due to the fact that the summation and integration over final states is not 
complete,
prohibiting the formal use of the operator product expansion, we proceed
along the lines of the diagrammatic approach of Refs.\ \cite{RS-79,EFP-83}, 
based on nonlocal operators. 
The quark and gluon
lines connected to the soft parts represent matrix elements of (nonlocal) 
quark and gluon operators.

The hadron tensor is calculated for current fragmentation
in deep inelastic scattering. In that case the exchanged
momentum $-q^2 \equiv Q^2$ is large and one has for the
target momentum $P$ and the produced hadron momentum $P_h$ the conditions
that $P\cdot q$, $P_h\cdot q$ and $P\cdot P_h$
are large, of ${\cal O}(Q^2)$.
One is able to make a systematic expansion in orders of $1/Q$, of
which we will only consider the first two terms, $(1/Q)^0$ and $(1/Q)^1$.
In this situation one uses the scaling variables
\bea
\xbj & = & \frac{Q^2}{2P\cdot q} \approx -\frac{P_h\cdot q}{P_h\cdot P},
\\
z_h & = & \frac{P\cdot P_h}{P\cdot q} \approx -\frac{2P_h\cdot q}{Q^2}
\eea
where the approximate sign indicates equalities up to $1/Q^2$ (mass)
corrections.
It is convenient to introduce lightlike vectors $n_+$ and $n_-$  satisfying
$n_+\cdot n_- = 1$ along the hadron momenta writing
\bea
P^\mu & = &
\frac{ \xi\,M^2}{\tilde Q\sqrt{2}}\,n_-^\mu
+ \frac{\tilde Q}{\xi\sqrt{2}}\,n_+^\mu,
\label{sudakov-1}
\\
P_h^\mu & = &
\frac{\zeta\,\tilde Q}{\sqrt{2}}\,n_-^\mu
+ \frac{M_h^2}{\zeta \tilde Q\sqrt{2}}\,n_+^\mu,
\label{sudakov-2}
\\
q^\mu & = &
\frac{\tilde Q}{\sqrt{2}}\,n_-^\mu
- \frac{\tilde Q}{\sqrt{2}}\,n_+^\mu
+ q_\st^\mu,
\label{sudakov-3}
\eea
with $q_\st^2 \equiv -Q_\st^2$ and $\tilde Q^2 = Q^2 + Q_\st^2$.
These equations define the lightcone coordinates $a^\pm \equiv a\cdot n_\mp$
and the transverse projector $g_\st^{\mu\nu} = g^{\mu\nu} - n_+^{\{\mu}
n_-^{\nu\}}$. In our treatment of the 1-particle inclusive process
we will consider $Q_\st^2 \ll Q^2$, hence $\tilde Q^2 \approx Q^2$, while
$\xi \approx \xbj$ and $\zeta \approx z_h$ up to mass corrections of
order $1/Q^2$. The vector $q_\st^\mu \approx  q^\mu + x\,P^\mu
- P_h^\mu/z$ determines the off-collinearity in the process.
In principle, mass corrections can
straightforwardly be incorporated. Important to note is that the
lightlike directions
$n_\pm$ = $n_\pm(P,P_h)$ are determined by the hadron momenta
$P$ and $P_h$.

\begin{figure}[t]
\begin{center}
\epsfig{file=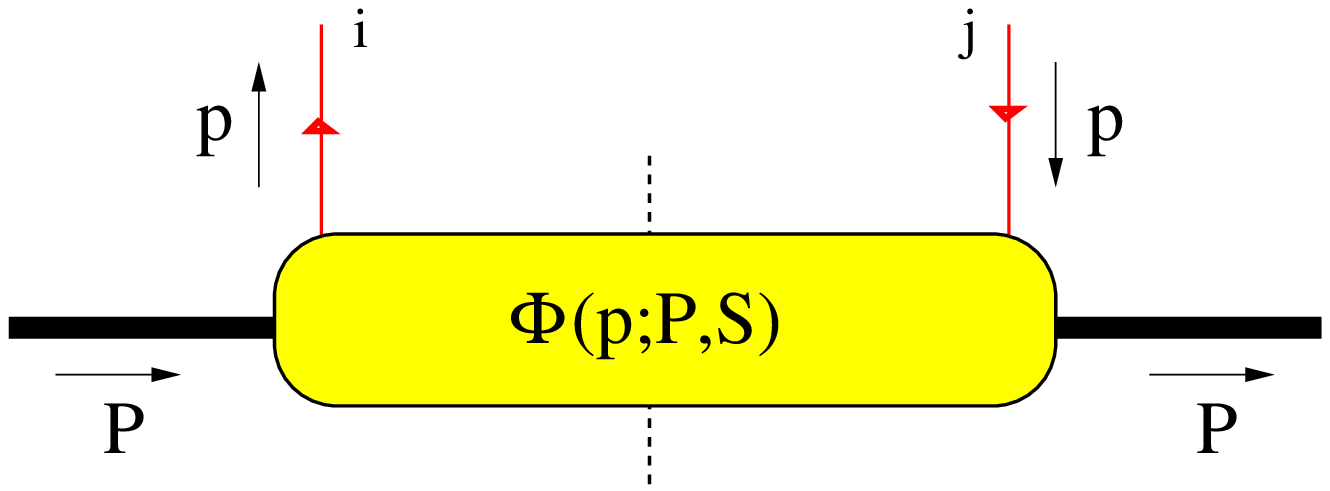, width=5.5cm}
\hspace{2.0cm}
\epsfig{file=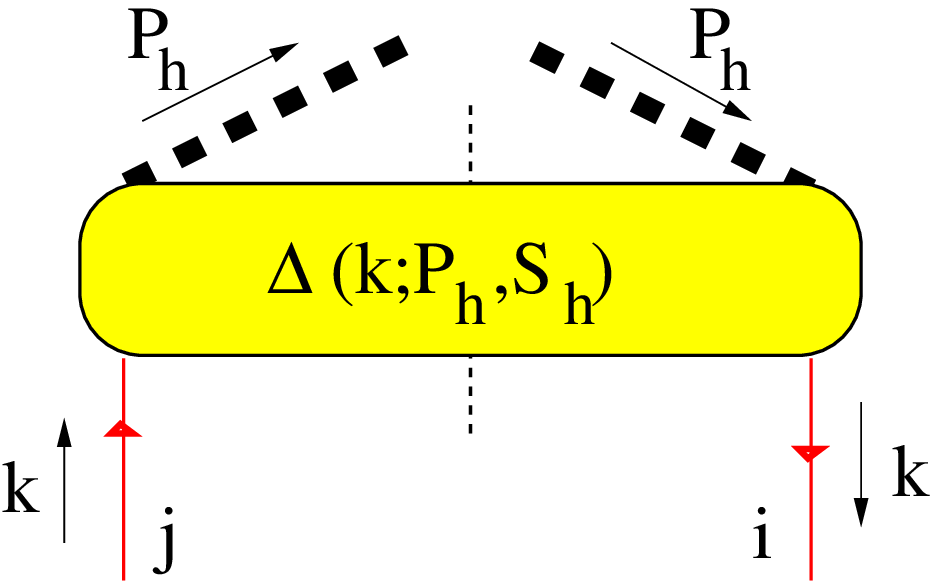, width=3.5cm}\qquad\mbox{}
\\
(a) \hspace{6.5cm} (b)
\\[0.3cm]
\epsfig{file=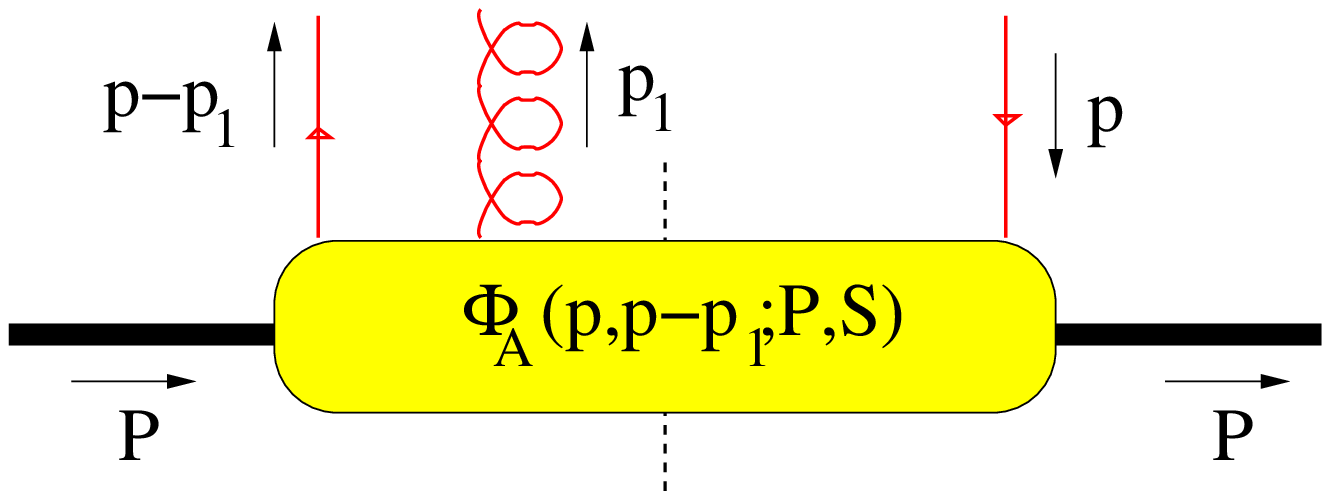, width=5.5cm}
\hspace{2.0cm}
\epsfig{file=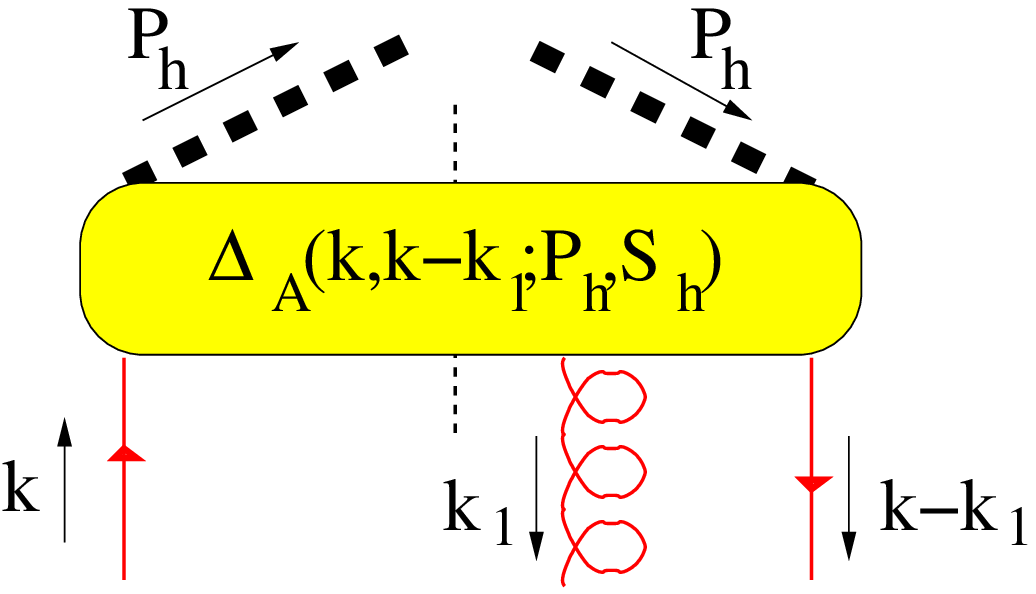, width=4cm}
\\
(c) \hspace{6.5cm} (d)
\\ \mbox{}
\caption{\protect
Soft parts representing the quark-quark and quark-quark-gluon
matrix elements
%$\Phi(P;P,S)$, $\Delta(k;P_h,S_h)$, $\Phi_A^\alpha(p,p-p_1;P,S)$
%and $\Delta_A(k,k-k_1;P_h,S_h)$
used in Eqs.~\ref{soft1} - \ref{soft4}}
\label{fig3}
\end{center}
\end{figure}

From the diagrammatic expansion (see Fig.~\ref{fig2}a-e) one obtains
up to ${\cal O}(g)$,
\begin{eqnarray}
2 M {\cal W}_{\mu\nu}(q; P,S;P_h,S_h)&=&
\int {\rm d}^4p\ {\rm d}^4 k\ \delta^4(p+q-k)\Biggl\{
\,{\rm Tr}(\Phi(p)\gamma_\mu \Delta(k) \gamma_\nu) \nonumber\\
& &   - \int {\rm d}^4 p_1\ {\rm Tr}\left(\gamma_\alpha
\frac{\slash k - \slash p_1 + m}{(k-p_1)^2 - m^2 +i\epsilon}
\gamma_\nu \Phi_A^\alpha(p,p-p_1)\gamma_\mu \Delta(k)\right)
\qquad [{\rm term} \, 1] \nonumber\\
& &   -  \int {\rm d}^4 p_1\ {\rm Tr}\left(\gamma_\mu
\frac{\slash k - \slash p_1 + m}{(k-p_1)^2 - m^2 -i\epsilon}
\gamma_\alpha\Delta(k)\gamma_\nu\Phi_A^\alpha(p-p_1,p)\right) \qquad [{\rm term} \, 2] \nonumber\\
& &     -  \int {\rm d}^4 k_1\ {\rm Tr}\left(\gamma_\nu
\frac{\slash p - \slash k_1+m}{(p-k_1)^2-m^2+i\epsilon}\gamma_\alpha
\Phi(p)\gamma_\mu\Delta_A^\alpha(k-k_1,k)\right)
\qquad [{\rm term} \, 3] \nonumber\\
& &     -  \int {\rm d}^4 k_1\ {\rm Tr}\left(\gamma_\alpha
\frac{\slash p - \slash k_1+m}{(p-k_1)^2-m^2-i\epsilon}\gamma_\mu
\Delta_A^\alpha(k,k-k_1)\gamma_\nu\Phi(p)\right)
\Biggr\}
\ \quad [{\rm term} \, 4]
\label{eqn1}
\end{eqnarray}
where
\bea
&&\Phi_{ij}(p;P,S) = \int \frac{{\rm d}^4 \xi}{(2\pi)^4}\ e^{i\,p\cdot\xi}
\langle P,S| \overline\psi_j (0) \psi_i (\xi) |P,S\rangle,
\label{soft1} \\
&&\Delta_{ij}(k;P_h,S_h) =   \sum_X \int \frac{{\rm d}^4 \xi}{(2\pi)^4}
\ e^{i\,k\cdot\xi}
\,<0| \psi_i (\xi) |P_h,X><P_h,X| \overline\psi_j (0) |0>,
\label{soft2}\\
&&\Phi^\alpha_{A\,ij}(p,p-p_1;P,S) = \int \frac{{\rm d}^4 \xi}{(2\pi)^4}
\,\frac{{\rm d}^4 \eta}{(2\pi)^4}\ e^{i\,p\cdot\xi}
\,e^{i\,p_1\cdot(\eta-\xi)}
\,\langle P,S| \overline\psi_j (0) g A^\alpha(\eta) \psi_i (\xi) |P,S\rangle ,
\label{soft3}\\
&& \Delta^\alpha_{A\, ij}(k,k-k_1;P_h,S_h) = \sum_X \int
\frac{{\rm d}^4 \xi}{(2\pi)^4}\,\frac{{\rm d}^4 \eta}{(2\pi)^4}
\ e^{i\,k\cdot\xi}\,e^{i\,k_1\cdot (\eta-\xi)} \,\langle 0 \vert
\psi_i(\xi)\, gA^\alpha (\eta)|P_h,X><P_h,X|
\overline \psi_j(0)\vert 0\rangle .
\label{soft4}
\eea
illustrated in Fig.~\ref{fig3}.
In the above expression we have omitted the contributions with the
opposite direction on the fermion line. It adds to the result in
Eq.~\ref{eqn1} terms with $q \leftrightarrow -q$ and
$\mu \leftrightarrow \nu$. In cross sections it will always lead
to extending a sum over contributions from quarks to the sum over
quarks {\em and} antiquarks.

The aim of the calculation is an expansion in powers of $1/Q$. For this
a number of considerations are important.
First, the matrix elements represented by blobs in the diagrammatic expansion
should vanish fast enough when any of the products of
momenta involved becomes large, e.g.~the virtualities of
the quarks or gluons. To be precise, in Fig.~\ref{fig3} the products
$p^2 \sim p_1^2 \sim p\cdot p_1 \sim p\cdot P \sim p_1\cdot P
\sim P^2 = M^2 \ll Q^2$.
With the choice of parametrization
in Eqs.~\ref{sudakov-1}-\ref{sudakov-3}, this implies that for the
momenta in Figs~\ref{fig3}a and \ref{fig3}c one has for the plus-components
$p^+$, $p_1^+$, $P^+ \sim Q$, while the minus-components
$p^-$, $p_1^-$, $P^- \sim 1/Q$.
For the fragmentation parts (Figs~\ref{fig3}b and \ref{fig3}d) one
has minus-components $k^-$, $k_1^-$, $P_h^- \sim Q$, while
for the plus-components $k^+$, $k_1^+$, $P_h^+ \sim 1/Q$.
The transverse momenta are of ${\cal O}(M)$. Introducing momentum
fractions $x = p^+/P^+$ and $z=P_h^-/k^-$, writing
\bea
&&p^\mu = p^-\,n_-^\mu + x\,P^+\,n_+^\mu + p_\st^\mu,
\\
&&k^\mu = \frac{P_h^-}{z}\,n_-^\mu + k^+\,n_+^\mu + k_\st^\mu,
\eea
one finds, when neglecting ${\cal O}(1/Q^2)$ contributions, that
$\delta^4(p+q-k)$ $\longrightarrow$ $\delta(x-\xbj)\,\delta(z-z_h)
\,\delta^2(p_\st+q_\st-k_\st)$, thus identifying the scaling
variables and momentum fractions, $x = \xbj$ and $z = z_h$.

Thus, in a calculation up to ${\cal O}(1/Q^2)$, the
integration over the minus-components of momenta in the
matrix elements $\Phi$ and $\Phi_A$ can be performed, restricting
them to the lightfront,
\be
\Phi_{ij}(x,p_\st) = \int {\rm d}p^-\ \Phi_{ij}(p;P,S) =
\left.\int \frac{{\rm d}\xi^-\,{\rm d}^2\xi_\st}{(2\pi)^3}\ e^{i\,p\cdot\xi}
<P,S| \overline\psi_j (0) \psi_i (\xi) |P,S>\right|_{\xi^+ = 0} ,
\ee
while $\Phi_A^\alpha(p^+,p_\st,p_1^+,p_{1\st}) = \int {\rm d}p^-\,{\rm d}p_1^-
\ \Phi_A^\alpha(p,p-p_1;P,S)$ involves two integrations over the
minus-components of the parton momenta. We will occasionally also use the
variable $x_1$ defined via $p_1^+ = x_1\,P^+$. The integrations over
minus-components are sufficient to render the time-ordering
in these matrix elements
superfluous, which can be proven completely analogous to the proof
for the matrix elements in which also the integration over transverse
momenta is performed, in that case restricting them to the
lightcone \cite{Jaffe-83,Diehl-98}. In the matrix elements of the
types $\Delta$ and $\Delta^\alpha_A$ the integrations over the
plus-components of the quark and gluon momenta can be performed,
leading to lightfront correlation functions $\Delta(z,k_\st)$,
\be
\Delta_{ij}(z,k_\st) = \int dk^+\ \Delta_{ij}(k;P_h)
= \left. \sum_X\int \frac{{\rm d}\xi^+\,{\rm d}^2\xi_\st}{(2\pi)^3}\ e^{i\,k\cdot\xi}
<0| \psi_i (\xi) |P_h,X><P_h,X| \overline\psi_j (0) |0>\right|_{\xi^- = 0} .
\ee
and $\Delta_A^\alpha(k^-,k_\st,k_1^-,k_{1\st})
= \int dk^+\,dk_1^+\ \Delta_A^\alpha(k,k-k_1;P_h)$.

Matrix elements like $\Phi(x,p_\st)$ have a particular Dirac structure,
Lorentz structure, and canonical dimension, which must be visible in the
parametrization of $\Phi$ through the dependence on
non-integrated parton momenta and the hadron momentum and spin vectors.
One deduces immediately the Dirac structure $\int {\rm d}p^-\ \Phi
\sim \slash n_+ = \gamma^-$ giving a leading contribution and the
Dirac structure involving the unit matrix in Dirac space requiring in
addition a factor $P^-$ in the parametrization,
which will lead in the calculation of Eq.~\ref{eqn1} to a suppression
factor $1/Q$. The leading structure of
$\int {\rm d}p^-\,{\rm d}p_1^-\ \Phi_A^\alpha$
matrix elements involving {\em two} integrations over minus components
gives for $\int {\rm d}p^-\,{\rm d}p_1^-\ \Phi_A^+ \sim \gamma^-$,
similar to $\int {\rm d}p^-\ \Phi$, but for a transverse gluon one gets
$\int {\rm d}p^-\,{\rm d}p_1^-\ \Phi_{A_\st}^\alpha \sim
P^-\gamma^-$ leading to a $1/Q$ suppression
in the calculation of Eq.~\ref{eqn1} (apart from the subtlety with the
boundary terms, where the role of $P^-$ is taken over by $\delta(p_1^+)$,
to be elaborated upon below). Of course in a parametrization
of the latter matrix element also the transverse index must appear, e.g.\
a non-integrated parton transverse momentum $p_\st$ or the
spin vector $S_\st$ in case of a transversely polarized hadron, but
these are not relevant for an expansion in powers of $1/Q$. The
matrix element $\int {\rm d}p^-\,{\rm d}p_1^-\ \Phi_A^-$ will always
appear suppressed by at least $(P^-)^2 \rightarrow 1/Q^2$.

For the fragmentation parts one has after integration over plus-components
$\int {\rm d}k^+\ \Delta \sim
\int {\rm d}k^+\,{\rm d}k_1^+\ \Delta_A^- \sim \gamma^+$, while
$\int {\rm d}k^+\,{\rm d}k_1^+\ \Delta_{A_\st}^\alpha \sim
P_h^+\,\gamma^+$, becoming suppressed by $1/Q$ (again apart from the subtlety
with boundary terms). The matrix element
$\int {\rm d}k^+\,{\rm d}k_1^+\ \Delta_A^+$ will always appear suppressed
by at least $(P_h^+)^2 \rightarrow 1/Q^2$.
The explicit parametrizations for
the matrix elements in terms of distribution and fragmentation functions
have been extensively discussed in many papers
\cite{LM94,Tangerman-Mulders-95a,MT96,Boer-Mulders-98,Boer-97}
and will be summarized in section VII.

In order to find the leading contributions,
we need in the calculation of Eq.~\ref{eqn1} not only
the first term (diagram in Fig.~\ref{fig2}a), but also the terms
involving $\Phi_A^+$ and $\Delta_A^-$ (Figs.~\ref{fig2}b-e)
and even multiple-gluon matrix elements of the form $\Phi_{AA}^{++}$
(Fig.~\ref{fig2}f), etc.
Such a resumming of multiple-gluon matrix elements can be easily
performed in DIS, where
the integration over transverse momenta of partons can always be performed
in addition to the minus-integration. The resummation leads to a modified
first term in Eq.~\ref{eqn1} with in the $\Phi(x) =
\int {\rm d}p^-\,{\rm d}^2p_\st\ \Phi(p;P,S)$ matrix element the inclusion
of a gauge link
\bea
U^-_{[a,\xi]} & = &
\left.{\cal P}\,\exp\left(-ig\int_a^\xi {\rm d}\zeta^-\,A^+(\zeta)\right)
\right|_{\zeta^+ = \xi^+ = a^+,\,\zeta_\st=\xi_\st=a_\st}
\nonumber \\
& = &\left. \sum_{N=0}^\infty (-ig)^N \int_{a^-}^{\xi^-} d\zeta_1^-\,A^+(\zeta_1)
\ldots \int_{\zeta_{N-1}^-}^{\xi^-} d\zeta_N^-\,A^+(\zeta_N)
\right|_{\zeta_i^+ = \xi^+ = a^+,\,\zeta_{i\st}=\xi_\st=a_\st},
\label{simplelink}
\eea
connecting the quark fields, rendering the object color-gauge invariant 
\cite{Rad-Efr}.
We will discuss the full procedure to obtain a color-gauge invariant
object in 1-particle inclusive leptoproduction in the next section,
following in part Ref.\ \cite{Boer-Mulders-00} and recent work by 
Belitsky, Ji and Yuan \cite{Belitsky}.

\section{Color gauge invariance in SIDIS}

In this section we will discuss the resummation of
contributions in SIDIS coming from
diagrams in Figs~\ref{fig2}b-f.

At orders $(1/Q)^0$ (leading) and $(1/Q)^1$ (first sub-leading)
the integrations over $p^-$, $p_1^-$, and $k^+$
in the corresponding soft parts can be performed.
The result of the first term of four quark-quark-gluon
contributions in Eq.~\ref{eqn1} is
\bea
\mbox{[term 1]} &=&
-\int {\rm d}^4p\ {\rm d}^4 k\ \delta^4(p+q-k)\Biggl\{
\int {\rm d}^4 p_1\ {\rm Tr}\left(\gamma_\alpha
\frac{\slash k - \slash p_1 + m}{(k-p_1)^2 - m^2 +i\epsilon}
\gamma_\nu \Phi_A^\alpha(p,p-p_1)\gamma_\mu \Delta(k)\right)
\Biggr\}
\nonumber \\
& = & -\int {\rm d}^2p_\st\ {\rm d}^2 k_\st\ \delta^2(p_\st+q_\st-k_\st)
\,{\rm d}p_1^+\,{\rm d}^2 p_{1\st}
\int \frac{{\rm d}\xi^-\,{\rm d}^2 \xi_\st}{(2\pi)^3}
\frac{{\rm d}\eta^-\,{\rm d}^2 \eta_\st}{(2\pi)^3}
e^{i p\cdot\xi} e^{i\,p_1\cdot(\eta - \xi)}
\nonumber \\ &&
\mbox{}\hspace{1cm}\times
\left.\langle P,S| \overline\psi(0) \gamma_\mu \Delta(z,k_\st)
\gamma_\alpha\,\frac{\slash k - \slash p_1 + m}{(k-p_1)^2 - m^2 +i\epsilon}
\,\gamma_\nu
g A^\alpha(\eta) \psi(\xi)|P,S\rangle\right|_{\xi^+=\eta^+ = 0} ,
\label{start1}
\eea
where $\Phi_A^\alpha$ is made explicit (Eq.~\ref{soft3})
and the minus- and plus-integrations are performed. In the
expression after the second equal sign, it is understood that in
the integrand $p^- = p_1^- = k^+ = 0$ while $p^+ = x\,P^+$
and $k^- = P_h^-/z$.

Next we will split off from the quark propagator those parts that are relevant
at leading and first sub-leading order in $1/Q$. These parts depend on whether
the index $\alpha$ of the gluon field is plus, transverse or minus. For the 
$1/Q$ order, we will restrict ourselves to obtaining a color
gauge invariant expression for the hadron tensor integrated over the
transverse momentum $q_\st$ of the photon. For this result one first needs to
consider the leading order term unintegrated over $q_\st$. The end results for
the hadron tensor in several cases are summarized in the next section.

One finds for the quark propagator explicitly
(with $k^- \approx q^- = Q/\sqrt{2}$)
\begin{equation}
\frac{\slash k - \slash p_1 + m}{(k-p_1)^2-m^2+i \epsilon} \approx
\frac{(\slash k + m) -\slash n_+\,p_1^+  - \slash p_{1\st}}
{-p_1^+\,Q\sqrt{2} + (k_\st-p_{1\st})^2 -m^2 +i\epsilon}.
\label{quarkprop1}
\end{equation}

Obvious contributions at leading order are, the $k^-\slash n_-$ term of
the quark propagator, which in combination with $A^+$
gluons leads to the link operator in the $\eta^-$ direction. Less
obvious are the contributions from fields that are independent of $\eta^-$
which, as can be seen from Eq.~\ref{start1}, lead to a delta-function
$\delta(p_1^+)$. In that case other leading contributions appear.
In particular, a contribution coming from the last (transverse) term will
lead to (leading) link contributions in the transverse direction.

Contributions at order $1/Q$ are coming from the $\slash n_+$ term of the
quark propagator in combination with the transverse gluons, and the transverse
part of the quark propagator, $\slash p_{1\st}$, in combination with the
$A^+$ gluons. These contributions can be combined into a color gauge
invariant matrix element containing the field strength tensor.
A summary is given in Table~\ref{table1}.
\begin{table}
\caption{\label{table1}
The color gauge covariant objects into which the gluon fields
in SIDIS are combined depending on the Dirac structure of specific terms
in the hard quark propagator to which the gluon couples}
\mbox{}\hspace{4cm}
\begin{tabular}{c|ccc}
& $\slash n_+$ & $\slash n_-$ & $\gamma_\st$ \\
\hline
$A^+$ & - & $U^-$ & $G^{+\alpha}$ \\
$A_\st^\alpha$ & $G^{+\alpha}$ &- & $U^T$
\end{tabular}
\hspace{4cm}\mbox{}
\end{table}

The leading contribution in Eq.~\ref{start1} comes from $\Phi_A^+$. We use
that $\gamma^-(\slash k+ m) = 2k^- - (\slash k- m)\gamma^-$,
the fact that $\Delta(k)\,(\slash k- m) \sim \Delta_A \sim 1/Q$ (QCD
equations of motion) to obtain
\bea
\Delta(k)\,\gamma^-
\,\frac{\slash k- \slash p_1 + m}{(k-p_1)^2-m^2+i \epsilon}\,A^+(\eta)
& \approx &
-\Delta(k)\,\gamma^-
\,\frac{\slash k- \slash p_1 + m}{2k^-\,(p_1^+ -i \epsilon)}\,A^+(\eta)
\nonumber \\
& \approx &
-\Delta(k)\,\frac{A^+(\eta)}{p_1^+ -i \epsilon}
+\Delta(k)\,\frac{\gamma^-\,\slash p_{1}\,A^+(\eta)}
{2k^-\,(p_1^+ -i \epsilon)}
+\frac{\Delta(k)\,(\slash k- m)\,\gamma^-\,A^+(\eta)}
{2k^-\,(p_1^+ -i \epsilon)}
\nonumber \\
& \approx &
-\Delta(k)\,\frac{A^+(\eta)}{p_1^+ -i \epsilon}
+\Delta(k)\,\frac{\gamma^-}{Q\sqrt{2}}\,\frac{\slash p_{1\st}\,A^+(\eta)}
{(p_1^+ -i \epsilon)}
\label{calc1}
\eea
with omitted parts being of ${\cal O}(1/Q^2)$. The first term inserted
in Eq.~\ref{start1} gives a leading contribution,
\bea
\mbox{[term 1.1]} &=&
\int {\rm d}^2p_\st\ {\rm d}^2 k_\st\ \delta^2(p_\st+q_\st-k_\st)
\,{\rm d}p_1^+\,{\rm d}^2 p_{1\st}
\int \frac{{\rm d}\xi^-\,{\rm d}^2 \xi_\st}{(2\pi)^3}
\frac{{\rm d}\eta^-\,{\rm d}^2 \eta_\st}{(2\pi)^3}
e^{i p\cdot\xi} e^{i\,p_1\cdot(\eta - \xi)}
\nonumber \\ &&
\mbox{}\hspace{1cm}\times
\left. \langle P,S| \overline\psi(0) \gamma_\mu \Delta(z,k_\st)
\,\gamma_\nu\,\frac{gA^+(\eta)}{p_1^+ -i \epsilon}
\psi(\xi)|P,S\rangle \right|_{\xi^+=\eta^+ = 0}
\nonumber \\ &=&
\int {\rm d}^2p_\st\ {\rm d}^2 k_\st\ \delta^2(p_\st+q_\st-k_\st)
\int \frac{{\rm d}\xi^-\,{\rm d}^2 \xi_\st}{(2\pi)^3}
e^{i p\cdot\xi}
\nonumber \\ &&
\mbox{}\hspace{1cm}\times
\left. <P,S| \overline\psi(0) \gamma_\mu \Delta(z,k_\st)
\gamma_\nu
\ (-ig)\int^{\xi^-}_\infty {\rm d} \eta^-\ A^+(\eta)
\psi(\xi)|P,S>\right|_{\xi^+=\eta^+ = 0;\ \eta_\st = \xi_\st}  .
\eea
This is precisely the ${\cal O}(g)$ term in the expansion of
$U^-_{[\infty,\xi]}$ multiplying $\psi(\xi)$. The result of
the diagram in Fig.~\ref{fig2}f with two $A^+$-gluons
gives the ${\cal O}(g^2)$ term, etc.
From the second term in Eq.~\ref{start1} (diagram in Fig.~\ref{fig2}c)
one obtains the ${\cal O}(g)$ term in the expansion of
$U^-_{[0,\infty]}$ following $\psi(0)$.

The $A^-$-gluons in the other diagrams in Fig.~\ref{fig2}d and e and
corresponding higher orders
can all be absorbed into link operators in modified soft parts of the form
\bea
&&\Phi_{ij}(x,p_\st) \Rightarrow
\left.\int \frac{{\rm d}\xi^-\,{\rm d}^2\xi_\st}{(2\pi)^3}\ e^{i\,p\cdot\xi}
<P,S| \overline\psi_j (0)\,U^-_{[0,\infty]}\,U^-_{[\infty,\xi]}
\,\psi_i (\xi) |P,S>\right|_{\xi^+ = 0} ,
\label{pgi-1}\\
&&\Delta_{ij}(z,k_\st) \Rightarrow
\left. \sum_X\int \frac{{\rm d}\xi^+\,{\rm d}^2\xi_\st}{(2\pi)^3}\ e^{i\,k\cdot\xi}
<0| U^+_{[-\infty,\xi]}\psi_i (\xi) |P_h,X><P_h,X| \overline\psi_j (0)
U^+_{[0,-\infty]} |0>\right|_{\xi^- = 0} ,
\label{pgi-2}
\eea
where $U^+_{[-\infty,\xi]}$ indicates a link along the lightcone
plus-direction running from $-\infty$ to $\xi^+$.
These quantities, however, are not color gauge invariant, although
we note that upon integration over $p_\st$ and $k_\st$ one
obtains color gauge invariant
lightcone correlators $\Phi(x)$ and $\Delta(z)$, in
which the two links merge into one connecting the lightlike
separated points $0$ and $\xi$. These are e.g.~important in
$q_\st$-integrated SIDIS cross sections at leading order.
For the transverse momentum dependent
functions, however, we are still missing a transverse piece that
leads to color gauge invariant definitions. It has to come from
transverse gluons, which are next to be investigated.

Since the dominant part of $\int dk^+\,\Delta$ is proportional to $\gamma^+$,
one finds (naively) for the transverse gluons in term 1 (Eq.~\ref{start1}),
\bea
\Delta(k)\,\gamma_\alpha
\,\frac{\slash k- \slash p_1 + m}{(k-p_1)^2-m^2+i \epsilon}\,A_\st^\alpha(\eta)
& \approx &
\Delta(k)\,\gamma_\alpha
\,\frac{p_1^+\gamma^-}{2k^-\,(p_1^+ -i \epsilon)}\,A_\st^\alpha(\eta)
\nonumber \\
& \approx &
-\Delta(k)\,\frac{\gamma^-}{Q\sqrt{2}}\,\gamma_\alpha\,A_\st^\alpha(\eta) .
\label{calc2}
\eea
The (remaining) second term in Eq.~\ref{calc1} and the
result of Eq.~\ref{calc2} give as ${\cal O}(1/Q)$ contribution in term 1,
\bea
\mbox{[term 1.2]} &=&
\int {\rm d}^2p_\st\ {\rm d}^2 k_\st\ \delta^2(p_\st+q_\st-k_\st)
\,{\rm d}p_1^+\,{\rm d}^2 p_{1\st}
\int \frac{{\rm d}\xi^-\,{\rm d}^2 \xi_\st}{(2\pi)^3}
\,\frac{{\rm d}\eta^-\,{\rm d}^2 \eta_\st}{(2\pi)^3}
e^{i p\cdot\xi} e^{i\,p_1\cdot(\eta - \xi)}
\nonumber \\ &&
\mbox{}\hspace{1cm}\times \frac{1}{Q\sqrt{2}}
\left.\,<P,S| \overline\psi(0) \,\gamma_\mu \Delta(z,k_\st)
\,\gamma^-\left(\gamma_\alpha\,A_\st^\alpha(\eta)
-\frac{\slash p_{1\st}\,A^+(\eta)}{(p_1^+ -i \epsilon)}\right)\,\gamma_\nu
\,\psi(\xi)|P,S>\right|_{\xi^+=\eta^+ = 0} .
\nonumber \\ & = &
\int {\rm d}^2p_\st\ {\rm d}^2 k_\st\ \delta^2(p_\st+q_\st-k_\st)
\,{\rm d}^2 p_{1\st}
\int \frac{{\rm d}\xi^-\,{\rm d}^2\! \xi_\st}{(2\pi)^3}
\ e^{i p\cdot\xi}
\nonumber \\ &&
\mbox{}\hspace{1cm}\times \frac{1}{Q\sqrt{2}}
\,<P,S| \overline\psi(0) \,\gamma_\mu \Delta(z,k_\st)
\,\gamma^-\gamma_\alpha
\left.\left(A_\st^\alpha(\xi)- \int_\infty^{\xi^-} {\rm d}\eta^-
\,\partial_\st^\alpha A^+(\eta^-)\right)\,\gamma_\nu
\,\psi(\xi)|P,S>\right|_{\xi^+ = 0} .
\nonumber \\ &&
\eea
Including in addition all diagrams with longitudinal $A^+$ gluon fields,
all colored fields become linked along the minus-direction, with the same
link directions for $\Phi$ and $\Phi_A$.
Using the relation between $G^{+\alpha}$ and $A_\st^\alpha$, outlined in the
Appendix~\ref{appendix-1} including all minus links, we find
(suppressing the links $U^-$)
\be
A_\st^\alpha(\xi) - \int_\infty^{\xi^-}
{\rm d}\eta^-\ \left(\partial_\st^\alpha A^+(\eta^-)\right)
= \int_\infty^{\xi^-} {\rm d}\eta^-\ G^{+\alpha}(\eta)
+ A_\st^\alpha(\infty^-),
\ee
with the points $\eta^- = (\eta^-,\xi^+,\xi_\st)$ and
$\infty^- = (\infty,\xi^+,\xi_\st)$. The part of term 1.2 containing
$G^{+\alpha}$ is
\bea
\mbox{[term 1.2a]} &=&
\int {\rm d}^2p_\st\ {\rm d}^2 k_\st\ \delta^2(p_\st+q_\st-k_\st)
\int \frac{{\rm d}\xi^-\,{\rm d}^2 \xi_\st}{(2\pi)^3}
\ e^{i p\cdot\xi}
\nonumber \\ &&
\mbox{}\hspace{1cm}\times \frac{1}{Q\sqrt{2}}
\left.\,<P,S| \overline\psi(0) \,\gamma_\mu \Delta(z,k_\st)
\,\gamma^-\gamma_\alpha
\int_\infty^{\xi^-} {\rm d}\eta^-\ G^{+\alpha}(\eta)\,\gamma_\nu
\,\psi(\xi)|P,S>\right|_{\eta^+ = \xi^+ = 0;\ \eta_\st=\xi_\st} .
\nonumber \\ &&
\eea
Upon integrating the hadron tensor over transverse momenta $q_\st$, the
above convolution factorizes and produces a
color-gauge invariant ${\cal O}(1/Q)$ term,
\bea
&&
\int d^2q_\st \ \mbox{[term 1.2a]} =
\mbox{Tr}\left(
\,\frac{\gamma^-\gamma_\alpha}{Q\sqrt{2}}
\,\gamma_\nu
\int_{-\infty}^\infty {\rm d}p_1^+\ \frac{i}{p_1^+ - i\epsilon}
\,\Phi_{G}^\alpha(p^+,p^+-p_1^+)\,\gamma_\mu \Delta(z)
\right) ,
\label{relation-tw3}
\eea
where (including link operators)
\bea
\Phi_{G\,ij}^\alpha (p^+, p^+-p_1^+) &=&
\left.\int \frac{{\rm d}\xi^-}{2\pi}\,\frac{{\rm d}\eta^-}{2\pi}
\ e^{i\,p\cdot\xi}\,e^{i\,p_1\cdot(\eta-\xi)} \langle P,S \vert
\overline \psi_j(0)\,U^-_{[0,\eta]}\,gG^{+\alpha} (\eta)\,U^-_{[\eta,\xi]}
\, \psi_i(\xi) \vert P,S \rangle\right|_{LC} ,
\label{defphiG}
\eea
also denoted $\Phi_G^\alpha(x,x-x_1)$ with $x_1 = p_1^+/P^+$, and
with $LC$ denoting
$\{ \xi^+ = \eta^+ = \xi_\st = \eta_\st = 0\}$.

\begin{figure}[t]
\begin{center}
\epsfig{file=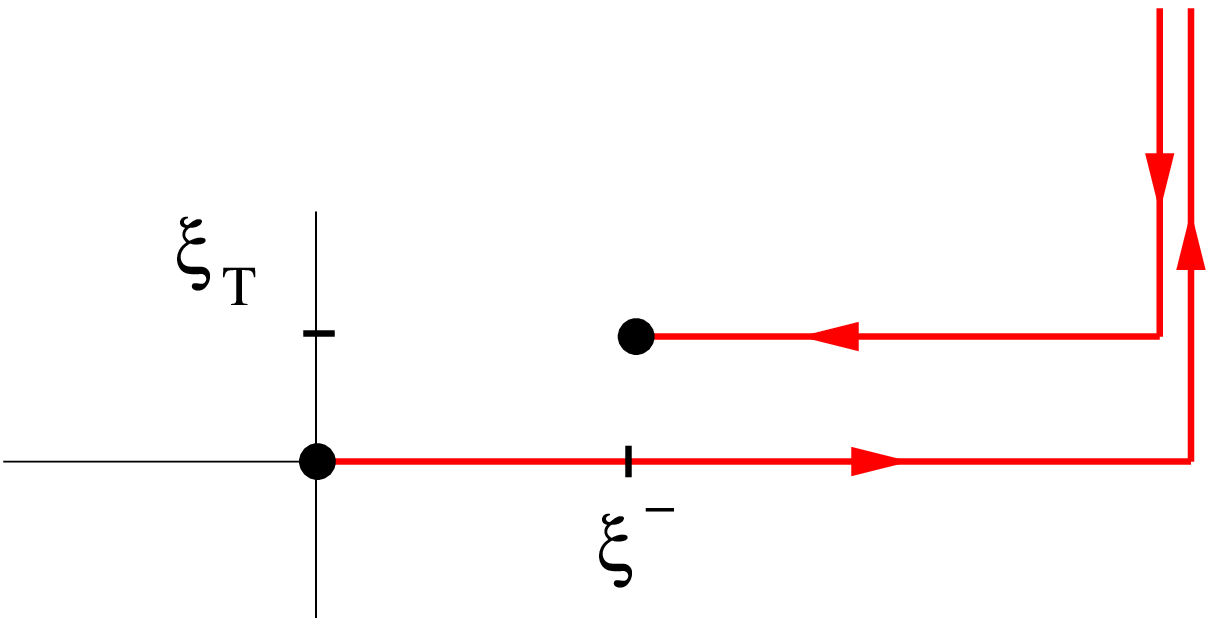, width=4.3cm}
\hspace{0.5cm}$\equiv$\hspace{0.5cm}
\epsfig{file=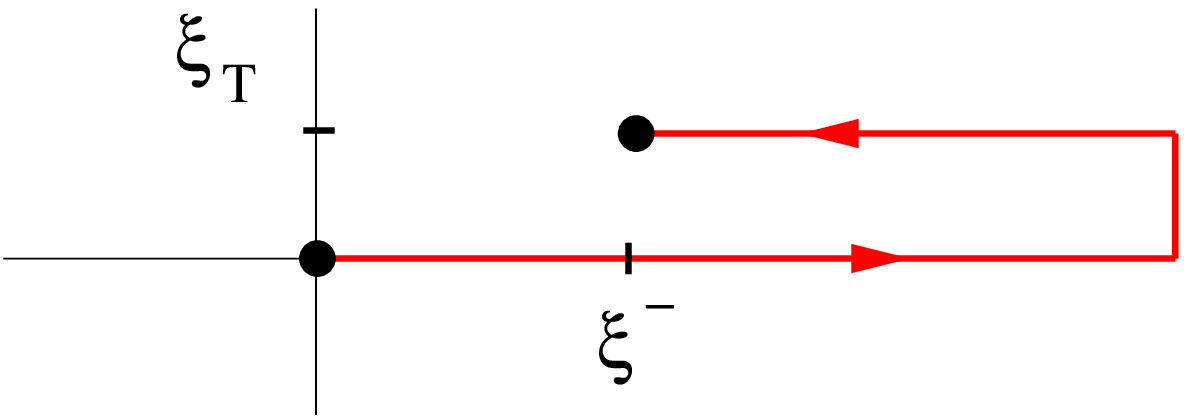, width=4.3cm}
\hspace{1cm}
\epsfig{file=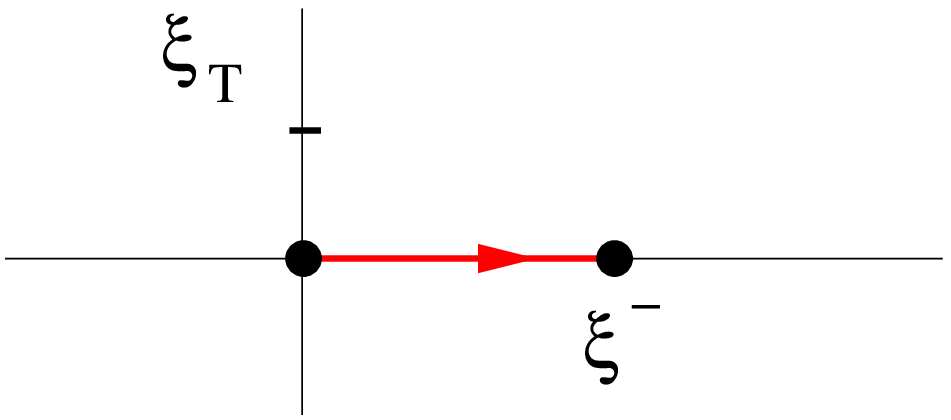, width=3.7cm}
\\
(a) \hspace{6cm} (b) \hspace{5cm} (c)
\\ \mbox{}
\caption{
Link structure for $\Phi^{[+]}(x,p_\st)$ (a and b) and for $\Phi(x)$ (c)
after integration over transverse momenta.}
\label{fig4}
\end{center}
\end{figure}

We are left with a boundary term containing $A_\st(\infty^-)$,
which needs special care. The argument of the transverse field in the
boundary term is fixed by the link direction in $U^-$.
The consequence is that the $\eta^-$ dependence disappears.
We note that the integration over $\eta^-$ thus can simply be
performed, showing that one deals with a matrix element that is
proportional to $\delta(p_1^+)$ in momentum space \cite{Boer4},
\bea
&& \delta(p_1^+)\,\Phi^\alpha_{A(\infty)\,ij}(p,p-p_1)
\equiv \int \frac{{\rm d}^4\xi}{(2\pi)^4}\,\frac{{\rm d}^4\eta}{(2\pi)^4}\ 
e^{i\,p\cdot \xi} \,e^{i\,p_1\cdot (\eta - \xi)}
\langle P,S\vert\overline \psi_j(0)\,gA_\st^\alpha(\infty,\eta^+,\eta_\st)
\psi_i(\xi)\vert P,S \rangle.
\label{boundary}
\eea
Because $\delta(p_1^+) \sim 1/Q$ one finds that
$\int dp^-\,dp_1^-\ \Phi^\alpha_{A(\infty)}
\sim \gamma^+$, i.e.\ it is not suppressed.
This means we have to revisit the approximations made to the fermion
propagator for the boundary term.
Going back to the starting point in Eq.~\ref{start1} we obtain for the
boundary contribution after integration over $\eta^-$,
\bea
\mbox{[term 1.2b]} &=&
-\int {\rm d}^2p_\st\ {\rm d}^2 k_\st\ \delta^2(p_\st+q_\st-k_\st)
\,{\rm d}p_1^+\,{\rm d}^2 p_{1\st}
\int \frac{{\rm d}\xi^-\,{\rm d}^2 \xi_\st}{(2\pi)^3}
\frac{{\rm d}^2 \eta_\st}{(2\pi)^2}
e^{i p\cdot\xi} e^{i\,p_{1\st}\cdot(\eta_\st-\xi_\st)} \ \delta(p_1^+)
\nonumber \\ &&
\mbox{}\hspace{1cm}\times
\left.<P,S| \overline\psi(0) \gamma_\mu \Delta(z,k_\st)
\gamma_\alpha\,\frac{\slash k- \slash p_1 + m}{(k-p_1)^2 - m^2 +i\epsilon}
\,\gamma_\nu
g A_\st^\alpha(\infty,\eta^+,\eta_\st)
\,\psi(\xi)|P,S>\right|_{\xi^+=\eta^+ = 0}.
\nonumber \\ &&
\eea
After a substitution for $A_T^\alpha$,
\begin{equation}
g\,A_\st^\alpha(\eta) = i\partial_{\eta}^\alpha (-ig)\int_{\infty_\st}^{\eta_\st}
{\rm d} \zeta_\st \cdot A_\st(\eta^-,\eta^+,\vec{\zeta_\st}),
\end{equation}
we do a partial integration. In the matrix element, we then encounter
the following part, which we need to consider in
the soft gluon limit ($p_1^+ = 0$) in which the denominator of the
quark propagator can no longer be approximated as in Eq.~\ref{calc1}
or \ref{calc2}. Realizing that $p_1^+ \sim Q$ and $p_1^- \sim 1/Q$
we obtain
\bea
\Delta(k)\,\slash p_{1\st}
\,\frac{\slash k- \slash p_1 + m}{(k-p_1)^2-m^2+i \epsilon}\,\delta(p_1^+)
& \approx &
\Delta(k)\,\slash p_{1}
\,\frac{\slash k- \slash p_1 + m}{(k-p_1)^2-m^2+i \epsilon}\,\delta(p_1^+)
\nonumber \\
& = &
\Delta(k)\,(-\slash k+ \slash p_{1} + m)
\,\frac{\slash k- \slash p_1 + m}{(k-p_1)^2-m^2+i \epsilon}\,\delta(p_1^+)
\nonumber \\
& = &  -\Delta(k)\,\delta(p_1^+) .
\eea
The result after integration over $\eta_\st$, $p_1^+$ and $p_{1\st}$
is a term
\bea
\mbox{[term 1.2b]} &=&
\int {\rm d}^2p_\st\ {\rm d}^2 k_\st\ \delta^2(p_\st+q_\st-k_\st)
\int \frac{{\rm d}\xi^-\,{\rm d}^2 \xi_\st}{(2\pi)^3}
e^{i p\cdot\xi}
\nonumber \\ &&
\mbox{}\hspace{1cm}\times
\left.<P,S| \overline\psi(0) \gamma_\mu \Delta(z,k_\st)
\,\gamma_\nu
(-ig)\int_{\infty_\st}^{\xi_\st}
{\rm d} \zeta_\st \cdot A_T(\infty,0,\zeta_\st)
\,\psi(\xi)|P,S>\right|_{\xi^+ = 0} ,
\eea
which gives precisely the first term of the transverse link that is
needed to modify Eq.~\ref{pgi-1} into a fully color gauge invariant
matrix element. Note that we did not assume a
specific pure gauge expression for the $A_T$ field at $\zeta^- = \infty$.
Furthermore we did not neglect the quark masses in the quark propagator.
For $N$ transverse gluons we have to work out the
boundary terms with more transverse gluons, for which we need the
following relation that also holds for nonabelian fields,
\bea
&&\left((-ig)^M\int_\infty^{\eta} {\rm d} \zeta_{1\st} \cdot A_\st(\zeta_1)
\ldots
\int^\eta_{\zeta_{M-1}}{\rm d} \zeta_{M\st} \cdot A_\st(\zeta_M)\right)
\ g\,A_T^\alpha(\eta) \nonumber \\ &&
\mbox{} \hspace{2cm} =
i\partial_{\eta}^\alpha \left((-ig)^{M+1}
\int_\infty^{\eta} {\rm d} \zeta_{1\st} \cdot A_\st(\zeta_1) \ldots
\int^\eta_{\zeta_{M}}{\rm d} \zeta_{M+1\,\st} \cdot A_\st(\zeta_{M+1})\right).
\eea
This indeed produces nicely the nested integrations needed in the
path-ordered exponential and we find that the terms 1.1 and 1.2b,
are taken care of by using in the ${\cal O}(g^0)$ part of Eq.~\ref{eqn1}
the color gauge invariant matrix element~\cite{Belitsky}
\bea
&&\Phi^{[+]}_{ij}(x,p_\st) =
\left.\int \frac{{\rm d}\xi^-\,{\rm d}^2\xi_\st}{(2\pi)^3}\ e^{i\,p\cdot\xi}
<P,S| \overline\psi_j (0)\,U^-_{[0,\infty]}\,U^\st_{[0_\st,\infty_\st]}
\,U^\st_{[\infty_\st,\xi_\st]}\,U^-_{[\infty,\xi]}
\,\psi_i (\xi) |P,S>\right|_{\xi^+ = 0} ,
\label{gi-1}
\eea
with
\bea
U^\st_{[a,\xi]} & = & \left.{\cal P}\,\exp\left(
-ig\int d\zeta_\st\cdot A_\st(\zeta)\right)\right|_{\zeta^+=\xi^+ =a^+,
\,\zeta^-=\xi^-=a^-}
\nonumber \\
& = & \left. \sum_{N=0}^\infty
(-ig)^N\int_{a_\st}^{\xi_\st} {\rm d} \zeta_{1\st} \cdot A_\st(\zeta_1) \ldots
\int^\xi_{\zeta_{N-1}}
{\rm d} \zeta_{N\st} \cdot A_\st(\zeta_N)\right|_{\zeta_i^+=\xi^+ =a^+,
\,\zeta_i^-=\xi^-=a^-}.
\eea
The $U^\st_{[0_\st,\infty_\st]}$-link of course comes from diagrams
with transverse gluons like in Fig.~\ref{fig2}c and higher orders.
The link structure for the soft part describing the distribution of
quarks in a hadron probed by a spacelike photon is illustrated in
Fig.~\ref{fig4}. The direction of the link, running to $+\infty$ along
the minus direction is indicated via the superscript $[+]$ in Eq.~\ref{gi-1}.
In other processes one will find that the link can also run in the
opposite direction to $-\infty$ along the minus direction. This will
be indicated with a superscript $[-]$.

Eq.~\ref{gi-1} is an important expression, since in its full generality it 
allows for certain
distribution functions, usually referred to as T-odd functions, that would be
absent in case one ignores the gauge links (cf.\ e.g.\
\cite{s90,Collins-93b,Boer-Mulders-98}). Without the transverse gauge
links, it may therefore seem that a choice of $A^+=0$ gauge would demonstrate
the absence of such T-odd functions. In the derivation of Eq.~\ref{gi-1} no
gauge was assumed (one can actually arrive at this result by first considering
the $A^-=0$ gauge as done in Ref.\ \cite{Boer-Mulders-00}), hence it should
not be viewed as one out of many ways to
``gauge-invariantize'' the matrix element \cite{Ji-gaugeinvariantize}. 
Up to the order we consider here, the result is derived rather than assumed.

We note that in leading results the color-gauge invariant object
$\Phi^{[+]}(x,p_\st)$ contracted with $\gamma^+$
still is a semi-positive definite matrix in Dirac space, which is the basis
for deriving positivity conditions, such as the Soffer bound~\cite{Soffer}
and many more~\cite{BBHM}.
Hence, we
disagree with the statement ``Structure functions are not parton
probabilities'' by Brodsky {\em et al.\/} \cite{BHMSS-02}. To be precise,
the distribution functions containing the transverse link are still
probability densities.

As mentioned already, when one considers the $q_\st$-integrated results
one obtains the lightcone quark-quark correlations with the link
in Eq.~\ref{simplelink}. The transverse link does not affect that result
and one has $\Phi^{[+]}(x) = \Phi^{[-]}(x) = \Phi(x)$ (see Fig.~\ref{fig4}c).
If one looks at azimuthal asymmetries or weighted cross sections one
needs to consider matrix elements weighted with transverse momentum.
In those cases one explicitly needs to take into account
the transverse part of the link. We define {\em transverse moments}
$\Phi_\partial^{[\pm]\,\alpha}(x)$,
\bea
\Phi_{\partial}^{[\pm]\,\alpha}(x)
& \equiv & \int {\rm d}^2p_\st \ p_\st^\alpha\,\Phi^{[\pm]}(x,\bm p_\st),
\label{transvmom}
\eea
that in a straightforward way can be related to color gauge invariant
quark-quark-gluon matrix elements $\Phi_G^\alpha$ and
$\Phi_D^\alpha$, the latter involving the covariant derivative,
\bea
\left(\Phi_{\partial}^{[\pm]\alpha}\right)_{ij}(x)
& = &
\left. \int {\rm d}^2p_\st \int \frac{d\xi^- d^2\bm \xi_\perp}{(2\pi)^3}
\ e^{i\,p\cdot \xi}\,\langle P,S \vert \overline \psi_j(0)\,U^-_{[0,\pm\infty]}
U^\st_{[0_\st,\pm\infty_\st]}
\,i\partial_\xi^\alpha\,U^\st_{[\pm\infty_\st,\xi_\st]}U^-_{[\pm\infty,\xi]}
\,\psi_i(\xi) \vert P,S \rangle
\right|_{\xi^+ \,=\, 0}
\nonumber \\ & = &
\int \frac{d\xi^-}{(2\pi)}\ e^{ip\cdot \xi}
\biggl\{ \langle P,S\vert \overline \psi_j(0)
\,U^-_{[0,\xi]}\, iD_\st^\alpha\psi_i(\xi) \vert P,S\rangle
\biggr|_{LC}
\nonumber \\ && \mbox{} \hspace{2cm}
- \langle P,S\vert \overline \psi_j(0)\,U^-_{[0,\pm\infty]}
\int_{\pm\infty}^{\xi^-}d\eta^- \,U^-_{[\pm\infty,\eta]}\,g\,G^{+\alpha}(\eta)
\,U^-_{[\eta,\xi]}\,\psi_i(\xi) \vert P,S\rangle
\biggr|_{LC}\biggr\} ,
\eea
or
\bea
\Phi_{\partial}^{[\pm]\,\alpha}(x)
& = &\Phi^\alpha_D(x) - \int_{-\infty}^\infty {\rm d}p_1^+
\ \frac{i}{p_1^+ \mp i\epsilon}\,\Phi_{G}^\alpha(p^+,p^+-p_1^+) ,
\label{relation-twist3}
\eea
where
\begin{eqnarray}
&&
\Phi^\alpha_{D\,ij}(x) =
\int {\rm d} p_1^+\ \Phi_{D\,ij}^\alpha (p^+, p^+-p_1^+)
=
\left.\int \frac{{\rm d}\xi^-}{2\pi}\ e^{i\,p\cdot \xi}
\langle P,S \vert \overline \psi_j(0)\,U^-_{[0,\xi]}
\,iD^\alpha (\xi) \psi_i(\xi)\vert P,S \rangle\right|_{LC} .
\label{phid1}
\end{eqnarray}
The important observation we want to make here is that the difference
between correlation functions with links running to $\pm \infty$,
respectively, is related to a quark-gluon correlator,
\be
\Phi_\partial^{[+]\alpha}(x) - \Phi_\partial^{[-]\alpha}(x)
= 2\pi\,\Phi_G^\alpha(x,x),
\label{gluonic pole}
\ee
the latter being given the name {\em gluonic-pole} matrix element
since it corresponds to the soft-gluon point $p_1^+ = 0$. Its consequences
have been studied for several processes
\cite{QS-91b,Hammon-97,qs,KK01,DBQiu,Boer4} and it is viewed as one of
the possible mechanism to generate single spin asymmetries. We will comment on
this further below, but already mention that the above relation between 
$\Phi_\partial^{[\pm]\alpha}(x)$ and $\Phi_G^\alpha(x,x)$ implies that the
Sivers effect \cite{s90} is directly related to the Qiu-Sterman mechanism (the
gluonic-pole matrix element), i.e.\ if one is nonzero, then the other also is.
We will make this relation more specific below. 
 
We further define
\bea
&&\Phi_\partial^\alpha(x)
\equiv \frac{1}{2}\left(\Phi_\partial^{[+]\alpha}(x)
+ \Phi_\partial^{[-]\alpha}(x)\right),
\label{Phid}
\\ &&
\tilde \Phi_A^\alpha(x) = {\rm PV} \int dx_1\ \frac{1}{x_1}
\,\Phi_G^\alpha(x,x-x_1),
\label{phidelta}
\eea
where we use $\tilde \Phi_A^\alpha(x)$ to distinguish the function
from the non-gauge-invariant $\Phi_A^\alpha(x)$.
These definitions imply
\bea
&&\Phi_\partial^{[\pm]\alpha}(x) = \Phi_\partial^\alpha(x)
\pm \pi\,\Phi_G^\alpha(x,x),
\\ &&
\Phi_{\partial}^{\alpha}(x)
= \Phi^\alpha_D(x) - \tilde \Phi_A^\alpha(x).
\label{dDA}
\eea
The relations in Eqs~\ref{phidelta} - \ref{dDA} are relations connecting
color gauge-invariant quantities.
We will return to the above functions and their properties in section
\ref{TRprop}.

\begin{figure}[b]
\begin{center}
\epsfig{file=Figures/linkdisspace2.eps, width=4.3cm}
\hspace{2cm}
\epsfig{file=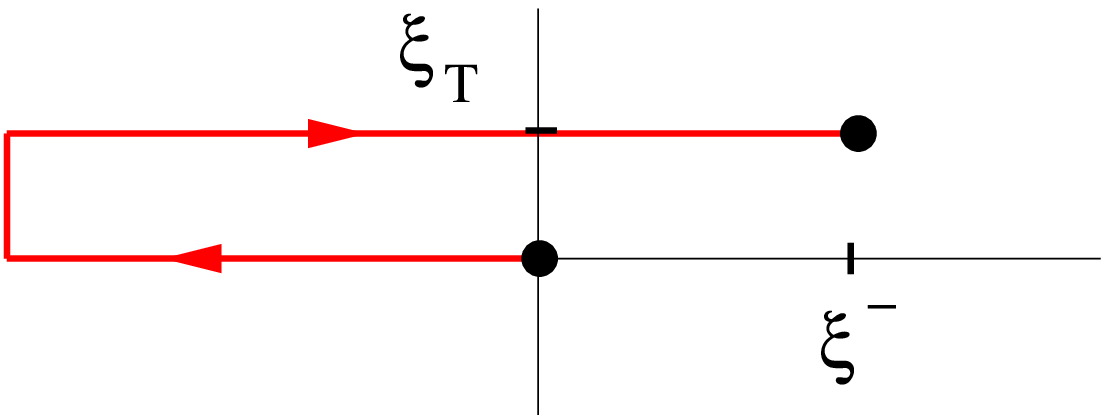, width=4.3cm}
\\
(a) spacelike distribution $\Phi^{[+]}$ \hspace{1cm}
(b) timelike distribution $\Phi^{[-]}$
\\
\epsfig{file=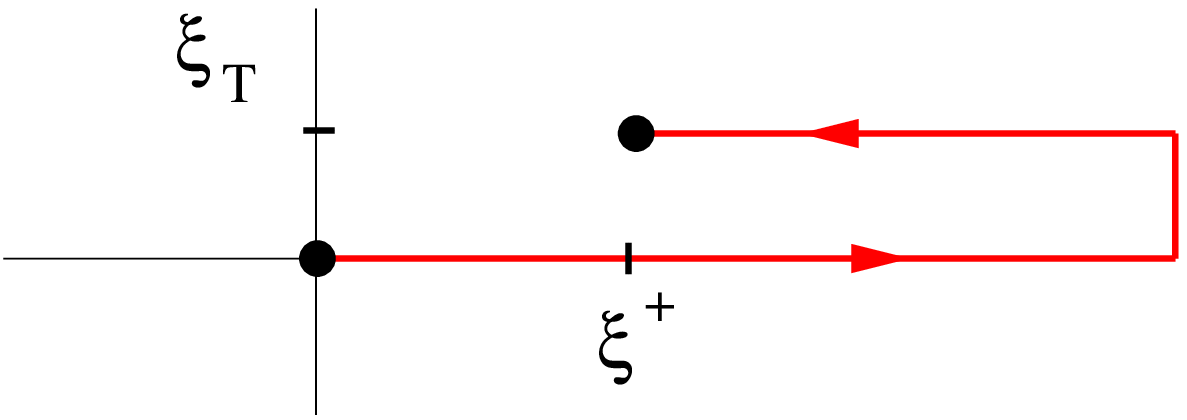, width=4.3cm}
\hspace{2cm}
\epsfig{file=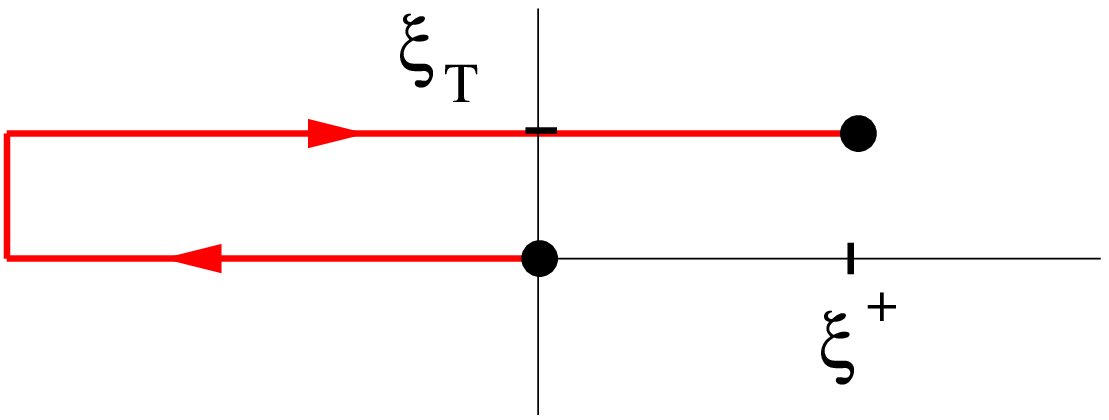, width=4.3cm}
\\
(c) timelike fragmentation $\Delta^{[+]}$ \hspace{1cm} (d)
spacelike fragmentation $\Delta^{[-]}$
\\ \mbox{}
\caption{
Link structure for $\Phi(x,p_\st)$ (a and b) and $\Delta(z,k_\st)$ (c and d).}
\label{fig5}
\end{center}
\end{figure}

We end this section by giving the remaining contributions in the
$q_\st$-integrated hadron tensor at order $1/Q$. We give a systematic summary
of the SIDIS hadron tensor for several cases in the next section. 
We can use Eq.~\ref{relation-twist3}
to rewrite the twist-3 contribution in Eq.~\ref{relation-tw3} obtained after
$q_\st$-integration into the form
\bea
&&
\int d^2q_\st \ \mbox{[term 1.2a]} =
\mbox{Tr}\left(
\,\frac{\gamma^-\gamma_\alpha}{Q\sqrt{2}}
\,\gamma_\nu\,\Bigl[\Phi_D^\alpha(x) - \Phi_\partial^{[+]\alpha}(x)\Bigr]
\,\gamma_\mu\,\Delta(z)
\right) .
\eea
The transverse gluons in the second term of Eq.~\ref{start1} (diagram
in Fig.~\ref{fig2}c) produces besides the transverse
link $U^\st_{[0_\st,\infty_\st]}$, already absorbed into Eq.~\ref{gi-1},
also a twist-three piece. For this
one needs matrix elements with interchanged arguments such as $\Phi_A(p-p_1,p)$.
The resulting twist-3 term after integration over transverse momenta is
\bea
\int d^2q_\st \ \mbox{[term 2.2a]} & = &
\mbox{Tr}\left(
\,\gamma_\nu
\int_{-\infty}^\infty {\rm d}p_1^+\ \frac{-i}{p_1^+ + i\epsilon}
\,\Phi_{G}^\alpha(p^+-p_1^+,p^+)
\,\gamma_\mu\,\frac{\gamma_\alpha\gamma^-}{Q\sqrt{2}}
\,\Delta(z)
\right)
\nonumber \\ & = &
\mbox{Tr}\left(
\,\gamma_\nu
\Bigl[\gamma_0\Phi_D^{\alpha\dagger}(x)\gamma_0 -
\Phi_\partial^{[+]\alpha}(x)\Bigr]
\gamma_\mu\,\frac{\gamma_\alpha\gamma^-}{Q\sqrt{2}}
\,\Delta(z)
\right) ,
\eea
where we have used that $\Phi_\partial^{[+]}(x) =
\gamma_0\,\Phi_\partial^{[+]\alpha \dagger}(x)\,\gamma_0$.
The third and fourth ${\cal O}(g)$ terms in Eq.~\ref{start1} (diagrams
in Fig.~\ref{fig2}d and e) need the (direct and conjugate) quark propagators
(with $p^+ \approx -q^+ = Q/\sqrt{2}$)
\be
\frac{\slash p - \slash k_1 + m}{(p-k_1)^2-m^2\pm i \epsilon} \approx
\frac{(\slash p + m) -\slash n_-\,k_1^- - \slash k_{1\st}}
{-k_1^-\,Q\sqrt{2} + (p_\st-k_{1\st})^2 -m^2 \pm i\epsilon}.
\ee
The calculation yields the links $U^+_{[-\infty,\xi]}$ and $U^+_{[0,-\infty]}$
running along the plus-direction in the matrix elements
$\Delta(z,k_\st)$ in Eq.~\ref{pgi-2} for the fragmentation part.
Including the transverse gluons we get the fully color-gauge invariant
matrix element indicated as the spacelike fragmentation
$\Delta^{[-]}(z,k_\st)$ with the link as indicated in Fig.~\ref{fig5}.
Furthermore one obtains twist-3 contributions containing
$\Delta_G^\alpha(k-k_1,k)$ and $\Delta_G^\alpha(k, k-k_1)$,
which after integration over $q_\st$ yield
\bea
\int d^2q_\st \ \mbox{[term 3.2a]} & = &
\mbox{Tr}\left(
\,\gamma_\mu
\int_{-\infty}^\infty {\rm d}k_1^-\ \frac{-i}{k_1^- - i\epsilon}
\,\Delta_{G}^\alpha(k^--k_1^-,k^-)
\,\gamma_\nu\,\frac{\gamma_{\alpha}\,\gamma^+}{Q\sqrt{2}}
\,\Phi(x)\right)
\nonumber \\ & = &
\mbox{Tr}\left(
\,\gamma_\mu\Bigl[\gamma_0\Delta_D^{\alpha\dagger} (z)\gamma_0
- \Delta_\partial^{[-]\alpha}(z)\Bigr]
\gamma_\nu\,\frac{\gamma_{\alpha}\,\gamma^+}{Q\sqrt{2}}
\,\Phi(x)\right) ,
\\
\int d^2q_\st \ \mbox{[term 4.2a]} & = &
\mbox{Tr}\left(
\frac{\gamma^+\,\gamma_{\alpha}}{Q\sqrt{2}}\,\gamma_\mu
\int_{-\infty}^\infty {\rm d}k_1^-\ \frac{i}{k_1^- + i\epsilon}
\,\Delta_{G}^\alpha(k^-,k^--k_1^-)\,\gamma_\nu \Phi(x)
\right)
\nonumber \\ & = &
\mbox{Tr}\left(
\frac{\gamma^+\,\gamma_{\alpha}}{Q\sqrt{2}}\,\gamma_\mu
\,\Bigl[\Delta_D^\alpha (z) - \Delta_\partial^{[-]\alpha}(z)\Bigr]
\,\gamma_\nu \Phi(x)
\right) .
\eea
In deriving these results we stepped over some subtleties involving
the inclusion of diagrams with crossed gluon lines and the use of the
full QCD equations of motion, but for this we refer to
Ref.~\cite{Boer-Mulders-00}.

\section{SIDIS and DIS cross sections}

The basic expression for ${\cal W}_{\mu\nu}(q;P,S;P_h,S_h)$ contains
a convolution of the transverse momentum dependent functions. We have
seen that upon integration over $q_\st$,
\be
\int {\rm d}^2q_\st {\rm d}^2p_\st\,{\rm d}^2 k_\st
\ \delta^2(p_\st+q_\st-k_\st)\ldots
= \int {\rm d}^2p_\st\,{\rm d}^2 k_\st\ \ldots ,
\label{integrated-1}
\ee
the integral can be deconvoluted. This is also true for azimuthal
asymmetries constructed by weighting with $q_\st^\alpha$,
\be
\int {\rm d}^2q_\st \ q^\alpha_\st \ {\rm d}^2p_\st\,{\rm d}^2 k_\st
\ \delta^2(p_\st+q_\st-k_\st)\ldots
=  \int {\rm d}^2p_\st\,{\rm d}^2 k_\st\ (k_\st^\alpha - p_\st^\alpha)
\ldots .
\ee
If one calculates the ${\cal O}(1/Q)$ result, one has to be careful,
however. One cannot simply perform the integration over $q_\st$ in
the hadron tensor, since the lepton tensor involves $q$.
To proceed, one starts with a (Cartesian) set of vectors, starting
with $q$ defining in SIDIS a spacelike direction while the other
external vectors are used to define orthogonal directions. In
particular, it is convenient to start with $q$ and
$\tilde P$ = $P - (P\cdot q/q^2)q$,
\bea
&&\hat z^\mu \equiv -\frac{q^\mu}{Q} = - \hat q^\mu,
\\
&&\hat t^\mu = \frac{\tilde P^\mu}{\sqrt{\tilde P^2}} \approx
\frac{q^\mu + 2\xbj\,P^\mu}{Q} ,
\eea
with $\hat z^2 = -1$ and $\hat t^2 = 1$.
These two vectors can be used both for inclusive and semi-inclusive
leptoproduction.
The lepton tensor can be expressed in the vectors $\hat t$ and $\hat z$
and a perpendicular vector (equal for the lepton in initial and final state)
which defines the azimuthal angle of the lepton scattering plane.
From the Cartesian vectors two new lightlike vectors
$n_\pm^\prime = (\hat t \pm \hat z)/\sqrt{2} = n_\pm^\prime(P,q)$ can
be constructed as well as a perpendicular tensor
$g_\perp^{\mu\nu} \equiv g^{\mu\nu} + \hat q^\mu \hat q^\nu
-\hat t^\mu \hat t^\nu
= g^{\mu\nu} - n_+^{\prime \{\mu}n_-^{\prime\,\nu\}}$.
Since the lightlike directions $n_\pm^\prime$ are determined
by $P$ and $q$, instead of $P$ and $P_h$, the momentum of the produced
hadron $P_h$ will have in general a nonvanishing perpendicular component
enabling us to define a vector $X^\mu \equiv -P^\mu_{h\perp}/z_h$,
which defines the azimuthal angle of the hadron production plane.

It is straightforward to see that up to ${\cal O}(1/Q^2)$ corrections,
the previously defined set $\{ n_+,\ n_-,\ q_\st\}$ is related to
the set $\{ \hat t,\ \hat z,\ X\}$ or $\{ n_+^\prime,\ n_-^\prime,
\ X\}$ via
\bea
&&n_-^\mu\ \approx\  n_-^{\prime \mu}
- 2\,\frac{X^\mu}{Q\sqrt{2}} ,
\label{transf-1}
\\
&&n_+^\mu\ \approx\ n_+^{\prime\mu},
\label{transf-2}
\\
&&q_\st^\mu\ \approx\ X^\mu
- \sqrt{2}\,\frac{Q_\st^2}{Q}\,n_+^{\prime\,\mu} ,
\label{transf-3}
\eea
and $X^2 \approx -Q_\st^2$.
We note that the leptonic tensor is independent of $X$. The $1/Q$
term appearing on the righthandside of Eq.~\ref{transf-3} is
irrelevant in our calculations. Hence for experimental
cross sections up to that order we can use for integrated and weighted
cross sections the replacements
$\int d^2X \ \ldots \ \rightarrow \ \int d^2q_\st\ \ldots$ and
$\int d^2X\,X^\alpha\ \ldots \ \rightarrow \ \int d^2q_\st\,q_\st^\alpha
\ \ldots$\ .
We now consider separately integrated SIDIS, DIS and azimuthal asymmetries
in SIDIS. We will also consider a few special cases.
If one measures in the final state the jet-direction, this can be
considered as a measurement of $p_\st$,
i.e.\ $q_\st = -p_\st$. This is referred to as JET SIDIS.

\subsection{Integrated SIDIS cross section at leading order}

We have seen that inclusion of appropriate quark-gluon matrix elements
make the ${\cal O}(g^0)$ result in Eq.~\ref{eqn1} color-gauge invariant,
\bea
2 M {\cal W}_{\mu\nu}^{(0)}(q;P,S;P_h,S_h) &=&
\int {\rm d}^2p_\st\ {\rm d}^2 k_\st\ \delta^2(p_\st+q_\st-k_\st)
\,{\rm Tr}\left(\Phi^{[+]}(x,p_\st)\gamma_\mu
\Delta^{[-]}(z,k_\st) \gamma_\nu\right) .
\label{cross-section-1}
\eea
At leading order the integration over transverse momenta of the produced
hadrons simply can be performed and one obtains the basic result
\bea
&&\int d^2X\ 2M {\cal W}_{\mu\nu}^{(0)}(q;P,S;P_h,S_h)
\ = \ {\rm Tr}\Bigl(\Phi(x)\,\gamma_\mu\,\Delta(z)\,\gamma_\nu\Bigr)
\Biggr|_{n_\pm \rightarrow n_\pm^\prime}
\ +\ {\cal O}\left(\frac{1}{Q}\right).
\label{leading}
\eea

\subsection{Azimuthal asymmetries in SIDIS at leading order}

We consider here cross sections obtained after integration over $X$ and
explicit weighting with $X^\alpha$. In practice this means measurement
of the azimuthal angle of the produced hadron and compare it with other
azimuthal angles, such as that of the lepton scattering plane, the
(transverse) spin of the target hadron or the (transverse) spin of the
produced hadron. For our purposes it implies calculation of
$\int d^2X\ X^\alpha\,2M {\cal W}_{\mu\nu}$, which at leading
order gives
\bea
&&\int d^2X\ X^\alpha\,2M {\cal W}_{\mu\nu}^{(0)}(q;P,S;P_h,S_h)
\nonumber \\ && \qquad \mbox{}
 = {\rm Tr}\Bigl(\Phi(x)\,\gamma_\mu
\,\Delta_\partial^{[-]\alpha}(z)\,\gamma_\nu
\Bigr)\Biggr|_{n_\pm \rightarrow n_\pm^\prime}
- {\rm Tr}\Bigl(\Phi_\partial^{[+]\alpha}(x)\,\gamma_\mu
\,\Delta(z)\,\gamma_\nu\Bigr)\Biggr|_{n_\pm \rightarrow n_\pm^\prime}
+ {\cal O}\left(\frac{1}{Q}\right).
\label{leadingazimuthal}
\eea
In these cross sections one finds for instance the Collins and
Sivers effects~\cite{Kotzinian95,Anselmino,Boer-Mulders-98,BoM-99}.

\subsection{Integrated SIDIS cross section at ${\cal O}(1/Q)$}

As outlined one must be careful in integrating over transverse
momenta. At ${\cal O}(1/Q)$ the differences between using $n_\pm(P,P_h)$
or $n_\pm^\prime(P,q)$ matter. In particular the correlator $\Delta
\propto \slash n_-$ will lead to terms proportional to
$\slash q_\st/Q\sqrt{2}$. To find the $\slash n_-$ dependence in
a Dirac space correlator, we use the projectors
$P_\pm = \gamma^\mp\gamma^\pm/2 = \slash n_\pm\slash n_\mp/2$.
We use that the leading term in
$\Delta$ satisfies $P_-\,\Delta = \Delta\,P_+ = P_-\,\Delta\,P_+$
to write
\bea
\Delta(z,k_\st) & \ =\ & \frac{1}{4}\Bigl(
\slash n_-\,\slash n_+\,\Delta(z,k_\st)\,\slash n_+\,\slash n_-\Bigr)
\nonumber \\ & \approx &
\Delta(z,k_\st)\Biggr|_{n_\pm \ \rightarrow\ n_\pm^\prime}
- \frac{1}{Q\sqrt{2}}\Bigl(
\slash q_\st\,\slash n_+ \,\Delta(z,k_\st)
+ \Delta(z,k_\st)\,\slash n_+\,\slash q_\st\Bigr) ,
\eea
and we obtain the $1/Q$ contribution coming from Eq.~\ref{cross-section-1},
\bea
&&
\int d^2X\ 2M {\cal W}_{\mu\nu}^{(0)}(q;P,S;P_h,S_h)\ =
\ {\cal O}(1)\ \mbox{result [Eq.~\ref{leading}]}
\nonumber \\ && \hspace{1cm}
-\frac{1}{Q\sqrt{2}}\int d^2q_\st
\int {\rm d}^2p_\st\ {\rm d}^2 k_\st\ \delta^2(p_\st+q_\st-k_\st)
\Biggl\{{\rm Tr}\Bigl(\Phi^{[+]}(x,p_\st)\,\gamma_\mu
\,\slash q_\st\,\slash n_+ \,\Delta^{[-]}(z,k_\st)\,\gamma_\nu\Bigr)
\nonumber \\ && \hspace{8.3cm} \mbox{}
+ {\rm Tr}\Bigl(\Phi^{[+]}(x,p_\st)\,\gamma_\mu
\,\Delta^{[-]}(z,k_\st)\,\slash n_+\,\slash q_\st\,\gamma_\nu\Bigr)
\Biggr\}
\nonumber \\ && \hspace{1cm} = \ {\cal O}(1)\ \mbox{result [Eq.~\ref{leading}]}
+ {\rm Tr}\Bigl(\Phi_\partial^{[+]\alpha}(x)\,\gamma_\mu
\,\frac{\gamma_\alpha\,\gamma^-}{Q\sqrt{2}}\,\Delta(z)
\,\gamma_\nu\Bigr)
- {\rm Tr}\Bigl(\Phi(x)\,\gamma_\mu\,\frac{\gamma_\alpha\,\gamma^-}{Q\sqrt{2}}
\,\Delta_\partial^{[-]\alpha}(z)
\,\gamma_\nu\Bigr)
\nonumber \\ && \hspace{4.5cm} \mbox{}
+ {\rm Tr}\Bigl(\Phi_\partial^{[+]\alpha}(x)\,\gamma_\mu
\,\Delta(z)\,\frac{\gamma^-\gamma_\alpha}{Q\sqrt{2}}
\,\gamma_\nu\Bigr)
- {\rm Tr}\Bigl(\Phi(x)\,\gamma_\mu
\,\Delta_\partial^{[-]\alpha}(z)\,\frac{\gamma^-\gamma_\alpha}{Q\sqrt{2}}
\,\gamma_\nu\Bigr).
\label{lH-overflow}
\eea
Including these $1/Q$-terms and the four contributions from quark-gluon
correlators with transverse gluons, one obtains the full integrated
SIDIS cross section up to ${\cal O}(1/Q)$,
\bea
&&\int d^2X\ 2M {\cal W}_{\mu\nu}^{(0+1)}(q;P,S;P_h,S_h)
= \ {\cal O}(1)\ \mbox{result [Eq.~\ref{leading}]}
\nonumber \\ && \hspace{1.2cm}
+ {\rm Tr}\Bigl(\Phi_\partial^{[+]\alpha}(x)\,\gamma_\mu
\,\frac{\gamma_\alpha\,\gamma^-}{Q\sqrt{2}}\,\Delta(z)
\,\gamma_\nu\Bigr)
- {\rm Tr}\Bigl(\Phi(x)\,\gamma_\mu\,\frac{\gamma_\alpha\,\gamma^-}{Q\sqrt{2}}
\,\Delta_\partial^{[-]\alpha}(z)
\,\gamma_\nu\Bigr)
\nonumber \\ && \hspace{1.2cm} \mbox{}
+ {\rm Tr}\Bigl(\Phi_\partial^{[+]\alpha}(x)\,\gamma_\mu
\,\Delta(z)\,\frac{\gamma^-\gamma_\alpha}{Q\sqrt{2}}
\,\gamma_\nu\Bigr)
- {\rm Tr}\Bigl(\Phi(x)\,\gamma_\mu
\,\Delta_\partial^{[-]\alpha}(z)\,\frac{\gamma^-\gamma_\alpha}{Q\sqrt{2}}
\,\gamma_\nu\Bigr)
\nonumber \\ && \hspace{1.2cm} \mbox{}
+ \mbox{Tr}\left(
\,\frac{\gamma^-\gamma_\alpha}{Q\sqrt{2}}
\,\gamma_\nu\,\Bigl[\Phi_D^\alpha(x) - \Phi_\partial^{[+]\alpha}(x)\Bigr]
\,\gamma_\mu\,\Delta(z)
\right)
+ \mbox{Tr}\left(
\,\gamma_\nu
\Bigl[\gamma_0\Phi_D^{\alpha\dagger}(x)\gamma_0 -
\Phi_\partial^{[+]\alpha}(x)\Bigr]
\gamma_\mu\,\frac{\gamma_\alpha\gamma^-}{Q\sqrt{2}}
\,\Delta(z)
\right)
\nonumber \\ && \hspace{1.2cm} \mbox{}
+ \mbox{Tr}\left(
\,\gamma_\mu
\Bigl[\gamma_0\Delta_D^{\alpha\dagger} (z)\gamma_0
- \Delta_\partial^{[-]\alpha}(z)\Bigr]
\gamma_\nu\,\frac{\gamma_{\alpha}\,\gamma^+}{Q\sqrt{2}}
\,\Phi(x)\right)
+ \mbox{Tr}\left(
\frac{\gamma^+\,\gamma_{\alpha}}{Q\sqrt{2}}\,\gamma_\mu
\,\Bigl[\Delta_D^\alpha (z) - \Delta_\partial^{[-]\alpha}(z)\Bigr]
\,\gamma_\nu \Phi(x)
\right)
\nonumber
\\ && \hspace{0.8cm} \mbox{}
= \ {\cal O}(1)\ \mbox{result [Eq.~\ref{leading}]}
\nonumber \\ && \hspace{1.2cm} \mbox{}
+ \mbox{Tr}\left(
\,\frac{\gamma^-\gamma_\alpha}{Q\sqrt{2}}
\,\gamma_\nu\,\Phi_D^\alpha(x)\,\gamma_\mu\,\Delta(z)
\right)
%+ \mbox{Tr}\left(
%\,\frac{\gamma^-\gamma_\alpha}{Q\sqrt{2}}
%\,\gamma_\mu\,\Phi_D^\alpha(x)\,\gamma_\nu\,\Delta(z)
%\right)^\ast
%\nonumber \\ && \hspace{1.2cm} \mbox{}
+ \mbox{Tr}\left(
\frac{\gamma^+\,\gamma_{\alpha}}{Q\sqrt{2}}\,\gamma_\mu
\,\Delta_D^\alpha (z)\,\gamma_\nu \Phi(x)\right)
%+ \mbox{Tr}\left(
%\frac{\gamma^+\,\gamma_{\alpha}}{Q\sqrt{2}}\,\gamma_\nu
%\,\Delta_D^\alpha (z)\,\gamma_\mu \Phi(x)\right)^\ast
\nonumber \\ && \hspace{1.2cm} \mbox{}
- \mbox{Tr}\left(
\frac{\gamma^+\,\gamma_{\alpha}}{Q\sqrt{2}}\,\gamma_\mu
\,\Delta_\partial^{[-]\alpha}(z)\,\gamma_\nu \Phi(x)
\right)
%- \mbox{Tr}\left(
%\frac{\gamma^+\,\gamma_{\alpha}}{Q\sqrt{2}}\,\gamma_\nu
%\,\Delta_\partial^{[-]\alpha}(z)\,\gamma_\mu \Phi(x)
%\right)^\ast
%\nonumber \\ && \hspace{1.2cm} \mbox{}
- {\rm Tr}\Bigl(\gamma_\mu\,\frac{\gamma_\alpha\,\gamma^-}{Q\sqrt{2}}
\,\Delta_\partial^{[-]\alpha}(z)
\,\gamma_\nu\,\Phi(x)\Bigr)
%- {\rm Tr}\Bigl(\gamma_\nu\,\frac{\gamma_\alpha\,\gamma^-}{Q\sqrt{2}}
%\,\Delta_\partial^{[-]\alpha}(z)
%\,\gamma_\mu\,\Phi(x)\Bigr)^\ast
+ \left( \mu \leftrightarrow \nu\right)^\ast,
\label{lH-total}
\eea
where the hermiticity properties of the various matrix elements have
been used (see section \ref{TRprop}). We note that
the $1/Q$ part in Eq.~\ref{lH-overflow} is by itself not electromagnetically
gauge invariant, but together with the parts arising from quark-gluon
matrix elements, leading to the result in Eq.~\ref{lH-total}, it is
gauge invariant.

\subsection{Integrated DIS cross section at ${\cal O}(1/Q)$}

The hadron tensor for this case is obtained by integrating the
SIDIS result over $z$ and using for the fragmentation part the
free (massless) quark $\rightarrow$ quark result,
$\Delta(z) = \slash n_-\delta(1-z)$
and $\Delta_\partial^{[-]\alpha}(z) = \Delta_D^\alpha(z) = 0$.
This gives the well-known result \cite{EFP-83},
\bea
2M\,W_{\mu\nu}(q;P,S)
& = & \ {\rm Tr}\Bigl(\Phi(x)\,\gamma_\mu
\,\gamma^+\,\gamma_\nu\Bigr)
\nonumber \\ && \mbox{}
+ \mbox{Tr}\left(
\,\frac{\gamma^-\gamma_\alpha}{Q\sqrt{2}}
\,\gamma_\nu\,\Phi_D^\alpha(x)\,\gamma_\mu\,\gamma^+
\right)
+ \mbox{Tr}\left(
\,\frac{\gamma^-\gamma_\alpha}{Q\sqrt{2}}
\,\gamma_\mu\,\Phi_D^\alpha(x)\,\gamma_\nu\,\gamma^+
\right)^\ast .
\eea

\subsection{Azimuthal asymmetries in JET SIDIS at leading order}

Azimuthal asymmetries in JET DIS are obtained at measured
$q_\st = -P_{{\rm jet}\,\perp}$ and with
in addition still fixed $P$ and $q$. We simply replace $\Delta(z,k_\st)
= \slash n_-\,\delta(1-z)\,\delta^2(k_\st)$. Hence we start with
\bea
2 M {\cal W}_{\mu\nu}^{(0)}(q;P,S;q_\st) &=&
{\rm Tr}\left(\Phi^{[+]}(x,-q_\st)\gamma_\mu
\gamma^+ \gamma_\nu\right) .
\eea
In this situation one will
have to be careful with Sudakov effects~\cite{Boer-Sudakov}, but the
following
weighted azimuthal asymmetry is free of these effects,
\be
\int {\rm d}^2q_\st\ q_\st^\alpha
\ 2 M {\cal W}_{\mu\nu}^{(0)}(q;P,S;q_\st)
= -{\rm Tr}\left(\Phi_\partial^{[+]\alpha}(x)\gamma_\mu
\gamma^+ \gamma_\nu\right).
\ee
This result is also a direct consequence of Eq.~\ref{leadingazimuthal},
taking for $\Delta^{[-]}(z,k_\st)$ the quark $\rightarrow$ quark limit.

%\subsection{Azimuthal asymmetries in JET+hadron SIDIS at leading order}
%
%For the 2-particle inclusive leptoproduction tensor
%in the particular case that the first hadron is the jet
%direction, i.e.\ one obtains in essence $p_\st$, and the second
%hadron defines a transverse momentum $k_\st = P_{h\perp}/z_h$
%(with respect to the jet-direction), one finds just the full
%result for $2 M {\cal W}_{\mu\nu}^{(0)}(q;P,S;P_h,S_h)$ without
%integration. Upon averaging over jet-direction and weighting with
%transverse momentum of a hadron within the jet one finds
%the first term in Eq.~\ref{leadingazimuthal},
%\bea
%&&\int {\rm d}^2P_{{\rm jet}\,\perp}\,{\rm d}^2P_{h\perp}\
%\frac{\vert P_{h\perp}\vert}{z_h}\ \,2M
%{\cal W}_{\mu\nu}(q;P,S;P_{\rm jet};P_h,S_h)
%%\nonumber \\ && \qquad \mbox{}
%= {\rm Tr}\Bigl(\Phi(x)\,\gamma_\mu
%\,\Delta_\partial^{[-]\alpha}(z)\,\gamma_\nu\Bigr) .
%\eea

\subsection{Other subleading cross sections}

Azimuthal asymmetries at ${\cal O}(1/Q)$, azimuthal asymmetries
involving higher weighting than with one power of the momentum $X^\alpha$
or SIDIS cross sections at ${\cal O}(1/Q^2)$, require considerably
more theoretical efforts than the one presented above. Moreover, it might
be impossible to unambiguously disentangle
hadrons within a `jet' from hadrons in other `jets', since the
`separation' of jets is merely an ${\cal O}(Q^2)$ effect, meaning
$P_{\rm jet\ 1}\cdot P_{\rm jet\ 2} \propto Q^2$. This also implies
that a factorization proof most likely cannot be given, just like the
failure of factorization in unpolarized processes at $1/Q^4$ \cite{Basu}.
As is well-known, these difficulties do not appear for the
inclusive DIS cross section involving just one (target) hadron
which allows for a rigorous treatment at any order in powers of $1/Q$.

\section{The Drell-Yan cross sections}

\begin{figure}[bbbb]
\begin{center}
\begin{minipage}{6cm}
\epsfig{file=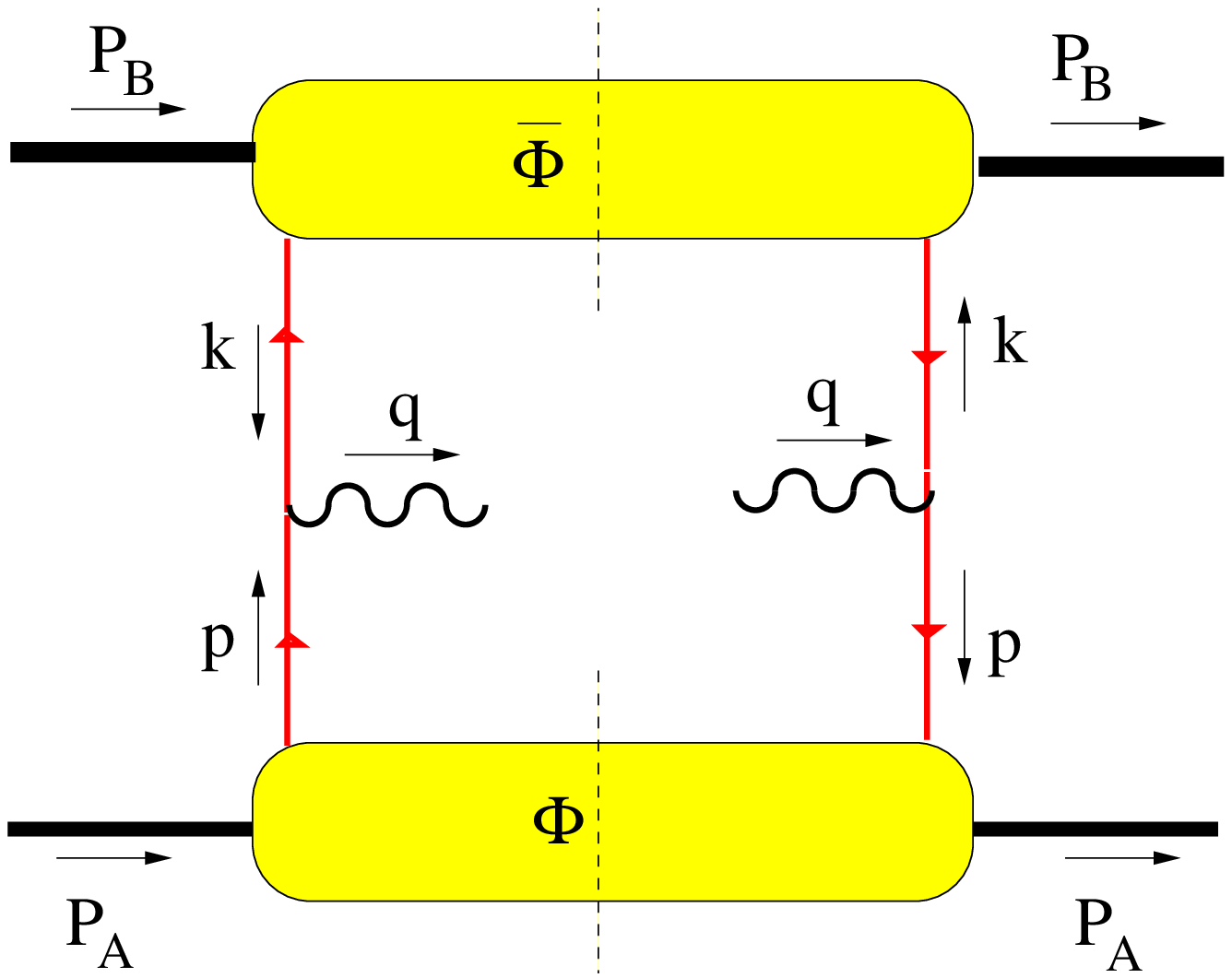, width=5.5cm}
\end{minipage}
\hspace{2cm}
\begin{minipage}{6cm}
\epsfig{file=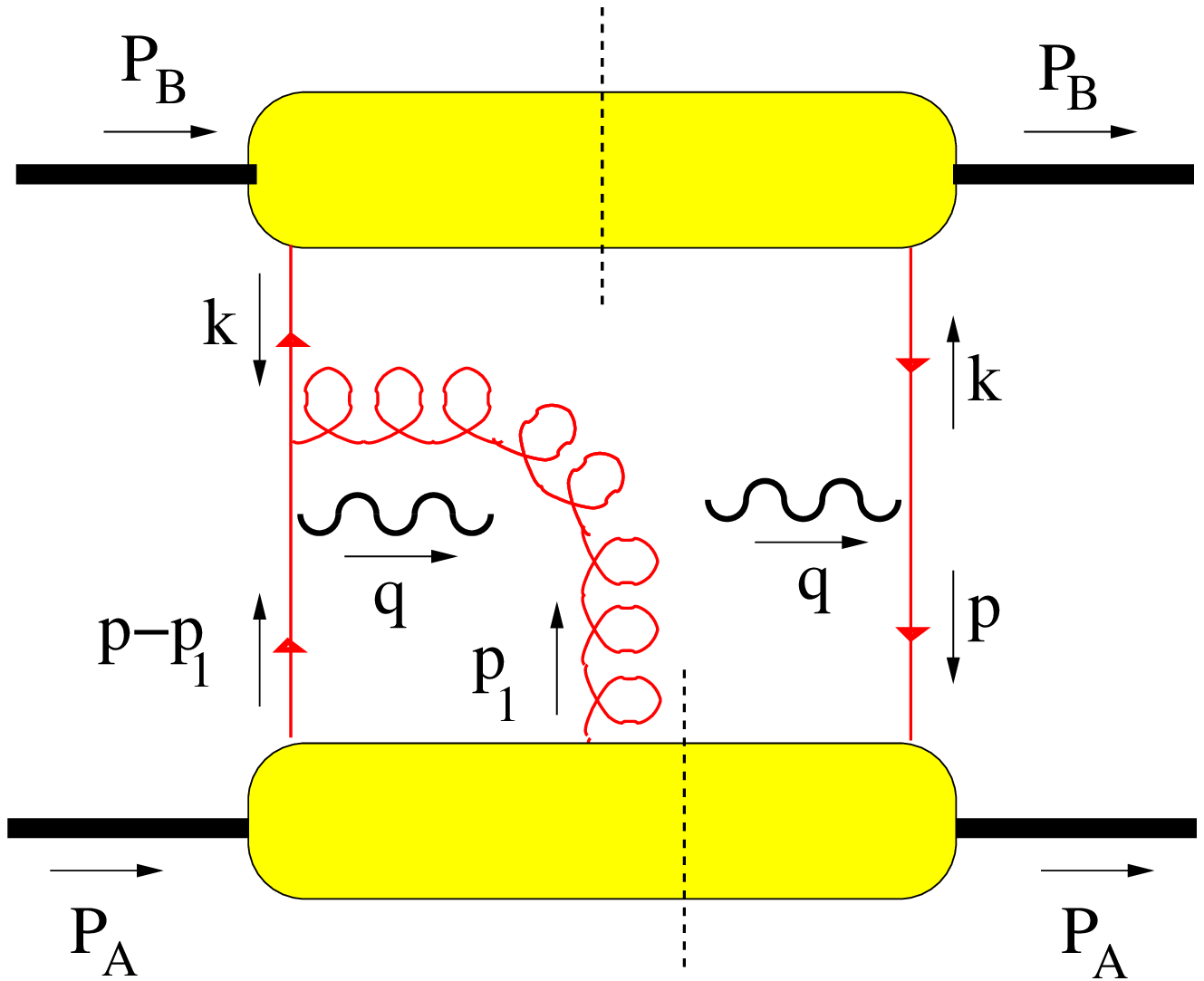, width=5.5cm}
\end{minipage}
\\
(a) \hspace{7.5cm} (b)
\\ \mbox{}
\caption{
Quark-quark (a) and one of the quark-quark-gluon (b) correlators
in tree-level diagrams for Drell-Yan scattering}
\label{fig6}
\end{center}
\end{figure}

For Drell-Yan, one has a similar treatment as for leptoproduction.
The calculation
involves now two soft distribution parts and annihilation of a quark-antiquark
pair into a gauge boson (we will only discuss the vector coupling here).
The handbag diagram is given in Fig.~\ref{fig6}a and an example
of a diagram with an additional gluon in Fig.~\ref{fig6}b.

A full calculation at tree level
including quark-gluon matrix elements as discussed for leptoproduction
gives in this case
\begin{eqnarray}
2 M {\cal W}_{\mu\nu}(q; P_A,S_A;P_B,S_B)&=&
\int {\rm d}^4p\ {\rm d}^4 k\ \delta^4(p+k-q)\Biggl\{
                \,{\rm Tr}(\Phi(p)\gamma_\mu \overline\Phi(k) \gamma_\nu) \nonumber\\
              & &   - \int {\rm d}^4 p_1\ {\rm Tr}\left(\gamma_\alpha
                  \frac{-\slash k - \slash p_1 + m}{(k+p_1)^2 - m^2 +i\epsilon}
                 \gamma_\nu \Phi_A^\alpha(p,p-p_1)\gamma_\mu \overline\Phi(k)\right) \nonumber\\
             & &   -  \int {\rm d}^4 p_1\ {\rm Tr}\left(\gamma_\mu
                  \frac{-\slash k - \slash p_1 + m}{(k+p_1)^2 - m^2 -i\epsilon}
                  \gamma_\alpha\overline\Phi(k)\gamma_\nu\Phi_A^\alpha(p-p_1,p)\right) \nonumber\\
           & &     -  \int {\rm d}^4 k_1\ {\rm Tr}\left(\gamma_\nu
                  \frac{\slash p + \slash k_1+m}{(p+k_1)^2-m^2+i\epsilon}\gamma_\alpha
                  \Phi(p)\gamma_\mu\overline\Phi_A^\alpha(k-k_1,k)\right)\nonumber\\
           & &     -  \int {\rm d}^4 k_1\ {\rm Tr}\left(\gamma_\alpha
                  \frac{\slash p + \slash k_1+m}{(p+k_1)^2-m^2-i\epsilon}\gamma_\mu
                  \overline\Phi_A^\alpha(k,k-k_1)\gamma_\nu\Phi(p)\right)\Biggr\}
           + \ldots,
\label{dy-eqn1}
\end{eqnarray}
where $\Phi(p)$ and $\Phi_A(p,p-p_1)$ are the same as in leptoproduction,
but the role of $\Delta$ and $\Delta_A$ is taken over by
\bea
&&\overline\Phi_{ij}(k;P_B,S_B) =  \int \frac{{\rm d}^4 \xi}{(2\pi)^4}
\ e^{-i\,k\cdot\xi}
\,\langle P_B,S_B| \psi_i (\xi) \overline\psi_j (0) |P_B,S_B\rangle,
\label{dysoft2}\\
&& \overline\Phi^\alpha_{A\, ij}(k,k-k_1;P_B,S_B) = \int
\frac{{\rm d}^4 \xi}{(2\pi)^4}\,\frac{{\rm d}^4 \eta}{(2\pi)^4}
\ e^{-i\,k\cdot\xi}\,e^{-i\,k_1\cdot (\eta-\xi)} \,\langle P_B,S_B\vert
\psi_i(\xi)\, gA^\alpha (\eta)\,\overline \psi_j(0)\vert P_B,S_B\rangle .
\label{dysoft4}
\eea
(note that this implies $\overline\Phi_\partial^\alpha(x,k_\st)
= - k^\alpha\,\overline\Phi(x,k_\st)$).

The important difference between DY and SIDIS turns out to be the
direction of the links. The result for the quark propagator in a
quark-gluon diagram as in Fig.~\ref{fig2}b, but then in the case of
DY with a timelike outgoing photon (cf. Fig.~\ref{fig6}a)
yields a propagator
\begin{equation}
\frac{-\slash k - \slash p_1 + m}{(k+p_1)^2-m^2+i \epsilon} \approx
\frac{-(\slash k - m) -\slash n_+\,p_1^+  - \slash p_{1\st}}
{p_1^+\,Q\sqrt{2} + (k_\st+p_{1\st})^2 -m^2 +i\epsilon}.
\end{equation}
(since $k^- \approx q^- = Q/\sqrt{2}$). The difference with
Eq.~\ref{quarkprop1} is the sign
with which $p_1^+$ appears in the denominator. For the $A^+$-gluons this
produces a link running along the minus direction to $-\infty$, i.e.\
one finds transverse momentum dependent distribution functions
$\Phi^{[-]}(x_A,p_\st)$, where $x_A \approx p^+/P_A^+ \approx q^+/P_A^+$.
Also in the antiquark matrix element the link
runs to $-\infty$ (along the plus direction with our choice of
lightlike vectors),
i.e.\ the $A^-$ gluons lead to the matrix element
$\overline\Phi^{[-]}(x_B, k_\st)$, where $x_B \approx k^-/P_B^-
\approx q^-/P_B^-$.

As in leptoproduction the tree-level calculation is most conveniently
done with lightlike directions $n_\pm(P_A,P_B)$ defined
via the hadron momenta.
In order to perform the transverse integration at the level of the
hadron tensor one
introduces \cite{Collins-Soper-frame,MOS,Tangerman-Mulders-95a} a
Cartesian set
$\{ \hat t,\ \hat z,\ X\}$ starting with $\hat t$ = $q/Q$. A symmetric choice
for $\hat z$ is $z = (x_A\,P_A - x_B\,P_B)/Q$ that defines the Collins-Soper
frame. Using $n_\pm^\prime = (\hat t \pm \hat z)/\sqrt{2}$, one has
\bea
&&n_-^\mu\ \approx\ \ n_-^{\prime\mu} - \frac{q_{\st}^\mu}{Q\sqrt{2}},
\label{dy-transf-1}
\\
&&n_+^\mu\ \approx\ n_+^{\prime\mu} - \frac{q_{\st}^\mu}{Q\sqrt{2}},
\eea
while as in leptoproduction one finds that for the calculation of the
cross section the orthogonal direction $X$
differs from $q_\st$ only by the
irrelevant ${\cal O}(1/Q)$ terms multiplying $n_\pm^\prime$, i.e.
for our purposes $X \approx q_\st \approx q - x_A\,P_A - x_B\,P_B$.

\subsection{Integrated DY cross section at leading order}

As indicated, the inclusion of appropriate quark-gluon matrix elements
make the ${\cal O}(g^0)$ result in Eq.~\ref{eqn1} color-gauge invariant
leading to
\bea
{\cal W}_{\mu\nu}^{(0)}(q;P_A,S_A;P_B,S_B) &=&
\int {\rm d}^2p_\st\ {\rm d}^2 k_\st\ \delta^2(p_\st+k_\st-q_\st)
\,{\rm Tr}\left(\Phi^{[-]}(x_A,p_\st)\gamma_\mu
\overline\Phi^{[-]}(x_B,k_\st) \gamma_\nu\right) .
\label{dy-cross-section-1}
\eea
This is analogous to the expression employed by Ralston and Soper
\cite{RS-79}, except that now the correlation functions are fully color gauge
invariant.
At leading order the integration over transverse momenta of the produced
hadrons simply can be done and one obtains the basic result for the DY process,
\bea
&&\int {\rm d}^2X\ {\cal W}_{\mu\nu}^{(0)}(q;P_A,S_A;P_B,S_B)
\ = \ {\rm Tr}\Bigl(\Phi(x_A)\,\gamma_\mu\,\overline\Phi(x_B)\,\gamma_\nu\Bigr)
\Biggr|_{n_\pm \rightarrow n_\pm^\prime}
\ +\ {\cal O}\left(\frac{1}{Q}\right).
\label{dy-leading}
\eea

\subsection{Azimuthal asymmetries in DY at leading order}

We consider here cross sections obtained after integration over $X$ and
explicit weighting with $X^\alpha$. In practice this means measurement
of the azimuthal angle of the leptonic plane with respect to that of the
hadron plane or relative to (transverse) spin azimuthal angles.
For our purposes it implies calculation of
$\int d^2X\ X^\alpha\ {\cal W}_{\mu\nu}$, which at leading
order gives
\bea
&&\int {\rm d}^2X\ X^\alpha\ {\cal W}_{\mu\nu}^{(0)}(q;P_A,S_A;P_B,S_B)
\nonumber \\ && \qquad \mbox{}
 = -{\rm Tr}\Bigl(\Phi(x_A)\,\gamma_\mu
\,\overline\Phi_\partial^{[-]\alpha}(x_B)\,\gamma_\nu\Bigr)
\Biggr|_{n_\pm \rightarrow n_\pm^\prime}
+ {\rm Tr}\Bigl(\Phi_\partial^{[-]\alpha}(x_A)\,\gamma_\mu
\,\overline\Phi(x_B)\,\gamma_\nu\Bigr)
\Biggr|_{n_\pm \rightarrow n_\pm^\prime}
+ {\cal O}\left(\frac{1}{Q}\right).
\label{dy-leadingazimuthal}
\eea

\subsection{Integrated DY cross section at ${\cal O}(1/Q)$}

As outlined, above, one must be careful in the DY process
in integrating over transverse
momenta. In this case both correlators $\Phi \propto \slash
n_+$ and $\overline \Phi \propto \slash n_-$ will lead to terms proportional
to $\slash q_\st/Q\sqrt{2}$. To find the $\slash n_-$ dependence in
a Dirac space correlator, we again use the projectors $P_\pm$.
We now get
\bea
\Phi(x_A,p_\st) & \ =\ & \frac{1}{4}\Bigl(
\slash n_+\,\slash n_-\,\Phi(x_A,p_\st)\,\slash n_-\,\slash n_+\Bigr)
\nonumber \\ & \approx &
\Phi(x_A,p_\st)\Biggr|_{n_\pm \ \rightarrow\ n_\pm^\prime}
- \frac{1}{2Q\sqrt{2}}\Bigl(
\slash q_\st\,\slash n_- \,\Phi(x_A,p_\st)
+ \Phi(x_A,p_\st)\,\slash n_-\,\slash q_\st\Bigr) ,
\\
\overline \Phi(x_B,k_\st) & \ =\ & \frac{1}{4}\Bigl(
\slash n_-\,\slash n_+\,\overline \Phi(x_B,k_\st)\,\slash n_+\,\slash n_-\Bigr)
\nonumber \\ & \approx &
\overline \Phi(x_B,k_\st)\Biggr|_{n_\pm \ \rightarrow\ n_\pm^\prime}
- \frac{1}{2Q\sqrt{2}}\Bigl(
\slash q_\st\,\slash n_+ \,\overline \Phi(x_B,k_\st)
+ \overline \Phi(x_B,k_\st)\,\slash n_+\,\slash q_\st\Bigr) .
\eea
Combining this $1/Q$ contribution coming from Eq.~\ref{dy-cross-section-1}
with the parts from the quark-gluon diagrams one obtains
\bea
&&
\int {\rm d}^2X\ {\cal W}_{\mu\nu}^{(0+1)}(q;P_A,S_A;P_B,S_B)\ =
\ {\cal O}(1)\ \mbox{result [Eq.~\ref{dy-leading}]}
\nonumber \\ && \hspace{1.2cm} \mbox{}
- \mbox{Tr}\left(
\,\frac{\gamma^-\gamma_\alpha}{Q\sqrt{2}}
\,\gamma_\nu\,\Phi_D^\alpha(x_A)\,\gamma_\mu\,\overline\Phi(x_B)
\right)
%- \mbox{Tr}\left(
%\,\frac{\gamma^-\gamma_\alpha}{Q\sqrt{2}}
%\,\gamma_\mu\,\Phi_D^\alpha(x_A)\,\gamma_\nu\,\overline\Phi(x_B)
%\right)^\ast
%\nonumber \\ && \hspace{1.2cm} \mbox{}
+ \mbox{Tr}\left(
\frac{\gamma^+\,\gamma_{\alpha}}{Q\sqrt{2}}\,\gamma_\mu
\,\overline\Phi_D^\alpha (x_B)\,\gamma_\nu \Phi(x_A)\right)
%+ \mbox{Tr}\left(
%\frac{\gamma^+\,\gamma_{\alpha}}{Q\sqrt{2}}\,\gamma_\nu
%\,\overline\Phi_D^\alpha (x_B)\,\gamma_\mu \Phi(x_A)\right)^\ast
\nonumber \\ && \hspace{1.2cm} \mbox{}
+ \frac{1}{2}\,\mbox{Tr}\Bigl(
\frac{\gamma^-\gamma_\alpha}{Q\sqrt{2}}
\,\gamma_\nu\,\Phi_\partial^{[-]\alpha}(x_A)
\,\gamma_\mu\,\overline\Phi(x_B)\Bigr)
%+ \frac{1}{2}\,\mbox{Tr}\Bigl(
%\frac{\gamma^-\gamma_\alpha}{Q\sqrt{2}}
%\,\gamma_\mu\,\Phi_\partial^{[-]\alpha}(x_A)
%\,\gamma_\nu\,\overline\Phi(x_B)\Bigr)^\ast
%\nonumber \\ && \hspace{1.2cm} \mbox{}
- \frac{1}{2}\,{\rm Tr}\Bigl(\gamma_\nu
\frac{\gamma_\alpha\,\gamma^+}{Q\sqrt{2}}
\,\Phi_\partial^{[-]\alpha}(x_A)\,\gamma_\mu\,\overline\Phi(x_B)\Bigr)
%- \frac{1}{2}\,{\rm Tr}\Bigl(\gamma_\mu
%\frac{\gamma_\alpha\,\gamma^+}{Q\sqrt{2}}
%\,\Phi_\partial^{[-]\alpha}(x_A)\,\gamma_\nu\,\overline\Phi(x_B)\Bigr)^\ast
\nonumber \\ && \hspace{1.2cm} \mbox{}
- \frac{1}{2}\,\mbox{Tr}\left(
\frac{\gamma^+\,\gamma_{\alpha}}{Q\sqrt{2}}\,\gamma_\mu
\,\overline\Phi_\partial^{[-]\alpha}(x_B)\,\gamma_\nu \Phi(x_A)\right)
%-\frac{1}{2}\,\mbox{Tr}\left(
%\frac{\gamma^+\,\gamma_{\alpha}}{Q\sqrt{2}}\,\gamma_\nu
%\,\overline\Phi_\partial^{[-]\alpha}(x_B)\,\gamma_\mu \Phi(x_A)
%\right)^\ast
%\nonumber \\ && \hspace{1.2cm} \mbox{}
+\frac{1}{2}\,{\rm Tr}\Bigl(\gamma_\mu\,\frac{\gamma_\alpha\,\gamma^-}{Q\sqrt{2}}
\,\overline\Phi_\partial^{[-]\alpha}(x_B)
\,\gamma_\nu\,\Phi(x_A)\Bigr)
%+\frac{1}{2}{\rm Tr}\Bigl(\gamma_\nu\,\frac{\gamma_\alpha\,\gamma^-}{Q\sqrt{2}}
%\,\overline\Phi_\partial^{[-]\alpha}(x_B)
%\,\gamma_\mu\,\Phi(x_A)\Bigr)^\ast
+ \left( \mu \leftrightarrow \nu\right)^\ast ,
\eea
where the hermiticity properties of the various matrix elements have
been used (see section \ref{TRprop}).

\section{Back-to-back jet production in electron-positron annihilation}

\begin{figure}[t]
\begin{center}
\begin{minipage}{5cm}
\epsfig{file=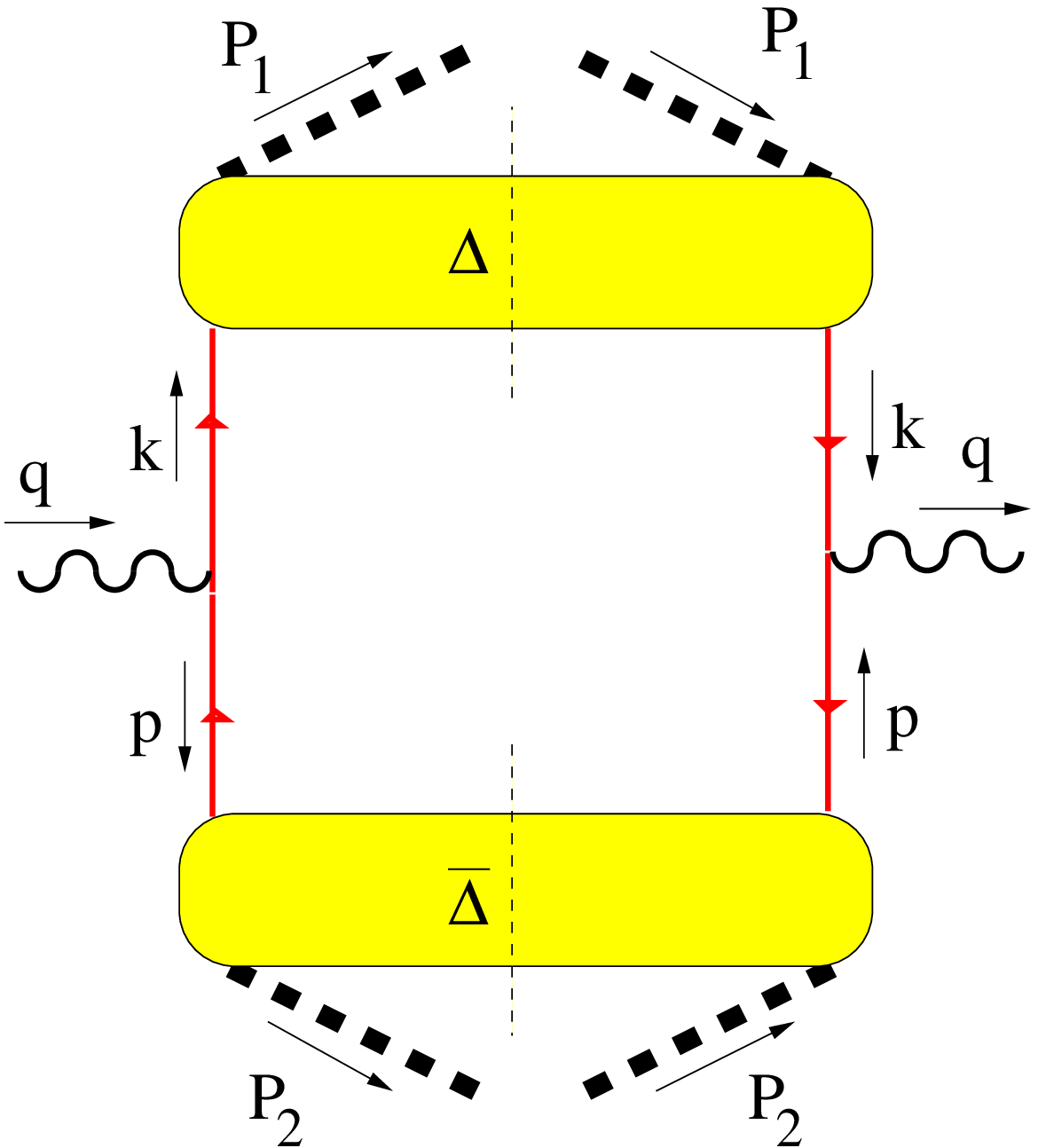, width=4.5cm}
\end{minipage}
\hspace{2cm}
\begin{minipage}{5cm}
\epsfig{file=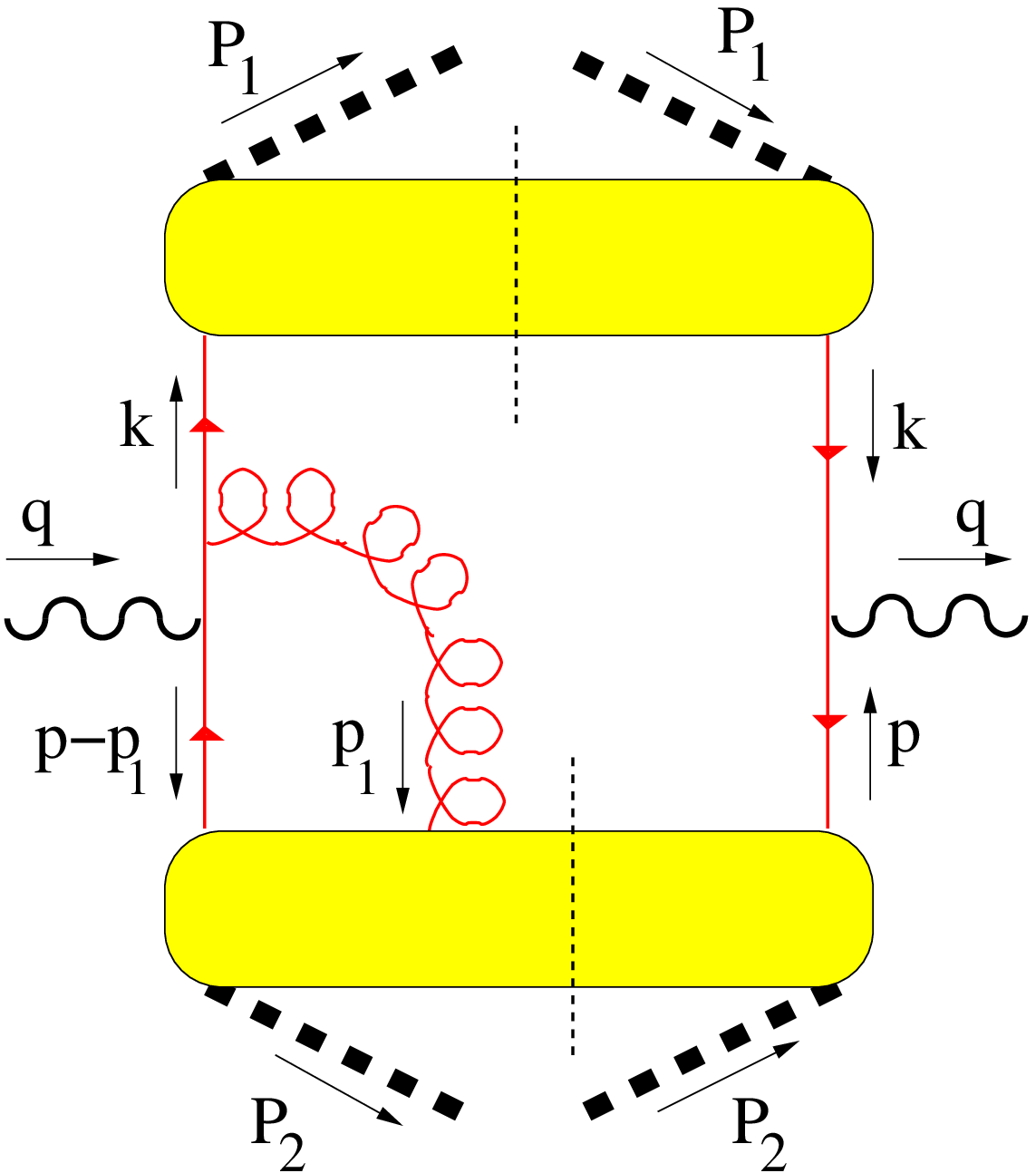, width=4.5cm}
\end{minipage}
\\
(a) \hspace{6.5cm} (b)
\\ \mbox{}
\caption{
Quark-quark (a) and one of the quark-quark-gluon (b) correlators
in tree-level diagrams for
back-to-back jet production in electron-positron annihilation}
\label{fig7}
\end{center}
\end{figure}

Also for 2-particle inclusive electron-positron annihilation we have
a quite similar procedure. The calculation involves two soft fragmentation
parts and the creation of a quark-antiquark pair. We will discuss
only the case of creation from a (timelike) photon. The
handbag diagram is given in Fig.~\ref{fig7}a and an example
of a diagram involving an additional gluon in Fig.~\ref{fig7}b.

The calculation of this tensor in a diagrammatic expansion proceeds
as in the case of leptoproduction and gives
\begin{eqnarray}
{\cal W}_{\mu\nu}(q; P_1,S_1;P_2,S_2)&=&
\int {\rm d}^4p\ {\rm d}^4 k\ \delta^4(p+k-q)\Biggl\{
                \,{\rm Tr}(\overline \Delta(p)\gamma_\mu \Delta(k) \gamma_\nu) \nonumber\\
              & &   - \int {\rm d}^4 p_1\ {\rm Tr}\left(\gamma_\alpha
                  \frac{\slash k + \slash p_1 + m}{(k+p_1)^2 - m^2 +i\epsilon}
                 \gamma_\nu \overline \Delta_A^\alpha(p,p-p_1)\gamma_\mu \Delta(k)\right) \nonumber\\
             & &   -  \int {\rm d}^4 p_1\ {\rm Tr}\left(\gamma_\mu
                  \frac{\slash k + \slash p_1 + m}{(k+p_1)^2 - m^2 -i\epsilon}
                  \gamma_\alpha\Delta(k)\gamma_\nu\overline \Delta_A^\alpha(p-p_1,p)\right) \nonumber\\
           & &     -  \int {\rm d}^4 k_1\ {\rm Tr}\left(\gamma_\nu
                  \frac{-\slash p - \slash k_1+m}{(p+k_1)^2-m^2+i\epsilon}\gamma_\alpha
                  \overline \Delta(p)\gamma_\mu\Delta_A^\alpha(k-k_1,k)\right)\nonumber\\
           & &     -  \int {\rm d}^4 k_1\ {\rm Tr}\left(\gamma_\alpha
                  \frac{-\slash p - \slash k_1+m}{(p+k_1)^2-m^2-i\epsilon}\gamma_\mu
                  \Delta_A^\alpha(k,k-k_1)\gamma_\nu\overline \Delta(p)\right)\Biggr\}
           + \ldots,
\label{ee-eqn1}
\end{eqnarray}
where
\bea
&&\overline \Delta_{ij}(p;P,S) = \sum_X\int \frac{{\rm d}^4 \xi}{(2\pi)^4}
\ e^{-i\,p\cdot\xi}\langle 0 | \overline\psi_j (0)
\vert P_2, X\rangle \langle P_2, X \vert \psi_i (\xi) |0\rangle,
\label{ee-soft1}
\\
&&\overline \Delta^\alpha_{A\,ij}(p,p-p_1;P,S) = \sum_X
\int \frac{{\rm d}^4 \xi}{(2\pi)^4}
\,\frac{{\rm d}^4 \eta}{(2\pi)^4}\ e^{-i\,p\cdot\xi}
\,e^{-i\,p_1\cdot(\eta-\xi)}
\,\langle 0| \overline\psi_j (0) g A^\alpha(\eta)\vert
P_2, X\rangle \langle P_2, X\vert \psi_i (\xi) |0\rangle .
\label{ee-soft3}
\eea
(note that $\overline \Delta_\partial^\alpha(z,p_\st) = -p_\st^\alpha
\,\overline\Delta(z,p_\st)$).

It turns out that the direction
of the links in the fragmentation functions changes in going from
SIDIS to the annihilation process, just as the direction of the links
in distribution functions changed in going from SIDIS to DY.
The $A^-$-gluons connected to
hadron 1, produce a link running along the plus direction to $+\infty$, i.e.\
one finds transverse momentum dependent fragmentation functions
$\Delta^{[+]}(z_1,k_\st)$, where $z_1 \approx P_1^-/k^- \approx P_1^-/q^-$.
Also in the antiquark matrix element the link
runs to $+\infty$ (along minus direction with our choice of lightlike vectors),
i.e.\ the inclusion of $A^+$ gluons lead to the matrix element
$\overline\Delta^{[+]}(z_2, p_\st)$, where
$z_2 \approx P_2^+/p^+ \approx P_2^+/q^+$.

Again, to perform transverse integrations at the level of the hadron
tensor one switches from $n_\pm(P_1,P_2)$ directions to directions
$n_\pm^\prime = (\hat t \pm \hat z)/\sqrt{2}$ with $\hat t = q/Q$. In order
to treat both 1-particle and 2-particle inclusive annihilation it is
convenient \cite{Boer-97} to fix $\hat z$ via one hadron momentum,
for which we will
choose $P_2$, i.e. $\hat z = (q - 2\,P_2/z_2)/Q$. The relation between
$n_\pm^\prime$ and $n_\pm$, then is the same as in leptoproduction.
The orthogonal
direction determining the azimuthal angle of the hadron production
plane, then is $X = -P_{1\perp}/z_1$ which for our purposes equals
$X \approx q_\st \approx q - P_1/z_1 - P_2/z_2$.

\subsection{Integrated annihilation cross section at leading order}

The inclusion of appropriate quark-gluon matrix elements
make the ${\cal O}(g^0)$ result in Eq.~\ref{ee-eqn1} color-gauge invariant,
\bea
{\cal W}_{\mu\nu}^{(0)}(q; P_1,S_1;P_2,S_2) &=&
\int {\rm d}^2p_\st\ {\rm d}^2 k_\st\ \delta^2(p_\st+k_\st-q_\st)
\,{\rm Tr}\left(\overline \Delta^{[+]}(z_2,p_\st)\gamma_\mu
\Delta^{[+]}(z_1,k_\st) \gamma_\nu\right) .
\label{ee-cross-section-1}
\eea
At leading order the integration over transverse momenta of the produced
hadrons can be performed and one obtains the basic result
\bea
&&\int {\rm d}^2X\ {\cal W}_{\mu\nu}^{(0)}(q; P_1,S_1;P_2,S_2)
\ = \ {\rm Tr}\Bigl(\overline\Delta(z_2)\,\gamma_\mu\,\Delta(z_1)\,\gamma_\nu\Bigr)
\Biggr|_{n_\pm \rightarrow n_\pm^\prime}
\ +\ {\cal O}\left(\frac{1}{Q}\right).
\label{ee-leading}
\eea
By taking the free (massless) quark result
$\overline \Delta(z_2) = \slash n_+\,\delta(1-z_2)$
we obtain the 1-particle inclusive result ($q_\st = 0$),
\be
W_{\mu\nu}^{(0)}(q; P_h,S_h)
\ = \ {\rm Tr}\Bigl(\Delta(z_h)
\,\gamma_\nu\gamma^-\gamma_\mu\Bigr)
+ {\cal O}\left(\frac{1}{Q}\right).
\label{ee-integrated-leading}
\ee

\subsection{Azimuthal asymmetries in the annihilation process at leading order}

We consider here cross sections obtained after integration over $X$ and
explicit weighting with $X^\alpha$, i.e.
$\int d^2X\ X^\alpha\,{\cal W}_{\mu\nu}$, which at leading
order gives
\bea
&&\int {\rm d}^2X\ X^\alpha\ {\cal W}_{\mu\nu}^{(0)}(q; P_1,S_1;P_2,S_2)
\nonumber \\ && \qquad \mbox{}
 = {\rm Tr}\Bigl(\overline\Delta(z_2)\,\gamma_\mu
\,\Delta_\partial^{[+]\alpha}(z_1)\,\gamma_\nu\Bigr)
\Biggr|_{n_\pm \rightarrow n_\pm^\prime}
- {\rm Tr}\Bigl(\overline\Delta_\partial^{[+]\alpha}(z_2)\,\gamma_\mu
\,\Delta(z_1)\,\gamma_\nu\Bigr)
\Biggr|_{n_\pm \rightarrow n_\pm^\prime}
+ {\cal O}\left(\frac{1}{Q}\right).
\label{ee-leadingazimuthal}
\eea
By taking $\overline \Delta(z_2) = \slash n_+\,\delta(1-z_2)$
and $\overline \Delta_\partial^{[+]\alpha}(z_2) = 0$, we get the
weighted $e^+e^- \rightarrow {\rm jet}(P_j) + h(P_h) + X$
hadron tensor
\be
\int {\rm d}^2X\ X^\alpha\ {\cal W}_{\mu\nu}^{(0)}(q; P_h,S_h;P_j)
=  {\rm Tr}\Bigl(
\Delta_\partial^{[+]\alpha}(z_h)\,\gamma_\nu\gamma^-\gamma_\mu\Bigr)
\Biggr|_{n_\pm \rightarrow n_\pm^\prime}
+ {\cal O}\left(\frac{1}{Q}\right).
\ee

\subsection{Integrated annihilation cross section at ${\cal O}(1/Q)$}

As before, one must be careful in integrating over transverse
momenta at ${\cal O}(1/Q)$. In particular the correlator $\Delta
\propto \slash n_-$ will lead to terms proportional to
$\slash q_\st/Q\sqrt{2}$. We obtain
\bea
\Delta(z_1,k_\st) & \ =\ & \frac{1}{4}\Bigl(
\slash n_-\,\slash n_+\,\Delta(z_1,k_\st)\,\slash n_+\,\slash n_-\Bigr)
\nonumber \\ & \approx &
\Delta(z_1,k_\st)\Biggr|_{n_\pm \ \rightarrow\ n_\pm^\prime}
- \frac{1}{Q\sqrt{2}}\Bigl(
\slash q_\st\,\slash n_+ \,\Delta(z_1,k_\st)
+ \Delta(z_1,k_\st)\,\slash n_+\,\slash q_\st\Bigr) ,
\eea
which in analogy to leptoproduction leads to the full, integrated
annihilation cross section up to ${\cal O}(1/Q)$,
\bea
&&\int {\rm d}^2X\ {\cal W}_{\mu\nu}^{(0+1)}(q; P_1,S_1;P_2,S_2)
= \ {\cal O}(1)\ \mbox{result [Eq.~\ref{ee-leading}]}
\nonumber \\ && \hspace{1.2cm} \mbox{}
+ \mbox{Tr}\left(
\,\frac{\gamma^-\gamma_\alpha}{Q\sqrt{2}}
\,\gamma_\nu\,\overline\Delta_D^\alpha(z_2)\,\gamma_\mu\,\Delta(z_1)
\right)
%- \mbox{Tr}\left(
%\,\frac{\gamma^-\gamma_\alpha}{Q\sqrt{2}}
%\,\gamma_\mu\,\overline \Delta_D^\alpha(z_2)\,\gamma_\nu\,\Delta(z_1)
%\right)^\ast
%\nonumber \\ && \hspace{1.2cm} \mbox{}
- \mbox{Tr}\left(
\frac{\gamma^+\,\gamma_{\alpha}}{Q\sqrt{2}}\,\gamma_\mu
\,\Delta_D^\alpha (z_1)\,\gamma_\nu \overline\Delta(z_2)\right)
%- \mbox{Tr}\left(
%\frac{\gamma^+\,\gamma_{\alpha}}{Q\sqrt{2}}\,\gamma_\nu
%\,\Delta_D^\alpha (z_1)\,\gamma_\mu \overline\Delta(z_2)\right)^\ast
\nonumber \\ && \hspace{1.2cm} \mbox{}
+ \mbox{Tr}\left(
\frac{\gamma^+\,\gamma_{\alpha}}{Q\sqrt{2}}\,\gamma_\mu
\,\Delta_\partial^{[+]\alpha}(z_1)\,\gamma_\nu \overline\Delta(z_2)
\right)
%+ \mbox{Tr}\left(
%\frac{\gamma^+\,\gamma_{\alpha}}{Q\sqrt{2}}\,\gamma_\nu
%\,\Delta_\partial^{[+]\alpha}(z_1)\,\gamma_\mu \overline\Delta(z_2)
%\right)^\ast
%\nonumber \\ && \hspace{1.2cm} \mbox{}
- {\rm Tr}\Bigl(\gamma_\mu\,\frac{\gamma_\alpha\,\gamma^-}{Q\sqrt{2}}
\,\Delta_\partial^{[+]\alpha}(z_1)\,\gamma_\nu\,\overline\Delta(z_2)\Bigr)
%- {\rm Tr}\Bigl(\gamma_\nu\,\frac{\gamma_\alpha\,\gamma^-}{Q\sqrt{2}}
%\,\Delta_\partial^{[+]\alpha}(z_1)\,\gamma_\mu\,\overline\Delta(z_2)
%\Bigr)^\ast
+ \left( \mu \leftrightarrow \nu\right)^\ast ,
\label{ee-total}
\eea
where the hermiticity properties of the various matrix elements have
been used (see section \ref{TRprop}).

With the choice of $\hat t$ and $\hat z$, the 1-particle inclusive result
is found
by taking $\Delta(z_1) = \slash n_-\,\delta(1-z_1)$,
$\Delta_\partial^{[+]\alpha}(z_1) = \Delta_D^\alpha(z_1) = 0$. After a
(for ease of comparison) change
of plus $\rightarrow$ minus and taking the crossed
(particle $\rightarrow$ antiparticle, $\mu \leftrightarrow \nu$) term
we obtain
\bea
W_{\mu\nu}(q;P_h,S_h)
& = & \ {\cal O}(1)\ \mbox{result [Eq.~\ref{ee-integrated-leading}]}
\nonumber \\ && \mbox{}
- \mbox{Tr}\left(
\,\frac{\gamma^+\gamma_\alpha}{Q\sqrt{2}}
\,\gamma_\mu\,\Delta_D^\alpha(z_h)\,\gamma_\nu\right)
- \mbox{Tr}\left(
\,\frac{\gamma^+\gamma_\alpha}{Q\sqrt{2}}
\,\gamma_\nu\,\Delta_D^\alpha(z_h)\,\gamma_\mu\right)^\ast   .
\eea

\section{Time-reversal properties}
\label{TRprop}

In order to discuss the possible relations between the expressions for the 
various processes given in the previous three sections, we first discuss the
time reversal properties of the correlation functions involved. After that we
will discuss the parametrizations, such that we are able to address the Sivers
effect, the Collins effect and the Qiu-Sterman mechanism (gluonic poles), 
explicitly.

\subsection{Distribution functions}

Hermiticity, parity and time reversal invariance yield conditions
for the correlator $\Phi$, that constrain its parametrization,
\begin{eqnarray}
& & \Phi^\dagger (p;P,S) = \gamma_0 \,\Phi(p;P,S)\,\gamma_0
\quad \hspace{2cm} \mbox{[Hermiticity]}, \\
& & \Phi(p,P,S) = \gamma_0 \,\Phi(\bar p;\bar P,-\bar S)\,
\gamma_0 \quad \hspace{1.8cm}\mbox{[Parity]}, \\
& & \Phi^\ast(p;P,S) = (-i\gamma_5 C)\,\Phi(\bar p;\bar P,
\bar S)\,(-i\gamma_5 C) \quad \mbox{[Time\ reversal]},
\end{eqnarray}
where $C$ = $i\gamma^2 \gamma_0$, $-i\gamma_5 C$= $i\gamma^1\gamma^3$
and $\bar p$ = $(p^0,-\bm p)$. Similar conditions arise for the
fragmentation matrix elements. Including link operators and for
quark-gluon matrix elements slightly different conditions apply.
For the gauge link one has
\be
U^\dagger_{[a,\xi]} = U_{[\xi,a]},
\qquad
{\cal P}\,U_{[a,\xi]}\,{\cal P}^\dagger = U_{[\bar a,\bar \xi]},
\qquad
{\cal T}\,U_{[a,\xi]}\,{\cal T}^\dagger = U_{[-\bar a,-\bar \xi]},
\ee
for which we used $A_\mu^\dagger = A_\mu$,
${\cal P}\,A_\mu(\xi)\,{\cal P}^\dagger = \bar A_\mu(\bar \xi)$ and
${\cal T}\,A_\mu(\xi)\,{\cal T}^\dagger = \bar A_\mu(-\bar \xi)$.
This means that the space-reversed (time-reversed) correlation
function has a different link structure, namely a link running
from $\bar a$ ($-\bar a$)
respectively. However, if the common point is defined with respect to
the two fields in the matrix element, no problem arises. For example the
straight line link with path $z^\mu(s) = (1-s)\,0^\mu + s\,\xi^\mu$ gives
a path $\bar z^\mu$ after applying parity, but after a change of
variables one ends up with the same path; similarly for time reversal.

For the transverse momentum dependent functions with links running along
the minus direction connected at infinity, the situation is
different. The point $\eta^- =
\infty$ is defined by $\eta\cdot n_+ = \infty$, which after parity
transforms into the point $\bar \eta\cdot \bar n_+ = \infty$, but
after time reversal transforms into the point $\bar \eta\cdot \bar n_+
= -\infty$. As a consequence one finds that for the $p^-$-integrated
functions
\bea
& & \Phi^{[+]\dagger} (x,p_\st) = \gamma_0 \,\Phi^{[+]}(x,p_\st)\,\gamma_0
\\
& & \Phi^{[+]}(x,p_\st) = \gamma_0 \,\Phi^{[+]}(x,-p_\st)\,\gamma_0 \\
& & \Phi^{[+]\ast}(x,p_\st)
= (-i\gamma_5 C)\,\Phi^{[-]}(x,-p_\st)\,(-i\gamma_5 C) ,
\label{trev}
\eea
where $\Phi^{[-]}$ is defined with the link running via $\eta^- = -\infty$,
referred to as {\em timelike distribution} in Fig.~\ref{fig5}.
Concluding, in the parametrization of $\Phi^{[+]}(x,p_\st)$
time reversal does not pose constraints. Application of
this operation transforms $\Phi^{[-]}$ into $\Phi^{[+]}$ and vice versa.
T-odd quantities will be defined as the ones that vanish when
$\Phi^{[-]} = \Phi^{[+]}$. Accounting for the transformation in Dirac
space and the sign change of $p_\st$, we define
\bea
2\,\Phi^{\rm [T-even]}(x,p_\st) & \equiv & \Phi^{[+]}(x,p_\st)
+ (-i\gamma_5 C)\,\Phi^{[+]\ast}(x,-p\st)\,(-i\gamma_5 C)
\nonumber \\ & = &
\Phi^{[+]}(x,p_\st) +  \Phi^{[-]}(x,p_\st),
\label{phieven}
\\[0.1cm]
2\,\Phi^{\rm [T-odd]}(x,p_\st) & \equiv & \Phi^{[+]}(x,p_\st)
- (-i\gamma_5 C)\,\Phi^{[+]\ast}(x,-p\st)\,(-i\gamma_5 C)
\nonumber \\ & = &
\Phi^{[+]}(x,p_\st) -  \Phi^{[-]}(x,p_\st).
\label{phiodd}
\eea
Note that the name `T-odd'
does {\em not\/} imply a violation of time reversal invariance.
For the integrated distributions the different links ($[\pm]$) merge into
one, $\Phi(x) = \Phi^{[+]}(x) = \Phi^{[-]}(x)$
and $\Phi(x)$ is T-even. For transverse momentum dependent correlations
the sum of $\Phi^{[+]}$ and $\Phi^{[-]}$, i.e.~$\Phi_\partial^\alpha(x)$,
is T-even, while the difference, related to $\Phi_G^\alpha(x,x)$
(see Eq.~\ref{gluonic pole}), is T-odd.
Summarizing, we have
\bea
\Phi^{[\pm]}(x,p_\st)
& = &
\Phi^{\rm [T-even]}(x,p_\st)
\pm \Phi^{\rm [T-odd]}(x,p_\st),
\label{Phikt}
\eea
and for the integrated and weighted distribution correlators
\bea
&&
\Phi(x)
= \Phi^{\rm [T-even]}(x),
\\ &&
\Phi_{\partial}^{[\pm]\,\alpha}(x) =
\Phi_{\partial}^{\alpha\,{\rm [T-even]}}(x)
\pm \pi\,\Phi_{G}^{\alpha\,{\rm [T-odd]}}(x,x) .
\label{Phix}
\eea
These results imply that T-odd distribution functions, e.g.~the
Sivers effect appearing in single spin azimuthal asymmetries
in leptoproduction ($\Phi_\partial^{[+]\alpha}$) and in Drell-Yan
scattering ($\Phi_\partial^{[-]\alpha}$) are opposite in
sign \cite{Collins-02}.
As shown in Ref.~\cite{BBHM} the T-odd functions can be considered
as imaginary parts in a helicity matrix representation, leading to
the representation in Fig.~\ref{fig8}a.
The behavior under time reversal of gluonic matrix elements,
like $\Phi_A$, $\Phi_D$ and $\Phi_G$, can also be studied separately.
It turns out that the quantity $\Phi_D(x)$ is T-even, while
$\Phi_G^\alpha(x,x)$ is T-odd.
The latter can also be seen in $A^+ = 0$ gauge. Using relation
Eq.~\ref{difference} for $\xi = - \infty^-$
in the matrix elements in Eq.~\ref{defphiG} and \ref{boundary} yields
$2\pi\,\Phi_G^\alpha(x,x)$ =
$\Phi^\alpha_{A(\infty)}(x,x) - \Phi^\alpha_{A(-\infty)}(x,x)$,
i.e.\ the gluonic pole matrix element is the difference of the
boundary terms that transform into each other under time reversal.

\subsection{Fragmentation functions}

Constraints on the correlator $\Delta$ come from hermiticity,
parity and time reversal invariance.
The essential difference with distribution functions
is that time reversal transforms
the out-states in the definition of fragmentation matrix elements into
in-states. Taking this into account and explicitly adding subscripts
in and out, one obtains the conditions,
\bea
& & \Delta_{\rm out}^{[-]\dagger} (z,k_\st)
= \gamma_0 \,\Delta_{\rm out}^{[-]}(z,k_\st)\,\gamma_0
\\
& & \Delta_{\rm out}^{[-]}(z,k_\st)
= \gamma_0 \,\Delta_{\rm out}^{[-]}(z,-k_\st)\,\gamma_0 \\
& & \Delta_{\rm out}^{[-]\ast}(z,k_\st)
= (-i\gamma_5 C)\,\Delta_{\rm in}^{[+]}(z,-k_\st)\,(-i\gamma_5 C) ,
\eea
where $\Delta^{[\pm]}$ are the
spacelike/timelike fragmentation
functions, illustrated in Fig.~\ref{fig5}.
Defining $\Delta_{\rm O}$ and $\Delta_{\rm FSI}$ as sum and differences
of matrix elements with out and in-states respectively, one has
\bea
4\,\Delta_{\rm O}^{\rm [T-even]}(z,k_\st) & = &
\Delta_{\rm out}^{[+]}(z,k_\st) +  \Delta_{\rm out}^{[-]}(z,k_\st)
+\Delta_{\rm in}^{[+]}(z,k_\st) +  \Delta_{\rm in}^{[-]}(z,k_\st),
\label{delta1}
\\[0.1cm]
4\,\Delta_{\rm O}^{[\rm T-odd]}(z,k_\st)\ & = &
\Delta_{\rm out}^{[+]}(z,k_\st) -  \Delta_{\rm out}^{[-]}(z,k_\st)
+\Delta_{\rm in}^{[+]}(z,k_\st) -  \Delta_{\rm in}^{[-]}(z,k_\st),
\label{delta2}
\\[0.1cm]
4\,\Delta_{\rm FSI}^{\rm [T-odd]}(z,k_\st)\ & = &
\Delta_{\rm out}^{[+]}(z,k_\st) +  \Delta_{\rm out}^{[-]}(z,k_\st)
-\Delta_{\rm in}^{[+]}(z,k_\st) -  \Delta_{\rm in}^{[-]}(z,k_\st),
\label{delta3}
\\[0.1cm]
4\,\Delta_{\rm FSI}^{\rm [T-even]}(z,k_\st) & = &
\Delta_{\rm out}^{[+]}(z,k_\st) -  \Delta_{\rm out}^{[-]}(z,k_\st)
-\Delta_{\rm in}^{[+]}(z,k_\st) +  \Delta_{\rm in}^{[-]}(z,k_\st),
\label{delta4}
\eea
which implies
\bea
\Delta_{\rm out}^{[\pm]}(z,k_\st)
& = &
\left[ \Delta_{\rm O}^{\rm [T-even]}(z,k_\st)
+ \Delta_{\rm FSI}^{\rm [T-odd]}(z,k_\st)\right]
\pm \left[ \Delta_{\rm O}^{\rm [T-odd]}(z,k_\st)
+ \Delta_{\rm FSI}^{\rm [T-even]}(z,k_\st)\right] .
\label{Deloutkt}
\eea
and for the integrated and weighted fragmentation functions
\bea
&&
\Delta_{\rm out}(z)
= \Delta_{\rm O}^{\rm [T-even]}(z) + \Delta_{\rm FSI}^{\rm [T-odd]}(z),
\label{Deloutz1}
\\ &&
\Delta_{\partial\,{\rm out}}^{[\pm]\,\alpha}(z) =
\underbrace{\bigl[ \Delta_{\partial\,{\rm O}}^{\alpha\,{\rm [T-even]}}(z)
+ \Delta_{\partial\,{\rm FSI}}^{\alpha\,{\rm [T-odd]}}(z)\bigr]}_{
\Delta_{\partial\,{\rm out}}^\alpha(z)}
\pm \underbrace{\bigl[
\pi\,\Delta_{G\ {\rm O}}^{\alpha\,{\rm [T-odd]}}(z,z)
+ \pi\,\Delta_{G\ {\rm FSI}}^{\alpha\,{\rm [T-even]}}(z,z)\bigr]}_{
\pi\,\Delta_{G\,{\rm out}}^\alpha(z,z)} .
\label{Deloutz}
\eea
The essential difference with the distribution functions is that for
the fragmentation functions the differences between in and out states
become relevant.
These constitute final state interactions {\em within\/} the soft part (here
labelled by `FSI'), decoupled from the quark and
gluon operators that make the connection to the hard scattering part. The
effects arising from the difference between $[+]$ and $[-]$ are labelled by a
subscript `O'. Note that in the literature \cite{BHMSS-02,BHS,BHS2} this is
also referred to as initial or final state interactions, depending on the
process under consideration.
The behavior of the various correlators is given in Table \ref{table2}.
Therefore, in contrast to the distribution functions, $\Delta_\partial$
and $\Delta_G$ contain T-even {\em and} T-odd parts.
Also the correlators $\Delta_D$ and $\tilde \Delta_A$ contain T-even
and T-odd parts.

\begin{table}[t]
\caption{Summary of time reversal allowed (yes) and forbidden (no) parts
in integrated or weighted distribution and fragmentation correlators.
\label{table2}}
\mbox{}
\begin{tabular}{l|cc||l|cc||l|cc||l|cc|}
& T-even & T-odd & & T-even & T-odd &
& T-even & T-odd & & T-even & T-odd \\
\hline
$\Phi$ & yes & no & $\Delta$ & yes & yes &
$\Delta_{\rm O}$ & yes & no &$\Delta_{\rm FSI}$ & no & yes \\
$\Phi_\partial$ & yes & no & $\Delta_\partial$ & yes & yes &
$\Delta_{\partial\,{\rm O}}$ & yes & no &
$\Delta_{\partial\,{\rm FSI}}$ & no & yes \\
$\Phi_D$ & yes & no & $\Delta_D$ & yes & yes &
$\Delta_{D\,{\rm O}}$ & yes & no &
$\Delta_{D\,{\rm FSI}}$ & no & yes \\
$\tilde\Phi_A$ & yes & no & $\tilde \Delta_A$ & yes & yes &
$\tilde\Delta_{A\,{\rm O}}$ & yes & no &
$\tilde \Delta_{A\,{\rm FSI}}$ & no & yes \\
$\Phi_G$ & no & yes & $\Delta_G$ & yes & yes &
$\Delta_{G\,{\rm O}}$ & no & yes &
$\Delta_{G\,{\rm FSI}}$ & yes & no \\
\end{tabular}
\end{table}

At first sight Eqs~\ref{Deloutkt} and \ref{Deloutz} seem to imply 
a breaking of
universality. Particular transverse momentum dependent functions
obtained as Dirac projections of $\Delta_{\rm out}^{[\pm]}(z,k_\st)$
give unrelated T-odd (and also T-even)
results for the two possible link configurations,
indicated with superscripts $[+]$ or $[-]$. In
$\Delta_{\partial\,{\rm out}}^{[\pm]\alpha}$ there arise T-odd parts
from $\Delta_{\partial\,{\rm FSI}}^{\alpha}$ {\em and} from
$\pi\Delta_{G\,{\rm O}}^{\alpha}$, the sign in front of the latter
coupled to the link structure.
Thus, the T-odd (Collins) effects in pion leptoproduction
and in electron-positron annihilation are a priori not identical.
The fact that a certain azimuthal asymmetry arises from a combination
of correlation functions ($\Delta_\partial$ and $\Delta_G$), however, need not
imply a breaking of universality. It is quite similar to the $q_\st$-integrated
structure functions of different processes which involve different flavor
weights. It seems possible that a factorization proof can be
established for the correlation functions $\Delta_\partial$ and
$\Delta_G$ separately.

\begin{figure}[bt]
\begin{center}
\epsfig{file=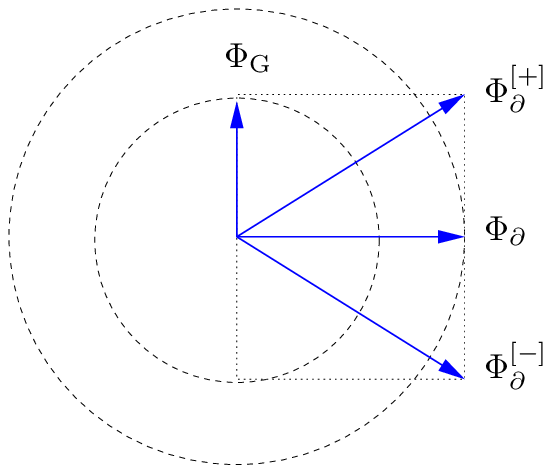, width=4.8cm}
\hspace{0.5cm}
\epsfig{file=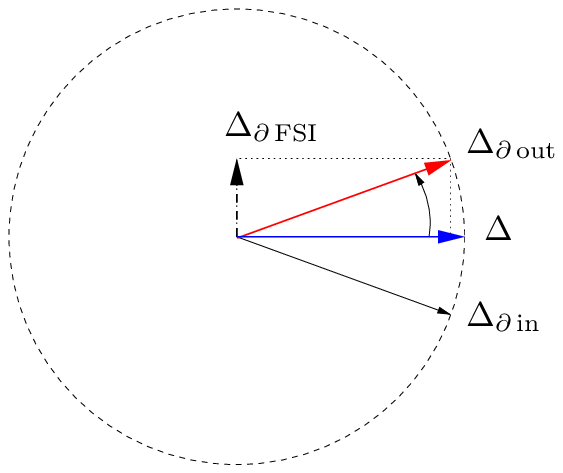, width=4.8cm}
\hspace{0.5cm}
\epsfig{file=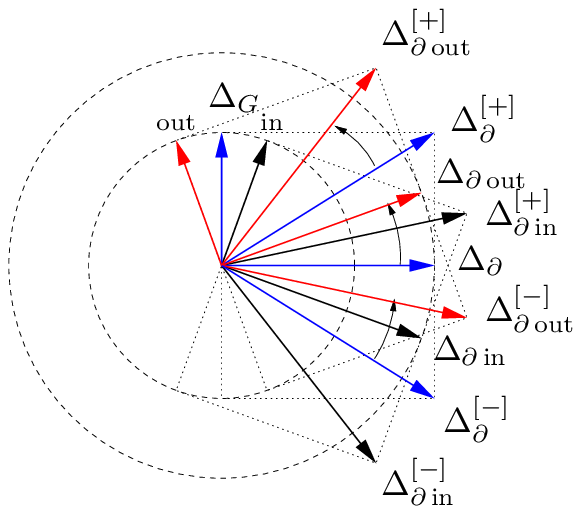, width=5.2cm}
\\[0.2cm]
(a) \hspace{4.3cm} (b) \hspace{5.3cm} (c)
\\ \mbox{}
\caption{
Illustration of the role of gauge links ($[+]$ versus $[-]$) and the role of
final states ([in] versus [out]) in T-odd effects for distribution and
fragmentation functions.}
\label{fig8}
\end{center}
\end{figure}

In Figs \ref{fig8} we illustrate the differences between $\Delta^{[\pm]}$
and those between the $\Delta_{\rm in/out}$. For the former one can
argue that
forgetting about the effect from in- and out-states one has the same
situation as for distribution functions where the T-odd parts can be
considered as imaginary parts of helicity amplitudes (shown as
the projections along the vertical axis in Fig.~\ref{fig8}a).
For $\Delta_{\rm in/out}$ one can use the fact that in- and out-states
can be obtained by different M\"oller operators, allowing a
connection of $\Delta_{\rm in}$ and $\Delta_{\rm out}$ via a unitary
operation, illustrated in Fig.~\ref{fig8}b.
The combined effect is illustrated in Fig.~\ref{fig8}c. It shows that 
the T-odd parts of 
$\Delta_{\partial \,{\rm out}}^{[+]}$ in $e^+ e^-$ and
$\Delta_{\partial \,{\rm
out}}^{[-]}$  in SIDIS are in general not equal in magnitude.

Actually, not only the T-odd effects acquire contributions from both terms
in $\Delta_\partial \pm \Delta_G$, but also the T-even effects, hence
affecting all comparisons of azimuthal asymmetries in leptoproduction
and electron-positron annihilation.

\subsection{Parametrization of quark and quark-gluon correlation functions}

Here we will discuss the parametrizations for transverse momentum
dependent distribution functions and fragmentation functions.
The parametrizations of $\Phi(x,p_\st)$ and $\Delta(z,k_\st)$ 
consistent with the conditions 
imposed by hermiticity and parity, including the
parts proportional to $(M/P^+)^0$ and $(M/P^+)^1$ are given in 
Appendix~\ref{appendix-3} \cite{MT96,Boer-Mulders-98}.
In principle the functions could differ depending on the $\pm$ gauge link
structure for both distribution and fragmentation functions
and also on the O/FSI characterization for fragmentation functions.
%Eqs \ref{phieven} and \ref{phiodd} for the distribution functions tell us
%that the T-even functions are $\pm$ independent, $f_1^{[\pm]}$
%= $f_1^{\rm [T-even]}$, etc., while the T-odd functions change sign,
%$f_{1T}^{[\pm]}$ = $\pm f_{1T}^{\rm [T-odd]}$, while for
%fragmentation functions, Eqs~\ref{delta1}-\ref{delta4}
%indicate which kinds of functions to expect.

We will first discuss the parametrizations for the transverse momentum
integrated and then the once-weighted functions (transverse moments).
As explained in Section IV-F, we do not address twice-weighted
functions (relevant for $k_\st$ broadening and the average $k_\st^2$ in jets)
in view of potential problems with factorization.
%expected of potential?

The result for the correlator $\Phi$ after integration is
\begin{eqnarray}
\Phi(x) & = &
\frac{1}{2}\,\Biggl\{
f_1(x)\,\slash n_+
+ S_{\sL}\,g_1(x)\, \gamma_5\,\slash n_+
+ h_1(x)\,\frac{\gamma_5\,[\slash S_\st,\slash n_+]}{2}\Biggr\}
\nonumber \\ & &
+ \frac{M}{2P^+}\Biggl\{
e(x) + g_T(x)\,\gamma_5\,\slash S_\st
+ S_{\sL}\,h_L(x)\,\frac{\gamma_5\,[\slash n_+,\slash n_-]}{2} \Biggr\}
\label{evenint}
%\\
%\Phi^{\rm [T-odd]}(x) & = &
%\frac{M}{2P^+}\Biggl\{
%-S_{\sL}\,e_L(x)\,i\gamma_5
%- f_T(x)\,\epsilon_\st^{\rho\sigma}\gamma_\rho S_{\st\sigma}
%+ h(x)\,\frac{i\,[\slash n_+,\slash n_-]}{2} \Biggr\} = 0.
\end{eqnarray}
For the distribution functions the T-odd part vanishes, i.e.\ 
$e_L(x) = f_T(x) = h(x) = 0$.
For the parametrization of fragmentation functions one has both T-even
and T-odd functions, i.e.~the functions $E_L(z)$, $D_T(z)$ and
$H(z)$ appear. Explicitly,
\begin{eqnarray}
\Delta_{\rm out}(z) & = &
zD_1(z)\,\slash n_-
+ S_{h\sL}\,zG_1(z)\, \gamma_5\,\slash n_+
+ zH_1(z)\,\frac{\gamma_5\,[\slash S_{h\st},\slash n_-]}{2}
\nonumber \\ & &
+ \frac{M_h}{P_h^-}\Biggl\{
zE(z) + zG_T(z)\,\gamma_5\,\slash S_{h\st}
+ S_{h\sL}\,zH_L(z)\,\frac{\gamma_5\,[\slash n_-,\slash n_+]}{2}
\nonumber \\ && \qquad\quad
-S_{h\sL}\,zE_L(z)\,i\gamma_5
+ zD_T(z)\,\epsilon_\st^{\rho\sigma}\gamma_\rho S_{h\st\sigma}
+ zH(z)\,\frac{i\,[\slash n_-,\slash n_+]}{2} \Biggr\} .
\end{eqnarray}
The T-even functions are `O'-type, the T-odd functions are `FSI'-type,
since 
$\Delta_{\rm out}^{\rm [T-even]}(z)  = \Delta_{\rm O}^{\rm [T-even]}(z)$
and
$\Delta_{\rm out}^{\rm [T-odd]}(x) = \Delta_{\rm FSI}^{\rm [T-odd]}(z)$,
while the link structure does not play a role. T-odd functions only
appear in sub-leading parts of the cross sections.

Next we consider the $p_\st$-weighted results
referred to as transverse moments. Restricting ourselves to the leading
$(M/P^+)^0$ part, we write for the correlators $\Phi_\partial^{\alpha [\pm]}$
(Eq.~\ref{transvmom}) 
\bea
\Phi_\partial^{\alpha [\pm]} (x) & = &
\frac{M}{2}\,\Biggl\{
g_{1T}^{(1)[\pm]}(x)\,S_\st^\alpha\,\gamma_5\,\slash n_+
-S_{\sL}\,h_{1L}^{\perp (1) [\pm]}(x)
\,\frac{\gamma_5\,[\gamma^\alpha,\slash n_+]}{2}
\nonumber \\ && \qquad
-f_{1T}^{\perp (1) [\pm]} (x)
\,\epsilon^{\alpha}_{\ \ \mu\nu\rho}\gamma^\mu n_+^\nu {S_\st^\rho}
- h_1^{\perp (1)[\pm]} (x)
\,\frac{i[\gamma^\alpha, \slash n_+]}{2}\Biggr\},
%\label{Phidodd}
\eea
where we have defined $\bm p_\st^2/2M^2$-moments (transverse moments) as
\be
g_{1T}^{(1)}(x) = \int d^2p_\st\ \frac{\bm p_\st^2}{2M^2}
\,g_{1T}(x,\bm p_\st^2),
\ee
and similarly for the other functions.

For the transverse moments, the average of the $\pm$ correlators
in Eq.~\ref{Phid} is T-even, i.e.
\bea
\Phi_\partial^{\alpha} (x) & = &
\frac{M}{2}\,\Biggl\{
g_{1T}^{(1)}(x)\,S_\st^\alpha\,\gamma_5\,\slash n_+
-S_{\sL}\,h_{1L}^{\perp (1)}(x)
\,\frac{\gamma_5\,[\gamma^\alpha,\slash n_+]}{2}
\Biggr\},
\label{Phideven}
\eea
while the gluonic pole contribution
$\Phi_G^\alpha(x,x)$ in Eq.~\ref{gluonic pole} is T-odd. Writing
\bea
\pi\,\Phi_G^{\alpha} (x,x) &= &
\frac{M}{2}\,\Biggl\{
-\tilde f_{1T}^{\perp (1)}(x)
\,\epsilon^{\alpha}_{\ \ \mu\nu\rho}\gamma^\mu n_+^\nu {S_\st^\rho}
- \tilde h_1^{\perp (1)}(x)
\,\frac{i[\gamma^\alpha, \slash n_+]}{2}\Biggr\} ,
\label{PhiGodd}
\eea
we have the relations
\bea
&&
g_{1T}^{(1)[\pm]}(x) = g_{1T}^{(1)}(x),
\\ &&
h_{1L}^{\perp (1) [\pm]}(x) = h_{1L}^{\perp (1)}(x) ,
\\ &&
f_{1T}^{\perp (1)[\pm]}(x) = \pm \tilde f_{1T}^{\perp (1)}(x) ,
\label{SiversQS}
\\ &&
h_1^{\perp (1)[\pm]}(x) = \pm \tilde h_1^{\perp (1)}(x) .
\label{SiversQS-chiralodd}
\eea
These results show that azimuthal spin asymmetries involving the
distribution functions $g_{1T}^{(1)}$ and
$h_{1L}^{(1)}$ are process independent,
while those involving the functions $f_{1T}^{\perp (1)}$ and
$h_1^{\perp(1)}$ change sign.
Eq.\ \ref{SiversQS} represents
the explicit connection between the Sivers effect
(l.h.s.) and the Qiu-Sterman effect (r.h.s., cf.\ Ref.\ \cite{Boer4}); Eq.\
\ref{SiversQS-chiralodd} is its chiral-odd counterpart.
We note that due to the presence of gluonic pole effects
the evolution equations of $f_{1T}^{\perp (1)}$ and
$h_1^{\perp(1)}$ \cite{Henneman} need to be reconsidered.
These functions originate solely from gluonic
pole effects.

For the fragmentation functions, not only the parametrization for
$\Delta_\partial^{\alpha [\pm]}$ contains both T-even and T-odd parts,
but also the average contains T-even and T-odd parts, i.e.
\bea
\Delta_\partial^{\alpha} (z) & = &
M_h\Biggl\{zG_{1T}^{(1)}(z)\,S_\st^\alpha\,\gamma_5\,\slash n_-
-S_{h\sL}\,zH_{1L}^{\perp (1)}(z)
\,\frac{\gamma_5\,[\gamma^\alpha,\slash n_-]}{2}
\nonumber \\ && \qquad\quad
-zD_{1T}^{\perp (1)} (z)
\,\epsilon^{\alpha}_{\ \ \mu\nu\rho}\gamma^\mu n_-^\nu {S_{h\st}^\rho}
- zH_1^{\perp (1)} (z)
\,\frac{i[\gamma^\alpha, \slash n_-]}{2}
\Biggr\},
\label{Deltad}
\eea
For the gluonic pole contribution one obtains
\bea
\pi\,\Delta_G^{\alpha} (z,z) &= &
M_h\,\Biggl\{
z\tilde G_{1T}^{(1)}(z)\,S_{h\st}^\alpha\,\gamma_5\,\slash n_-
-S_{h\sL}\,z\tilde H_{1L}^{\perp (1)}(z)
\,\frac{\gamma_5\,[\gamma^\alpha,\slash n_+]}{2}
\nonumber \\ && \qquad\quad
-z\tilde D_{1T}^{\perp (1)}(z)
\,\epsilon^{\alpha}_{\ \ \mu\nu\rho}\gamma^\mu n_-^\nu {S_{h\st}^\rho}
- z\tilde H_1^{\perp (1)}(z)
\,\frac{i[\gamma^\alpha, \slash n_-]}{2}\Biggr\} .
\eea
and we obtain for the transverse moment in $\Delta_\partial^{\alpha [\pm]}$,
\bea
&&
G_{1T}^{(1)[\pm]}(z) = G_{1T}^{(1)}(z) \pm \tilde G_{1T}^{(1)}(z),
\\ &&
H_{1L}^{\perp (1) [\pm]}(z) = H_{1L}^{\perp (1)}(z)
\pm \tilde H_{1L}^{\perp (1)}(z),
\\ &&
D_{1T}^{\perp (1)[\pm]}(z) = D_{1T}^{\perp (1)}(z)
\pm \tilde D_{1T}^{\perp (1)}(z) ,
\\ &&
H_1^{\perp (1)[\pm]}(z) = H_1^{\perp (1)}(z) \pm \tilde H_1^{\perp (1)}(z) .
\label{collinsf}
\eea
The occurrence of out-states in the fragmentation matrix elements
is responsible for the appearance of the T-even functions
$\tilde G_{1T}^{(1)}$ and $\tilde H_{1L}^{\perp (1)}$ and the
T-odd functions $D_{1T}^{\perp (1)}$ and $H_1^{\perp (1)}$.
We note that these results differ from Ref.\ \cite{MT96}.
For example, Eq.~\ref{collinsf} shows the explicit forms of the Collins
function in $e^+e^-$ (plus sign) and SIDIS (minus sign), respectively.
Furthermore, the above results imply that the evolution equations of
$G_{1T}^{(1)}, H_{1L}^{\perp (1)},
D_{1T}^{\perp (1)}$ and $H_1^{\perp (1)}$
\cite{Henneman} also need to be reconsidered.

For functions weighted twice with a transverse momentum in
$\Phi_{\partial\partial}^{\alpha\beta}$, one needs higher transverse
moments of the functions, such as $h_{1T}^{\perp (2)}(x)$. Relations
for these functions involve not only twist-three, but also twist-four
parts in the correlators. This includes the simple $p_\st^2$ average in
$\Phi_{\partial^2}(x)$.
As mentioned above, for these functions we do not expect a simple
process dependence as given for the once-weighted results, but this requires
further investigation.

In Refs.~\cite{Jaffe-Ji-92,Boer4} parametrizations have been given for
two-argument quark-gluon correlations. As shown
in Eq.~\ref{phid1} one finds after integration correlation functions
with $D^\alpha\psi(x)$ at the same point, for which the QCD equations
of motion can be used.
These relate $\Phi_D(x)$ (see Eq.~\ref{phid1})
and also $\gamma_0\,\Phi_D(x)\,\gamma_0$
to $\Phi(x)$; explicitly,
\bea
\Phi_D^{\alpha} (x) &= &
\frac{M}{2}\,\Biggl\{
\left(x\,g_{T}(x)- \frac{m}{M}\,h_1(x)\right)\,S_\st^\alpha\,\gamma_5\,\slash n_+
+S_{\sL}\left(x\,h_{L}(x)-\frac{m}{M}\,g_1(x)\right)
\,\frac{\gamma_5\,[\gamma^\alpha,\slash n_+]}{4}
\nonumber \\ && \qquad
-\left(x\,e(x) - \frac{m}{M}\,f_1(x)\right)
\,\frac{[\gamma^\alpha, \slash n_+]}{4}\Biggr\},
\eea
which contains only T-even functions.
For fragmentation functions one also finds T-odd functions,
\bea
\Delta_D^{\alpha} (z) &= &
M_h\,\Biggl\{
\left(G_{T}(z)- \frac{m}{M_h}\,zH_1(z)\right)\,S_{h\st}^\alpha\,\gamma_5\,\slash n_-
+S_{h\sL}\left(H_{L}(z)-\frac{m}{M_h}\,zG_1(z)\right)
\,\frac{\gamma_5\,[\gamma^\alpha,\slash n_-]}{4}
\nonumber \\ && \qquad\quad
-\left(E(z) - \frac{m}{M_h}\,zD_1(z)\right)
\,\frac{[\gamma^\alpha, \slash n_-]}{4}
\nonumber \\ && \qquad\quad
+D_T(z)
\,\epsilon^{\alpha}_{\ \ \mu\nu\rho}\gamma^\mu n_-^\nu {S_{h\st}^\rho}
+ H(z)
\,\frac{i[\gamma^\alpha, \slash n_-]}{4}
+E_L(z)
\,\frac{i\gamma_5\,[\gamma^\alpha, \slash n_-]}{4}\Biggr\}.
\eea

Besides the relations resulting from the equations of motion,
also Lorentz invariance may lead to relations between
correlators~\cite{Bukhvostov,MT96,Henneman}.
As can be seen from the explicit
treatment in Ref.~\cite{joao}, these relations are derived from
the Lorentz structure of non-integrated quark-quark correlators
as in Eq.~\ref{soft1}. At present it is not clear how matrix elements
of $A_\st$ fields at infinity and hence the link structure
play a role in these relations (see also Ref.~\cite{Goeke-Metz}).

\section{Summary and conclusions}

In this paper we have analyzed transverse momentum dependent distribution
and fragmentation functions appearing in several hard processes in which
at least two hadrons are involved in initial and/or final state. In these
processes one has besides a hard scale $Q$, a non-collinearity $q_\st$ which
is characterized by a hadronic scale $Q_\st$ and an azimuthal angle.
We have shown explicitly, using the results of Belitsky et al.~\cite{Belitsky}
how quark-quark-gluon matrix elements lead to fully color
gauge invariant definitions of the correlation functions that appear in
leading and first sub-leading order in $1/Q$ in the hadron tensor of
hard processes. The gluon fields appear in the gauge link connecting
the quark fields. The transverse gluons needed in the gauge link for
transverse momentum dependent functions involve gluon fields at lightlike
infinity. The fact that the gluonic effects can be cast into (conjugate)
links attached to the two (conjugate) quark fields still allows for an
interpretation of these functions as probability densities.

The structure of the gauge links in hard processes is not always the same.
In particular the gauge links in distribution functions in SIDIS and the
DY process run in opposite lightlike directions indicated with
indices $\pm$. The two different correlators are connected via a
time reversal operation.
Similarly, the gauge links in fragmentation functions
in SIDIS and electron-positron annihilation run in opposite lightlike
directions. At leading order, the difference between $\pm$ correlators
vanishes upon integration over transverse momenta.
In $q_\st$-weighted cross sections, one finds correlators weighted
with transverse momentum (transverse moments), which are dependent on
the ($\pm$) link structure. The same quantities appear in
subleading integrated cross sections.
The difference between transverse moments with different ($\pm$)
link structure corresponds to a (color gauge invariant)
gluonic pole matrix element, the word pole referring to the fact
that one deals with a zero-momentum gluon field. This matrix element
appears in different processes as the Qiu-Sterman effect.

Considering the behavior under time reversal for distribution functions,
it turns out that the gluonic pole contributions coincide with
T-odd effects, leading to single spin asymmetries, like the Sivers effect.
This establishes the direct connection between the Sivers effect and the
Qiu-Sterman effect.
Since for distribution functions gluonic poles are the only source
of T-odd effects, one finds that these effects have opposite signs in
SIDIS and the DY process.
For fragmentation functions, T-odd effects leading to single spin
asymmetries, like the Collins effect in pion leptoproduction, arise
not only from the gluonic
pole contribution, but also from final state interactions. The latter
are purely soft interactions in the fragmentation part and has the
same sign in different processes. Hence the T-odd effects in SIDIS and
electron-positron annihilation are not connected by a simple sign relation.
This is not a breaking of universality, but rather the appearance
of different combinations of fragmentation functions
in different processes. The T-odd effects
will not only appear in Collins asymmetries in SIDIS or
electron-positron annihilation, but likely also in other processes.
We emphasize the importance of an analysis of the link structure in
processes like $p\,p^\uparrow \rightarrow \pi\,X$. We note that also
for distribution functions T-odd contributions with the same sign in different
processes could arise if time reversal is realized in a nonstandard
way~\cite{Drago}. This would spoil the simple sign switches in
Eqs \ref{SiversQS} and \ref{SiversQS-chiralodd}.

Gluonic pole contributions appear in the azimuthal asymmetries, in
the processes that we have considered, with a particular sign. This
affects not only T-odd, but (in case of fragmentation)
also T-even azimuthal asymmetries.
The transverse moments with a different link structure differ by
`effective' twist-three functions. The consequences of these gluonic
pole contributions on the evolution of the transverse moments have
not been considered. Given the known operator structure of the
contributions, however, such a study ought to be doable.

Although in trying to model distribution or fragmentation
functions~\cite{joao,goeke}
one is always stuck with the problem of evolution, modelling
has been proven
useful to illustrate several effects~\cite{BKMM,BHS,BHS2,Metz}, such as
the sign change in the Sivers functions $f_{1T}^{\perp\,[\pm]}$ and the
sign behavior for the Collins function $H_1^{\perp\,[\pm]}$. As seen from
Eq.~\ref{collinsf}, the latter can have two contributions with
different signs. It is not clear whether both contributions are
present in the models studied.

Our general analysis of the various correlators in leading and subleading
order single spin and azimuthal asymmetries in hard processes may help to
analyze the results of the first generation of experiments that
presently are being performed or planned
by, for instance, HERMES (DESY), COMPASS (CERN), BELLE (KEK) and the
results that are obtained by looking at existing LEP data. We
hope to see the emergence of a coherent picture of these
asymmetries.

\acknowledgments
We acknowledge discussions with A.~Bacchetta, A.~Belitsky, S.J.~Brodsky,
J.C.~Collins,
D.S.~Hwang, P.~Hoyer, X.~Ji, A.~Metz, and J.G.~Milhano. The research
of D.B.~has been made possible by financial support from the Royal
Netherlands Academy of Arts and Sciences.

\appendix

\section{Transverse gluon fields and the field strength tensor}
\label{appendix-1}

In lightcone gauge $A^+ = 0$, the relation between $A_\st^\alpha$ and
$G^{+\alpha}$ becomes
\be
\bigl[\,\partial_x^+, A_\st^\alpha(x)\,\bigr]
= G^{+\alpha}(x),
\ee
which can be inverted to yield $A_\st^\alpha$ in terms of a boundary
term and an integral along the minus direction with $G^{+\alpha}$ in
the integrand.
Without gauge choice we have
\be
ig\,G^{+\alpha}(x) = \bigl[\,iD^+(x) , iD^\alpha(x)\,\bigr]
= \bigl[\,iD^+(x), gA_\st^\alpha(x)\,\bigr]
- ig\,\bigl[ \partial_{x}^\alpha , A^+(x)\,\bigr]
\ee
as our starting point. Next we multiply from left and right with link
operators $U^-_{[a,x]}$ and $U^-_{[x,a]}$ respectively,
built from $A^+$ fields, running along the minus
direction. They are denoted by
\be
U^-_{[a,x]} = {\cal P}\,\exp\left(-ig\int_{a^-}^{x^-} {\rm d}\zeta^-
\,A^+(\zeta^-,x^+,x_\st)\right),
\ee
and satisfy
\be
\partial_x^+\,U^-_{[a,x]} = U^-_{[a,x]}\,D^+(x).
\ee
We then obtain
\bea
ig\,U^-_{[a,x]}\,G^{+\alpha}(x)\,U^-_{[x,a]}  & = &
\bigl[\,U^-_{[a,x]}\,iD^+(x)\,U^-_{[x,a]} ,
U^-_{[a,x]}\,gA_\st^\alpha(x)\,U^-_{[x,a]}\,\bigr]
-ig\,U^-_{[a,x]}\,\bigl[ \partial_x^\alpha , A^+(x)\,\bigr]\,U^-_{[x,a]}
\nonumber
\\ && \nonumber \\
& = &
\bigl[\,i\partial_x^+ ,
U^-_{[a,x]}\,gA_\st^\alpha(x)\,U^-_{[x,a]}\,\bigr]
-ig\,U^-_{[a,x]}\,\bigl[ \partial_x^\alpha , A^+(x)\,\bigr]\,U^-_{[x,a]} .
\eea
Thus we find
\be
\bigl[\,\partial^+ ,
U^-_{[a,x]}\,A_\st^\alpha(x)\,U^-_{[x,a]}\,\bigr] =
\,U^-_{[a,x]}\,\left(G^{+\alpha}(x)
+ \bigl[ \partial_x^\alpha , A^+(x)\,\bigr]\right)
\,U^-_{[x,a]} ,
\ee
which is the relation needed to express the transverse gluon fields
in terms of the field strength. In particular one has
\be
U^-_{[\infty,\xi]}A_\st^\alpha(\xi)U^-_{[\xi,\infty]}
- A_\st^\alpha(\infty^-) =
\int_\infty^{\xi^-} d\eta^-\ U^-_{[\infty,\eta]}\left(G^{+\alpha}(\eta)
+ [\partial_\st^\alpha , A^+(\eta^-)]\right)U^-_{[\eta,\infty]} ,
\label{linkid}
\ee
or in $A^+ = 0$ gauge
\be
A_\st^\alpha(\xi)
- A_\st^\alpha(\infty^-) =
\int_\infty^{\xi^-} d\eta^-\ G^{+\alpha}(\eta) .
\label{difference}
\ee

\section{Parametrizations of transverse momentum dependent functions}
\label{appendix-3}

The parametrization of $\Phi(x,p_\st)$ for a spin 0 or
spin 1/2 target,
consistent with the conditions
imposed by hermiticity and parity, including the
parts proportional to $(M/P^+)^0$ and $(M/P^+)^1$ is given by
\begin{eqnarray}
\Phi(x,\bm p_\st) &\ =  \ &
\frac{1}{2} \Biggl\{
f_1(x,\bm p_\st^2)\,\slash n_+
+ g_{1s}(x,\bm p_\st)\,\gamma_5\,\slash n_+
\nonumber \\ & & \quad
+ h_{1T}(x,\bm p_\st^2)\,\frac{\gamma_5\,[\slash S_\st,\slash n_+]}{2}
+ h_{1s}^\perp(x,\bm p_\st)\,
\frac{\gamma_5\,[\slash p_\st,\slash n_+]}{2M}
\nonumber \\ & & \quad
+f_{1T}^\perp(x,\bm p_\st^2)\, \frac{\epsilon_{\mu \nu \rho \sigma}
\gamma^\mu n_+^\nu p_\st^\rho S_\st^\sigma}{M}
+ h_1^\perp(x,\bm p_\st^2)\,\frac{i\,[\slash p_\st,\slash n_+]}{2M}
\Biggr\}
\nonumber \\ & &
+ \frac{M}{2P^+} \Biggl\{
e(x,\bm p_\st^2)
+ f^\perp(x,\bm p_\st^2)\,\frac{\slash p_\st}{M}
+ g_{T}^\prime(x,\bm p_\st^2)\,\gamma_5\,\slash S_\st
\nonumber \\ & & \qquad\quad
+ g_{s}^\perp(x,\bm p_\st)\, \frac{\gamma_5\,\slash p_\st}{M}
+ h_T^\perp(x,\bm p_\st^2)
\,\frac{\gamma_5\,[\slash S_\st,\slash p_\st ]}{2M}
+ h_{s}(x,\bm p_\st)\,\frac{\gamma_5\,[\slash n_+,\slash n_-]}{2}
\nonumber \\ & & \qquad\quad
-f_T(x,\bm p_\st^2)\,\epsilon_\st^{\rho\sigma} \gamma_\rho S_{\st\sigma}
- S_{\sL}\,f_L^\perp(x,\bm p_\st^2)\,\frac{\epsilon_\st^{\rho\sigma}
\gamma_\rho p_{\st\sigma}}{M}
\nonumber \\ & & \qquad\quad
- e_s(x,\bm p_\st) \, i\gamma_5
+ h(x,\bm p_\st^2)\,\frac{i\,[\slash n_+,\slash n_-]}{2}
\Biggr\} .
\label{parametrization}
\end{eqnarray}
Here the spin vector is defined
\be
S = S_\sL\left(\frac{P^+}{M}\,n_+ - \frac{P^-}{M}\,n_-\right) + S_\st,
\ee
($S = 0$ for spin 0) and we have used the shorthand notation
\be
g_{1s}(x, \bm p_\st) \equiv
S_{\sL}\,g_{1L}(x ,\bm p_\st^2)
+ g_{1T}(x ,\bm p_\st^2)\,\frac{(\bm p_\st\cdot\bm S_\st)}{M} ,
\ee
and similarly for $h^\perp_{1s}$, $g_s^\perp$ and
$h_s$. We note that all noncontracted $p_\st$-dependence (including
appearance of dot products like $p_\st\cdot S_\st$) is treated
explicitly, leaving functions depending
on $p_\st^2$, which is important in distinguishing T-even
and T-odd behavior.
The structures multiplying the `twist-two'
functions $f_1$, $g_{1s}$, $h_{1T}$, $h_{1s}^\perp$ or the `twist-three'
functions $e$, $f^\perp$, $g_T^\prime$, $g_s^\perp$, $h_T^\perp$,
$h_s$ are T-even (satisfying Eq.~\ref{trev}). The structures multiplying
the `twist-two' functions $f_{1T}^\perp$, $h_1^\perp$ or the `twist-three'
functions $f_T$, $f_L^\perp$, $e_s$, $h$ are T-odd (satisfying Eq.~\ref{trev}
with an additional minus sign).

As notation in the parametrizations for fragmentation
we employ capital letters, to be precise
%$f \rightarrow 2z\,D$, $g \rightarrow 2z\,G$, $h \rightarrow 2z\,H$ and
%$e \rightarrow 2z\,E$.
\begin{eqnarray}
\Delta(z,k_\st) &\ = \ &
zD_1(z,-z\bm k_\st)\,\slash n_-
+ zG_{1s}(z,-z\bm k_\st)\,\gamma_5\,\slash n_-
\nonumber \\ & &
+ zH_{1T}(z,-z\bm k_\st)\,\frac{\gamma_5\,[\slash S_{h\st}, \slash n_- ]}{2}
+ zH_{1s}^\perp(z,-z\bm k_\st)\,
\frac{\gamma_5\, [\slash k_\st , \slash n_- ]}{2M_h}
\nonumber \\ & &
+zD_{1T}^\perp(z,-z\bm k_\st)\, \frac{\epsilon_{\mu \nu \rho \sigma}
\gamma^\mu n_-^\nu k_\st^\rho S_{h\st}^\sigma}{M_h}
+ zH_1^\perp(z,-z\bm k_\st)\,\frac{i\,[\slash k_\st ,\slash n_- ]}{2M_h}
\nonumber \\ & &
+ \frac{M_h}{P_h^-} \Biggl\{ zE(z,-z\bm k_\st)
+ zD^\perp(z,-z\bm k_\st)\,\frac{\slash k_\st}{M_h}
+ zG_{T}^\prime(z,-z\bm k_\st)\,\gamma_5\,\slash S_{h\st}
\nonumber \\ & & \qquad\quad
+ zG_{s}^\perp(z,-z\bm k_\st)\, \frac{\gamma_5\,\slash k_\st}{M_h}
+ zH_T^\perp(z,-z\bm k_\st)\,
\frac{\gamma_5\, [\slash S_{h\st} ,\slash k_\st ]}{2M_h}
+ zH_{s}(z,-z\bm k_\st)\,\frac{\gamma_5\,[\slash n_- ,\slash n_+ ]}{2}
\nonumber \\ & & \qquad\quad
+zD_T(z,-z\bm k_\st)\,\epsilon_\st^{\rho\sigma}\gamma_\rho S_{h\st\,\sigma}
+ S_{h\sL}\,zD_L^\perp(z,-z\bm k_\st)\,\frac{\epsilon_\st^{\rho\sigma}
\gamma_\rho k_{\st\,\sigma}}{M_h}
\nonumber \\ & & \qquad\quad
- zE_{s}(z,-z\bm k_\st) \, i\gamma_5
+ zH(z,-z\bm k_\st)\,\frac{i\,[\slash n_- , \slash n_+ ]}{2}
\Biggr\} .
\end{eqnarray}
Here the spin vector is defined
\be
S_h = S_{h\sL}\left(\frac{P_h^-}{M_h}\,n_- - \frac{P_h^+}{M}\,n_+\right)
+ S_{h\st},
\ee
($S_h = 0$ for spin 0) and we have used the shorthand notation
$G_{1s}$, etc.,
\be
G_{1s}(z,-z\bm k_\st) = S_{h\sL}\,G_{1L}(z,-z\bm k_\st)
+ G_{1T}(z,-z\bm k_\st)\,\frac{(\bm k_\st\cdot\bm S_{h\st})}{M_h} .
\ee
The second argument of the fragmentation functions is chosen to be
$k_\st^\prime = -z\,k_\st$, which is the transverse momentum of the
produced hadron with respect to the quark in a frame in which the
quark does not have transverse momentum. In fact the functions only
depend on $k\st^{\prime 2}$. As for the distribution functions a
division can be made into T-even functions ($D_1$, $G_{1s}$, $H_{1T}$,
$H_{1s}^\perp$, $E$, $D^\perp$, $G_T^\prime$, $G_s^\perp$, $H_T^\perp$,
$H_s$) and T-odd functions ($D_{1T}^\perp$, $H_1^\perp$, $D_T$, $D_L^\perp$,
$E_s$, $H$).


\begin{thebibliography}{99}

\bibitem{RS-79}
J.P. Ralston and D.E. Soper, Nucl.~Phys.~B 152 (1979) 109.

\bibitem{Tangerman-Mulders-95a}
R.D. Tangerman and P.J. Mulders, Phys.~Rev.~D 51 (1995) 3357.

\bibitem{s90}
D. Sivers, Phys. Rev. D 41, 83 (1990); Phys.~Rev.~D 43 (1991) 261.

\bibitem{Collins-93b}
J.C. Collins, Nucl.~Phys.~B 396 (1993) 161.

\bibitem{Anselmino}
M. Anselmino, M. Boglione and F. Murgia, Phys.~Lett.~B362 (1995) 165;
M. Anselmino and F. Murgia, Phys.~Lett.~B442 (1998) 470;
M. Anselmino, U. D'Alesio, F. Murgia, Phys.~Rev.~D 67 (2003) 074010.

\bibitem{Boer-Mulders-98}
D. Boer and P.J. Mulders, Phys.~Rev.~D 57 (1998) 5780.

\bibitem{BHS}
S.J.~Brodsky, D.S.~Hwang and I.~Schmidt, Phys.~Lett.~B 530 (2002) 99.

\bibitem{Collins-02}
J.C.~Collins, Phys.~Lett.~B 536 (2002) 43.

\bibitem{CS82}
J.C.~Collins and D.E.~Soper, Nucl.~Phys.~B 194 (1982) 445.

\bibitem{CSS83}
J.C. Collins, D.E. Soper and G. Sterman, Phys.~Lett.~B 109 (1982) 388;
Nucl.~Phys.~B 223 (1983) 381.

\bibitem{Rad-Efr}
A.V. Efremov and A.V. Radyushkin, Theor.~Math.~Phys.~44 (1981) 774.

\bibitem{Boer-Mulders-00}
D. Boer and P.J. Mulders, Nucl.~Phys.~B 569 (2000) 505.

\bibitem{Ji-Yuan}
X.~Ji and F.~Yuan, Phys.~Lett.~B 543 (2002) 66.

\bibitem{Belitsky}
A.V.~Belitsky, X.~Ji and F.~Yuan, Nucl.~Phys.~B 656 (2003) 165.

\bibitem{BHS2}
S.J.~Brodsky, D.S.~Hwang and I.~Schmidt, Nucl.~Phys.~B 642 (2002) 344.

\bibitem{Boer-Sudakov}
D. Boer, Nucl.~Phys.~B 603, 195 (2001).

\bibitem{EFP-83}
R.K.~Ellis, W.~Furmanski and R.~Petronzio, Nucl.~Phys.~B 212 (1983) 29;
Nucl.~Phys.~B 207 (1982) 1.

\bibitem{Efremov-Teryaev-84}
A.V.~Efremov and O.V.~Teryaev, Sov.~J.~Nucl.~Phys.~39 (1984) 962.

\bibitem{MT96}
P.J. Mulders and R.D. Tangerman, Nucl.~Phys.~B 461 (1996) 197;
Nucl.~Phys.~B 484 (1997) 538 (E).

\bibitem{Boer-97}
D. Boer, R. Jakob and P.J. Mulders, Nucl.~Phys.~B 504 (1997) 345;
Phys.~Lett.~B 424 (1998) 143.

\bibitem{Boer-Brodsky-Hwang}
D. Boer, S.J. Brodsky, D.S. Hwang, Phys.~Rev.~D 67 (2003) 054003.

\bibitem{Gamberg}
L.P.~Gamberg, G.R.~Goldstein, K.A.~Oganessyan,
Phys.~Rev.~D 67 (2003) 071504.

\bibitem{Jaffe-83}
R.L. Jaffe, Nucl.~Phys.~B 229 (1983) 205.

\bibitem{Diehl-98}
M. Diehl and T. Gousset,
Phys.~Lett.~B 428 (1998) 359.

\bibitem{LM94}
J. Levelt and P.J. Mulders, Phys. Lett. B 338 (1994) 357.

\bibitem{Boer4}
D. Boer, P.J. Mulders and O.V. Teryaev, Phys.~Rev.~D 57, 3057 (1998).

\bibitem{Ji-gaugeinvariantize}
X.~Ji, J.~P.~Ma and F.~Yuan, Nucl.~Phys.~B 652 (2003) 383.

\bibitem{Soffer}
J.~Soffer, Phys.~Rev.~Lett.~74, 1292 (1995).

\bibitem{BBHM}
A. Bacchetta, M. Boglione, A. Henneman and P.J. Mulders,
Phys. Rev. Lett.~85 (2000) 712

\bibitem{BHMSS-02}
S.J. Brodsky, P. Hoyer, N. Marchal, S. Peigne and
F. Sannino, Phys.~Rev.~D 65, 114025 (2002).

\bibitem{QS-91b}
J. Qiu and G. Sterman, Phys.~Rev.~Lett.~67, 2264 (1991);
Nucl.~Phys.~B 378, 52 (1992).

\bibitem{Hammon-97}
N. Hammon, O. Teryaev and A. Sch\"afer, Phys.~Lett.~B 390 (1997) 409.

\bibitem{qs}
J. Qiu and G. Sterman, Phys.~Rev.~D 59 (1999) 014004.

\bibitem{KK01}
Y. Kanazawa and Y. Koike, Phys.~Rev.~D 64 (2001) 034019.

\bibitem{DBQiu}
D. Boer and J. Qiu, Phys.~Rev.~D 65 (2002) 034008.

\bibitem{Kotzinian95}
A. Kotzinian, Nucl. Phys. B441 (1995) 234; A.M. Kotzinian and P.J. Mulders,
Phys. Lett.~B 406 (1997) 373

\bibitem{BoM-99}
M. Boglione and P.J. Mulders, Phys. Rev.~D 60 (1999) 054007.

\bibitem{Basu}
R. Basu, A.J. Ramalho and G. Sterman, Nucl.~Phys.~B 244 (1984) 221.

\bibitem{Collins-Soper-frame}
J.C.~Collins, D.E.~Soper, Phys.~Rev.~D 16 (1977) 2219.

\bibitem{MOS}
R. Meng, F.I. Olness and D.E. Soper, Nucl. Phys.~B 371 (1992) 79.

\bibitem{Henneman}
A.A.~Henneman, D.~Boer and P.J.~Mulders, Nucl.~Phys.~B 620, 331 (2002).

\bibitem{Jaffe-Ji-92}
R.L.~Jaffe and X.~Ji, Nucl.~Phys.~B 375 (1992) 527.

\bibitem{Bukhvostov}
A.P.~Bukhvostov, E.A.~Kuraev and L.N.~Lipatov,
Sov.~Phys.~JETP 60 (1984) 22.

\bibitem{joao}
R. Jakob, P.J. Mulders and J. Rodrigues, Nucl.~Phys.~A 626 (1997) 937.

\bibitem{Goeke-Metz}
K.~Goeke, A.~Metz, P.V.~Pobylitsa and M.V.~Polyakov, hep-ph/0302028.

\bibitem{Drago}
M.~Anselmino, V.~Barone, A.~Drago and F.~Murgia, hep-ph/0209073.

\bibitem{goeke}
A.V.~Efremov, K.~Goeke and P.V.~Pobylitsa, Phys. Lett. B 488 (2000) 182;
P. Schweitzer et al., Phys. Rev. D 64 (2001) 034013.

\bibitem{BKMM}
A.~Bacchetta, R.~Kundu, A.~Metz and P.J.~Mulders,
Phys. Rev.~D 65 (2002) 094021

\bibitem{Metz}
A.~Metz, Phys. Lett. B 549 (2002) 139.

\end{thebibliography}
\end{document}